\documentclass[preprint,superscriptaddress,floatfix]{revtex4-2}

\usepackage{graphicx,footmisc}
\usepackage[caption=false]{subfig}
\usepackage{color}
\usepackage{natbib,url}
\usepackage{amsmath}
\usepackage{float}
\usepackage[normalem]{ulem}
\usepackage{booktabs}
\usepackage{caption}
\usepackage{placeins}
\usepackage{extarrows}
\usepackage{epstopdf}
\usepackage{units}

\makeatletter
\newcommand{\raisemath}[1]{\mathpalette{\raisem@th{#1}}}
\newcommand{\raisem@th}[3]{\raisebox{#1}{$#2#3$}}
\makeatother

\def\be{\begin{equation}}
\def\ee{\end{equation}}

\def\begineqn{\begin{equation*}}
\def\endeqn{\end{equation*}}

\def\beginar{\begin{eqnarray}}
\def\endar{\end{eqnarray}}
\def\beginarn{\begin{eqnarray*}}
\def\endarn{\end{eqnarray*}}

\def\lb{\left ( }
\def\rb{\right ) }

\def\ub{\mathbf{u}}

\def\bb{\mathbf{b}}

\def\dst{{\partial_t}}

\def\dsz{{\partial_z}}

\renewcommand{\i}{^{(i)}}
\renewcommand{\o}{^{(o)}}

\def\mT{\overline{T}}

\newcommand{\ve}{\varepsilon}

\begin{document}

\title{Boundary layers in quasi-static magnetoconvection with a vertical field and their implications for heat transport}

\author{Talal AlRefae}
\affiliation{ 
Department of Physics, University of Colorado, Boulder, Colorado 80309, USA
}%

\author{Michael A. Calkins}
\affiliation{ 
Department of Physics, University of Colorado, Boulder, Colorado 80309, USA
}%

\begin{abstract}
Heat transport in quasi-static magnetoconvection with a vertical magnetic field in a plane layer geometry is investigated with direct numerical simulations and asymptotic theory in the limit of large Chandrasekhar number ($Q$). 
It is shown that thermal and magnetic boundary layers with thickness $O(Q^{-1/6})$, of the same order as the horizontal scale of the convection, are persistent over the range of investigated parameters. 
Thermal boundary layer control of the heat transport indicates that the Nusselt number depends on $Q$, suggesting that no asymptotic state of heat transport independent of $Q$ occurs in this system.
The magnetic boundary layers necessitate leading order modifications to the dominant vertical force balance as well as the amplitude of the horizontal induced magnetic field near the boundaries. The scaling behavior of the thermal boundary layer and resulting heat transport is found to be independent of the choice of mechanical and electromagnetic boundary conditions.
\end{abstract}

\maketitle

\section{Introduction}


Convection in the presence of magnetic field is important in many geophysical and astrophysical fluid systems. Indeed, most planets and stars possess electrically conducting fluid regions that, when driven sufficiently, support self-sustaining electromagnetic fields through the dynamo effect \citep{cJ11b,kS25}. 
In contrast to dynamos, magnetoconvection (MC) systems, whereby an externally imposed magnetic field permeates the fluid, allows for direct experimental control of the direction and magnitude of the magnetic field \citep[e.g.][]{sC00,jmA01,uB01,tZ16,wL18,mY19,rA20,tV21,yX23,sB23,iC23,shB23,aT24,jN22,chmc23}. With regard to applications to natural systems, MC in a plane layer geometry can be thought of as a local model for investigating the dynamics of large-scale dynamos that are characterized by a component of the magnetic field that varies on a system (i.e.~global) scale, and a component that varies on the same scale as the fluid motions \citep[][]{kM19}. 
A common difficulty encountered with both laboratory experiments and numerical simulations of MC, however, is that the accessible parameter space is typically limited to regimes that remain distant from those of natural systems. An overarching effort, then, is to study the behavior of the system as the parameters are made more extreme. Towards this end, one of the primary goals of the present work is to understand how heat transport in MC depends on the imposed field strength as it is made asymptotically large, as this is the regime of interest for planets and stars. Our results show that, at a fixed value of the thermal forcing, heat transport does not asymptote to a constant value independent of the imposed magnetic field strength  primarily due to the development of asymptotic thermal boundary layers which control heat transport.


In the present work we focus on the plane layer geometry in which an electrically conducting fluid layer of depth $H$ is confined between plane parallel boundaries. The strength of the imposed magnetic field is controlled with the non-dimensional Chandrasekhar number,
\be
Q = \frac{B^2 H^2}{\rho \mu \nu \eta},
\ee
where $B$ is the magnitude of the imposed magnetic field, $\rho$ is the fluid density, $\mu$ is the magnetic permeability, $\nu$ is the kinematic viscosity and $\eta$ is the magnetic diffusivity. Here we investigate the problem of a constant imposed magnetic field that points in the vertical (i.e.~parallel to gravity) direction, which we refer to as vertical-MC (VMC). 
The buoyancy force is controlled with the Rayleigh number,
\be
Ra = \frac{g \alpha \Delta T H^3}{\nu \kappa} ,
\ee
where $g$ is the magnitude of the (constant) gravitational field, $\alpha$ is the thermal expansion coefficient, $\Delta T$ is the temperature difference between the top and bottom boundaries and $\kappa$ is the thermal diffusivity. 
The electrically conducting regions of planets and stars are characterized by $Q \gtrsim 10^{16}$ and $Ra \gg 1$ \citep{mO03,kS25}, thus necessitating our need for understanding the asymptotic limit of $Q \rightarrow \infty$.

The linear asymptotic theory for VMC in a plane layer geometry shows that convection is stabilized by the imposed magnetic field, with the critical Rayleigh number scaling as $Ra_c = O(Q)$ \cite{sC61}. The curve characterizing the state of marginal convection exhibits a minimum for horizontal wavenumbers of size $k = O(Q^{1/6})$, implying that fluid motions become anisotropic. Moreover, there is a more stable, higher wavenumber, branch of the marginal curve in which $k = O(Q^{1/4})$. Asymptotic theory has also been applied to nonlinear VMC for both the $k = O(Q^{1/6})$ \cite{pM99} and $k = O(Q^{1/4})$ \cite{kJ99c} branches. Ref.~\cite{pM99} focused on single mode solutions in which the sole nonlinear term is the convective heat flux appearing in the horizontally averaged heat equation. In contrast, Ref.~\cite{kJ99c} first reduced the complete governing equations to a fully nonlinear model, but also presented results for single mode solutions only. These asymptotic models have therefore focused on simplified dynamics of VMC in specific regions of wavenumber space, though tests of this theory in fully nonlinear, multimodal simulations is lacking. Thus, a secondary goal of the present work is to characterize the asymptotic scaling behavior of VMC in the strongly driven regime that is accessible to direct numerical simulations.

%

One of the primary goals in convection studies is to quantify how heat transport depends on system input parameters. The Nusselt number, which is a non-dimensional measure of heat transport, is typically used for this purpose and defined by
\be \label{nudef}
Nu = 1 + Pr \langle w \theta \rangle_t,
\ee
where the angled brackets with subscript $t$ denote an average over volume and time, the Prandtl number is defined by $Pr = \nu / \kappa$, and the vertical velocity and fluctuating temperature are denoted by $w$ and $\theta$, respectively. In the present study we fix $Pr = 1$, and we consider the quasi-static limit of MC in which case the magnetic Prandtl number, $Pm = \nu / \eta$, is assumed to be asymptotically small such that it drops out of the governing equations \citep{roberts2007magnetohydrodynamics}. Note that in the quasi-static limit, the dynamics of VMC are not strongly dependent on the value of $Pr$ \citep[e.g.][]{mY19}. Thus, we are interested in how $Nu$ varies with $Q$ and $Ra$ only.

Numerous scaling theories have been proposed which aim to relate $Nu = Nu(Ra)$ in the context of turbulent convection, both in the absence and presence of global external stabilizing forces that arise from rotation or magnetic fields. Building on the Grossmann-Lohse scaling theory for turbulent convection \cite{sG00}, Ref.~\cite{zu16,zu20} extended this framework to the turbulent quasi-static VMC regime relevant for low-$Pr$, liquid-metal flows. The additional source of dissipation (ohmic) was incorporated and four distinct transport regimes were identified based on the relative dominance of bulk and boundary-layer dissipation, as distinguished by the kinetic Hartmann and thermal boundary layers. 

Previous work has shown that $Nu$ scales more strongly with $Ra$ as $Q$ is increased \citep{sC00,jmA01,uB01,mY19,rA20,shB23}. For instance, assuming a power law dependence of the form $Nu = (Ra/Ra_c)^{\gamma}$, where $Ra_c$ is the critical Rayleigh number, Ref.~\cite{mY19} found that $\gamma$ is an increasing function of $Q$. Some studies have argued that $\gamma \rightarrow 1$ as $Q \rightarrow \infty$ \citep[e.g.][]{sB06}. This value of $\gamma$ is appealing because it leads to a heat transport scaling law that is independent of viscosity, which is thought to be the relevant regime for natural systems since they are typically characterized by strongly turbulent dynamics. However, in the present study we argue that $\gamma$ is not approaching unity in the large-$Q$ limit. Rather, our findings show that a thermal boundary layer of thickness $O(Q^{-1/6})$ develops in this system, thus indicating that there is no bound on $\gamma$. An important consequence of this finding is that the heat transport remains controlled by viscosity in the strongly forced regime so long as the system remains magnetically constrained.

In the present work we use output from a broad suite of direction numerical simulations to understand how heat transport in VMC scales in the strong field limit, i.e.~$Q\rightarrow \infty$. The governing equations and numerical methods are discussed in section \ref{S:methods}. In section \ref{S:asymp} we discuss the asymptotic theory that is shown to be relevant to this system. Numerical results and comparison with asymptotic theory are presented in section \ref{S:results} and a discussion is given in section \ref{S:conclude}.

\section{Methods}
\label{S:methods}

In non-dimensional form, the equations governing quasi-static VMC for an Oberbeck-Boussinesq fluid are given by 
\begin{gather}
\partial_t \mathbf{u} + \mathbf{u} \cdot \nabla \mathbf{u}=-\nabla p +Q \partial_z \mathbf{b}+\frac{R a}{P r} T \hat{\mathbf{z}} + \nabla^2 \mathbf{u} \label{momeq} , \\ 
\dst T + \ub \cdot \nabla T = \frac{1}{Pr} \nabla^2 T \label{heateq}, \\ 
0 = \partial_z \mathbf{u} + \nabla^2 \mathbf{b}  \label{induceq}, \\
\nabla \cdot \ub =0, \quad \nabla \cdot \bb =0 \label{div}.
\end{gather}
In our Cartesian coordinate system $(x,y,z)$, $\ub = (u, v, w)$ is the velocity field, $p$ is the reduced pressure, $\bb =(b_x, b_y, b_z)$ is the induced magnetic field and $T$ is the temperature. The equations have been non-dimensionalized with the depth of the fluid layer $H$, viscous diffusion time $H^2/\nu$, magnetic field scale $B$, and temperature scale $\Delta T $. 

The thermal boundary conditions are isothermal on both boundaries such that 
\be \label{isotherm}
T =1 \quad \text { at } \quad z=0, \quad \text { and } \quad T =0 \quad \text { at } \quad z=1 .
\ee
The majority of the simulations employ impenetrable, stress-free mechanical boundary conditions given by
\begin{equation} \label{sf}
w= \partial_z u=\partial_z v=0 \quad \text { at } \quad z=0,1.
\end{equation}
Impenetrable, no-slip boundary conditions are also used in a small subset of simulations so that
\begin{equation} \label{ns}
\mathbf{u}= 0 \quad \text { at } \quad z=0,1.
\end{equation}
For most simulations we use electrically insulating boundary conditions in which the magnetic field is matched to an external potential field; details for implementing this approach are given in Ref.~\cite{cJ00b}. Another subset of simulations use boundary conditions such that the induced magnetic field is purely vertical at the boundaries,
\begin{equation} \label{vf}
b_x=b_y= \partial_z b_z = 0 \quad \text { at } \quad z=0,1 .
\end{equation}
The horizontal dimensions are periodic in all simulations.

The equations are solved numerically using a psuedo-spectral code which decomposes the dependent variables using Chebyshev polynomials in the vertical dimension and Fourier series in the horizontal dimensions. The velocity and magnetic field vectors are represented in terms of toroidal and poloidal scalars, thereby satisfying the solenoidal constraints exactly. A third-order mixed implicit/explicit timestepping scheme is employed. For more details on the numerical scheme, see \cite{pM16}.

The parameter space covered in the present work is similar to that in \cite{mY19}; the full details of the simulation parameters are provided as supplementary material. The range of input parameters covered is $Q \in [10^4, 10^8]$ and $Ra \in [2\times 10^5, 8 \times 10^{10}]$. For the majority of the simulations the horizontal dimensions of the domain are chosen such that $10 \lambda_c \times 10 \lambda_c$, where $\lambda_c = 2\pi/k_c$ is the critical wavelength. For the most extreme cases pertaining to $Q = 10^8$ we use a domain size of $5 \lambda_c \times 5 \lambda_c$; tests confirmed that this size was sufficient to obtain converged global statistics. We reiterate that $Pr$ is fixed at unity for all of the simulations presented.

\subsection{Definitions and notation}

Here we collect useful definitions and notation for the analysis presented below. We denote the decomposition of an arbitrary field $X$ into a mean and fluctuating component as
\be \label{mfdecomp}
X(x,y,z,t) = \overline{X}(z,t) + X'(x,y,z,t),
\ee
where an overline denotes an average over a horizontal plane (with area $A$) defined by
\be
\overline{X}(z,t) = \frac{1}{A}\iint_A X(x,y,z,t) \, \mathrm{d}A.
\ee
By definition the decomposition of (\ref{mfdecomp}) implies that $\overline{X'} = 0$. 

The root-mean-square (rms) of a scalar field $X$ is defined as
\be \label{rms}
X_{\text{rms}} (z) = \frac{1}{\mathcal{T}} \frac{1}{A} \int_{A, \mathcal{T}} \sqrt{ (X')^2 } \, \mathrm{d}A \, \mathrm{d}t
\ee
where $\mathcal{T}$ is the total integration time. Volumetric averages are defined as 
\be
\langle X_\text{rms} \rangle = \frac{1}{h} \int_0^{h} X_\text{rms} \, \mathrm{d}z.
\ee

\section{Asymptotic Theory}
\label{S:asymp}


Elements of the asymptotic theory for VMC, for both the linear \cite{sC61} and the nonlinear single mode \cite{pM99} systems, are applicable to the results of the simulations; a brief overview is given here. Stress-free mechanical boundary conditions are assumed, though, as shown below, the results are to leading order uninfluenced by this choice. By considering infinitesimal perturbations about a stationary conductive state, the Rayleigh number characterizing the marginal state defined by a zero growth rate is given by \cite{sC61} 
\be
Ra_m = \pi^4 Q k^{-2} + \pi^2 Q  + \lb \pi^2 + k^2 \rb^3  k^{-2} ,
\label{E:marg}
\ee
where $k = \sqrt{k_x^2+k_y^2}$ is the modulus of the horizontal wavenumber. The form of the marginal curve is independent of the electromagnetic boundary conditions  \cite{sC61}. Each of the three terms appearing on the right hand side of equation \eqref{E:marg} corresponds to a specific range of wavenumbers of the marginal curve. The first two terms arise in the equivalent inviscid problem, indicating that viscosity plays no role in determining the stability at leading order. It can be seen that in the $Q \rightarrow \infty$ limit the minimum value of $Ra_m$, or critical Rayleigh number $Ra_c$, becomes $Ra_c \rightarrow \pi^2 Q$, indicating that the marginal curve is independent of wavenumber at leading order. To determine the critical wavenumber $k_c$ it is necessary to investigate the next order correction in which case the viscous force becomes important and leads to the third term in \eqref{E:marg}. Minimization of \eqref{E:marg} leads to the condition
\be
2k_c^6 + 3 \pi^2 k_c^4 - \pi^6 - \pi^4 Q = 0 .
\label{E:kcrit}
\ee
In the limit $Q \rightarrow \infty$ a dominant balance between the first and last terms is achieved and $k_c \rightarrow \lb \pi^4 Q / 2\rb^{1/6}$. Thus, the large $Q$ asymptotic scalings for the critical parameters are given by
\be
Ra_c = O(Q), \quad k_c = O(Q^{1/6}).
\ee
The relations above, which are independent of boundary conditions, signify that the imposed magnetic field results in two primary effects: (1) the flow is stabilized relative to hydrodynamic convection since the critical Rayleigh number increases as a function of $Q$; and (2) convection becomes anisotropic and small-scale in the horizontal plane. As found in previous work \citep[e.g.][]{mY19,iC23}, and discussed here, this anisotropy persists in the strongly nonlinear regime.


Matthews \cite{pM99} extended the linear asymptotic theory to finite amplitude solutions consisting of a single horizontal wavenumber (i.e.~single-mode) using stress-free, isothermal, and vertical-field magnetic boundary conditions. Both steady and oscillatory solutions were investigated; here we discuss only the steady solutions given their applicability to the quasi-static approximation employed in the present work. The restriction to a single wavenumber limits nonlinearities to the convective heat flux. Moreover, Matthews showed that while the wavenumber is assumed to be $O(Q^{1/6})$, its numerical value can be scaled out of the problem, which is consistent with the leading order linear asymptotic theory. Matthews defined the relevant small parameter as
\be
\ve \equiv Q^{-1/6} ,
\ee
which we also employ throughout. The anisotropic flow structure is captured in spatial derivatives as
\be \label{ani}
\partial_x = \partial_y = O(\ve^{-1}), \quad \partial_z = O(1).
\ee
Matthews used the following asymptotic scalings
\be
\begin{gathered} \label{matscaling} 
u = v = O(1), \quad w = O(\ve^{-1}), \quad \overline{T} = O(1), \quad \theta = O(\ve), \\
b_x = b_y = O(\ve^2), \quad b_z = O(\ve), \quad Ra = O(\ve^{-6}),
\end{gathered}
\ee
where the temperature is decomposed into mean and fluctuating components according to 
\be \label{tempdecomp}
T(x,y,z,t) = \overline{T}(z,t) + \theta(x,y,z,t).
\ee
The force balances, including the pressure gradient, were not addressed directly by Matthews since the momentum equation was curled. Nevertheless, the asymptotic scaling for each force is easily deduced. In particular, the Lorentz force enters the leading order balance in all three components in the momentum equation; in the vertical component it is balanced by buoyancy,
\be
0 \approx Q \dsz b_z + Ra \theta . \label{E:zmom}
\ee 
In the horizontal components of the momentum equation only the pressure gradient is sufficiently large to balance the Lorentz force, implying $p = O \lb \ve^{-3} \rb$, so that
\be
0 \approx \nabla_{\perp} p + Q \dsz \mathbf{b}_{\perp}, \label{h:zmom}
\ee 
where $\nabla_{\perp} = (\partial_x, \partial_y, 0)$ and $\mathbf{b}_{\perp} = \lb b_x, b_y, 0 \rb$. The above balance holds throughout the fluid layer \citep{mY19}, though the presence of boundary layers requires modification to \eqref{E:zmom}.

It is instructive to consider the linear, leading order asymptotic solutions implied by (\ref{matscaling}). Assuming steady, normal-mode solutions of the form
\be \label{singmode}
w = W(z) f(x,y), \quad \theta = \Theta(z) f(x,y), \quad b_z = B(z) f(x,y),
\ee
where $f(x,y)$ satisfies the two-dimensional Helmholtz equation, $$\nabla_{\perp}^2 f(x,y) = -k^2 f(x,y),$$ and $\nabla_{\perp}^2 = \partial_x^2 + \partial_y^2$ is the horizontal Laplacian operator.
Under these assumptions, equation \eqref{E:zmom}, the vertical component of the induction equation, as well as the linearized heat equation, become, respectively, 
\be
\begin{gathered}
 \partial_z B + \Theta = 0 \label{matvmom}, \quad -k^2 B + \partial_z W = 0, \quad -k^2 \Theta + W = 0.
\end{gathered}
\ee
These balances are satisfied by  
\be \label{outsol}
W = \sin \pi z, \quad \Theta = \frac{1}{k^2} \sin \pi z, \quad B = \frac{ \pi}{k^2} \cos \pi z.
\ee
In producing these solutions we have normalized the amplitude of the rescaled vertical velocity to be unity. Solutions for the horizontal components of the velocity and magnetic fields can be derived by imposing the respective solenoidal constraints. 

With the vertical field boundary conditions of (\ref{vf}) these leading order asymptotic solutions are consistent with all of the imposed mechanical and thermal boundary conditions. In the case of electrically insulating plates, boundary layer modifications to the solutions of (\ref{outsol}) are necessary in order to satisfy all of the imposed boundary conditions. We briefly discuss only the leading order boundary modifications in order to motivate the asymptotic boundary layers observed in the full numerical solutions of section \ref{S:results} as well as their repercussions on the overall flow dynamics. 

A standard approach, and that which is used here, is to partition the flow into two regions: first, an anisotropic bulk interior flow which satisfy the solutions of (\ref{outsol}) and the asymptotic scalings of (\ref{matscaling}); second, a dynamically distinct boundary region where leading order modifications to the induced magnetic field are required to satisfy the electromagnetic boundary conditions. This boundary region is dynamically distinct in the sense that it is primarily isotropic, characterized by a different leading order force balance, and modifies the asymptotic scaling of the horizontal magnetic field amplitudes. In the context of boundary layer theory it is common to refer to the region near the boundary layers as the `inner' region while referring the bulk interior away from the boundary layer as the `outer' region \cite[e.g.][]{cB10}.

The clear partitioning of the flow dynamics into two distinct regions is appealing as it provides a clear physical interpretation of the asymptotic flow dynamics. This partitioning may be attributable to the fact that the primary instability at the onset of convection is independent of the imposed electromagnetic boundary conditions \citep{sC61} thereby necessitating the existence of magnetic boundary layers so as to satisfy the magnetic boundary conditions. Asymptotic boundary layer analysis of this kind has also been shown to be useful in modeling Ekman layers in the case of rotationally constrained convection \citep[e.g.][]{wH71,mC15c, kJ16}.

\subsection{Boundary Layer Modifications} \label{T: b_mod}

Here we present the modifications required to satisfy the boundary conditions. For simplicity we focus on the linear problem, though as mentioned below this procedure also applies to the nonlinear problem.
In the case of electrically insulating plates, the induced magnetic field must satisfy 
\be \label{elecin}
\partial_z B= \pm k B, \quad \text{for} \quad z = 0,1
\ee
Near the boundaries, (\ref{elecin}) admits exponential solutions where $B \sim \exp({\pm kz})$. Exponential solutions of this form are present in the homogenous solutions of the full induction equation though they are suppressed in the anisotropic singular asymptotic limit implied by (\ref{ani}) and (\ref{matscaling}). In attempt to recover this suppressed branch of solutions near the boundaries, thereby satisfying electrical insulation, we define the stretched boundary coordinate $\zeta = z/\ve$. Since $k = O(\ve^{-1})$, an application of the chain rule allows us to expand the derivative operator as $$\partial_z \rightarrow \partial_z + \ve^{-1} \partial_\zeta.$$

Supposing that the vertical velocity is purely an outer variable such that $W(z,\zeta) =W(z)$, the dominant balance of the vertical induction equation near the boundaries using the scalings of (\ref{matscaling}) and the additional vertical scale $\zeta$ becomes
\be \label{17}
(\partial_\zeta^2 - 1) B\i + \frac{1}{k^2}\partial_zW = 0.
\ee
where the superscript $(i)$  denotes the `inner' solution near the boundaries. We consider the behavior near the bottom plate at $z=0$, though the solution for the top plate can be obtained by symmetry of $B$ under the transformation $\zeta \mapsto (1-\zeta)$. Near the boundaries, the inner limit of the vertical velocity is 
$(\partial_z W)^{(i)} =  \pi \cos\pi =  \pi $ to leading order. Then, equation (\ref{17}) becomes  
\be \label{binduc}
(\partial_\zeta^2 - 1) B\i + \frac{\pi}{k^2} = 0,
\ee
satisfied by
\be \label{binsol}
B^{(i)}(\zeta) = \frac{\pi}{k^2} \left(1 - \frac{1}{2} e^{-\zeta}\right).
\ee
In order to satisfy the two degrees of freedom of equation (\ref{binduc}), we have enforced electrical insulation near the bottom plate and as well as the matching condition: 
\be \label{matchb}
\lim_{\zeta\rightarrow \infty}B^{(i)}(\zeta) = \lim_{z\rightarrow 0}B^{(o)}(z).
\ee
Here, the superscript $(o)$ denotes the `outer' solution where $z = O(1)$. The exponential boundary layer inner solution of (\ref{binsol}), coupled with the sinusoidal outer solution of (\ref{outsol}), constitutes an asymptotically leading order solution of the vertical magnetic field with electrically insulating boundary conditions. 

Although the magnetic field solutions discussed above are strictly relevant to the linear asymptotic regime, it is worth noting that, under the quasi-static approximation, the induction equation (\ref{induceq}) is a linear partial differential equation. As such, the solution to the induction equation is guaranteed to be composed of homogenous and particular solutions. The particular solution is proportional to $\partial_z \mathbf{u}$ and will be subject to nonlinear modifications as $Ra$ increases. The homogenous solutions that we present, however, will continue to satisfy the homogeneous component of the induction equation no matter the level of supercriticality of the flow. Although nonlinear modifications to the wavenumber are possible as $Ra$ increases, we verify \textit{a posteriori} that, for the induced magnetic field, $k$ is not significantly modified by the increase in supercriticality.

An immediate consequence of the isotropic vertical scale can be seen from the solenoidal constraint on the induced magnetic field, which we can rewrite as
\begin{equation} \label{solb}
\nabla_{\perp} \cdot \mathbf{b}_{\perp} + \partial_z b_z = 0.
\end{equation}
If $\nabla_{\perp} = O(\varepsilon^{-1})$ and $\partial_z = O(1)$, as is the case in the bulk interior, then the appropriate ratio of the amplitude between the horizontal and vertical fields is
\begin{equation}
b_{\perp} \sim \varepsilon b_z \quad \text{for} \quad z = O(1) ,
\end{equation}
and hence $b_{\perp} = O(\varepsilon^2)$ and $b_z = O(\varepsilon)$ in the interior. Near the boundaries, assuming we have a boundary layer such that $\partial_\zeta = O(\varepsilon^{-1})$, then the dominant balance of (\ref{solb}) requires
\begin{equation}
b_{\perp} \sim b_z, \quad \text{for} \quad z = O(\ve).
\end{equation}
The solution of (\ref{binsol}) requires that $b_z$ remain $O(\varepsilon)$ both near the boundaries and throughout the interior, otherwise the solution would violate the matching condition of (\ref{matchb}). Thus, the isotropic $b_{\perp} = O(\varepsilon)$ relation would follow as a consequence of the isotropic geometry near the boundaries as well as the solenoidal constraint. This modified amplitude of $b_\perp$ is supported by the numerical solutions presented in section \ref{S:results}. 


The magnetic boundary layer modification above pertains only to the case of electrically insulating plates. These magnetic boundary layers are linear in the sense that they are expected to develop even at the onset of convection \citep{sC61}. However, regardless of the electromagnetic boundary conditions, we observe an isothermalization of the fluid interior as $Ra$ increases, indicative of a well-mixed convective interior. As a result of the background temperature profile's saturation with depth, thin thermal boundary layers develop near the boundaries in order to satisfy all of the imposed boundary conditions. These thermal boundary layers are primarily conductive and laminar \cite[e.g.][]{chini2009large,long2020thermal}. 

Conventional arguments for the characteristic scale of the thermal boundary layer stem from a balance of advection and diffusion of heat near the boundaries. Extending this argument to the VMC system, a balance of these two terms yields the relation
\be
\mathbf{u}\cdot \nabla T \sim \frac{1}{Pr} \nabla^2 T \quad \Rightarrow \quad \delta_\theta \sim \frac{1}{w} = O(\ve),
\ee
where we have assumed $Pr = O(1)$ and denote the thermal boundary layer thickness height as $\delta_\theta$. Notably, these boundary layers are magnetically constrained and asymptotic, exhibiting the same scaling as the magnetic boundary layer thickness despite the difference in physical origin. 

The additional vertical scale $\zeta$ requires boundary layer modifications to the asymptotic scaling of the fluctuating temperature field. To see this, we refer to the mean heat equation given by
\be \label{mheat}
\dst \mT + \partial_z \overline{ \left(w\theta \right)} = \frac{1}{Pr} \partial_z^2 \overline{T} ,
\ee
Achieving a statistically steady state requires that the convective heat flux and thermal diffusion balance in \eqref{mheat} such that
\be
\partial_z \overline{ \left(w\theta \right)} \sim  \frac{1}{Pr} \partial_z^2 \overline{T} .
\label{mbal}
\ee 
This balance is achieved within the thermal boundary layer of thickness $\delta_\theta$ indicating
\be
\frac{ Pr \, \overline{ w\theta}}{\delta_\theta} \sim \frac{\mT}{\delta_\theta^2}, \qquad \Rightarrow \qquad  \delta_\theta^{-1} \sim  \frac{Pr \,\overline{ w\theta}}{\overline{T}}  
\ee 
Using the fluid interior scalings given in \eqref{matscaling} and assuming $Pr=O(1)$ would then imply that $\delta_\theta = O(1)$, which is not consistent with the results presented later. Moreover, the necessary correction is not expected from a change in scaling of the vertical velocity within the vicinity of the boundary layer since the results indicate it is an outer variable only.  
One possible resolution is that the fluctuating temperature changes scaling near the boundary such that $\theta\i = O(1)$, and the balance given in \eqref{mbal} is then achievable with $\delta_\theta = O(\ve)$.  
In the limit of an isothermal core, the change in mean temperature occurs across the thin conductive thermal boundary layers so that
\be
Nu \sim \frac{1}{\delta_\theta} = O(\ve^{-1}).
\ee
The above relation implies that heat transport cannot be rendered independent of thermal and viscous diffusion as was previously conjectured. 




The scalings given above have consequences on the dominant force balances near the boundaries. The dominant vertical balance of (\ref{E:zmom}) cannot be directly extended to the boundaries without violating electrical insulation. In order to resolve this, we consider the dominant horizontal force balance. Expressing the dominant horizontal interior force balance in terms of the outer variables gives
\be \label{o1hforce}
0 = -\nabla_\perp p\o + Q \partial_z b_\perp\o .
\ee
which is consistent with all of the imposed boundary conditions and is therefore expected to hold near the boundaries as well as the fluid interior \citep{mY19}. As such, the amplitude of the pressure field will need to dynamically adjust so as to match the increased amplitude of the horizontal induced magnetic field in the vicinity of the boundaries. This balance will imply that, for $z = O(\ve)$,
\be
\nabla_\perp p\i \sim Q \partial_\zeta b_\perp\i \quad \Rightarrow \quad p\i = O(\ve^{-5}).
\ee
This increased amplitude of the pressure field causes it to enter into the leading order vertical force balance near the boundaries. This dominant vertical force balance near the boundaries is
\be
0 = -\partial_\zeta p\i  + Q \partial_\zeta b_z\i + Ra \theta\i 
\ee 
where we have assumed the modified scaling $\theta\i = O(1)$ near the boundaries. The addition of the pressure gradient into the dominant triad of vertical forces near the boundaries provides the extra degree of freedom necessary for the flow field to satisfy all of the imposed electromagnetic and thermal boundary conditions simultaneously.

\section{Results} 
\label{S:results}

\subsection{Heat transport and thermal boundary layers}

\begin{figure}
\centering
\subfloat[][]{\includegraphics[width=0.49\textwidth]{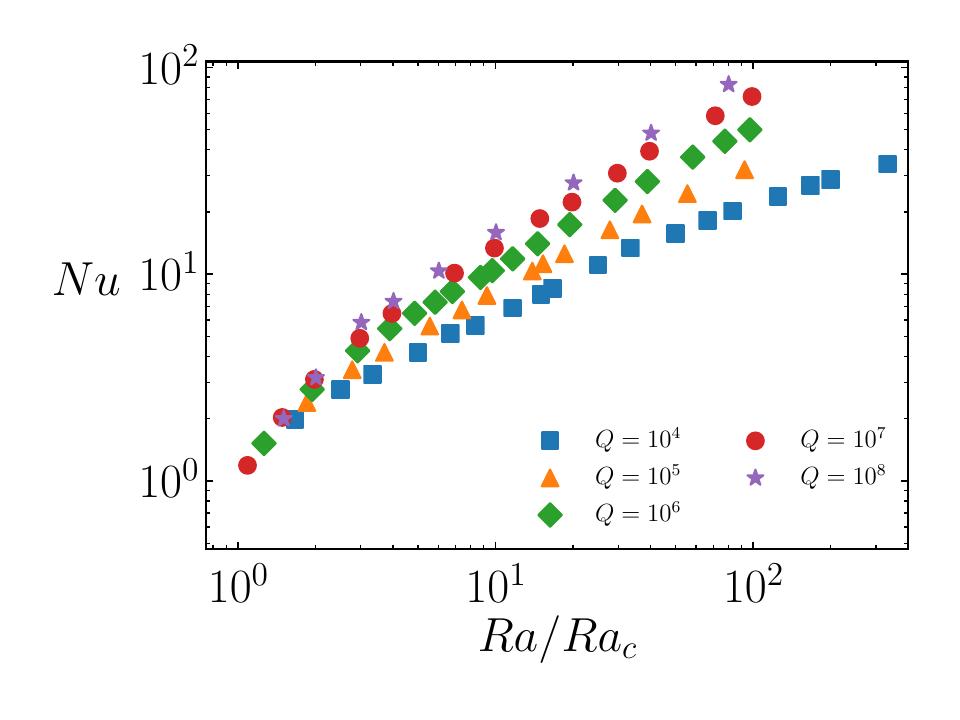}}
\subfloat[][]{\includegraphics[width=0.49\textwidth]{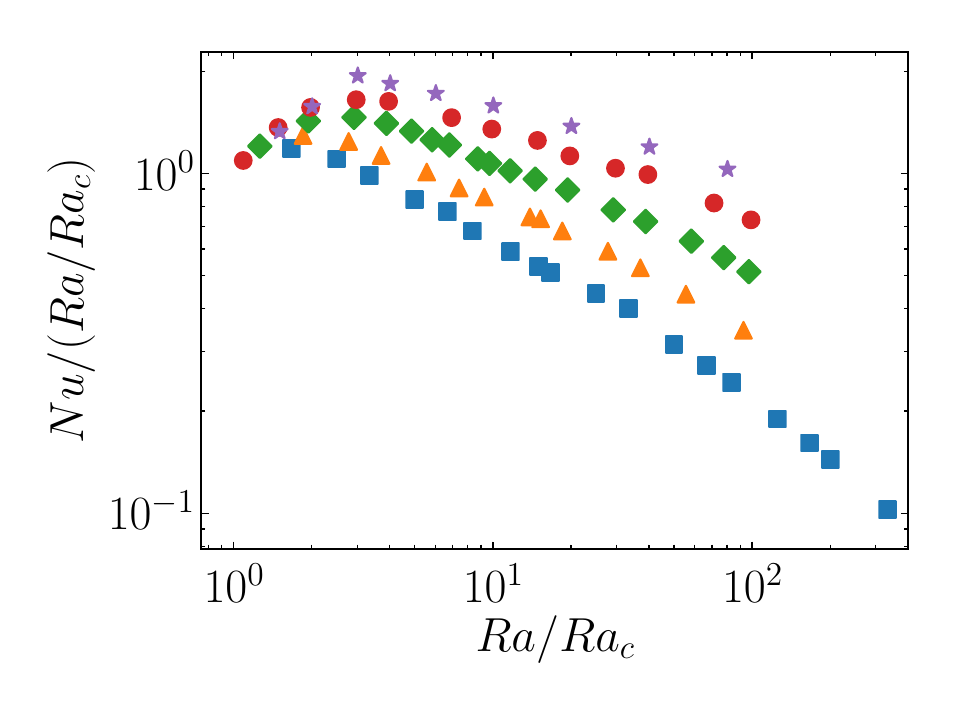}} \\
\subfloat[][]{\includegraphics[width=0.49\textwidth]{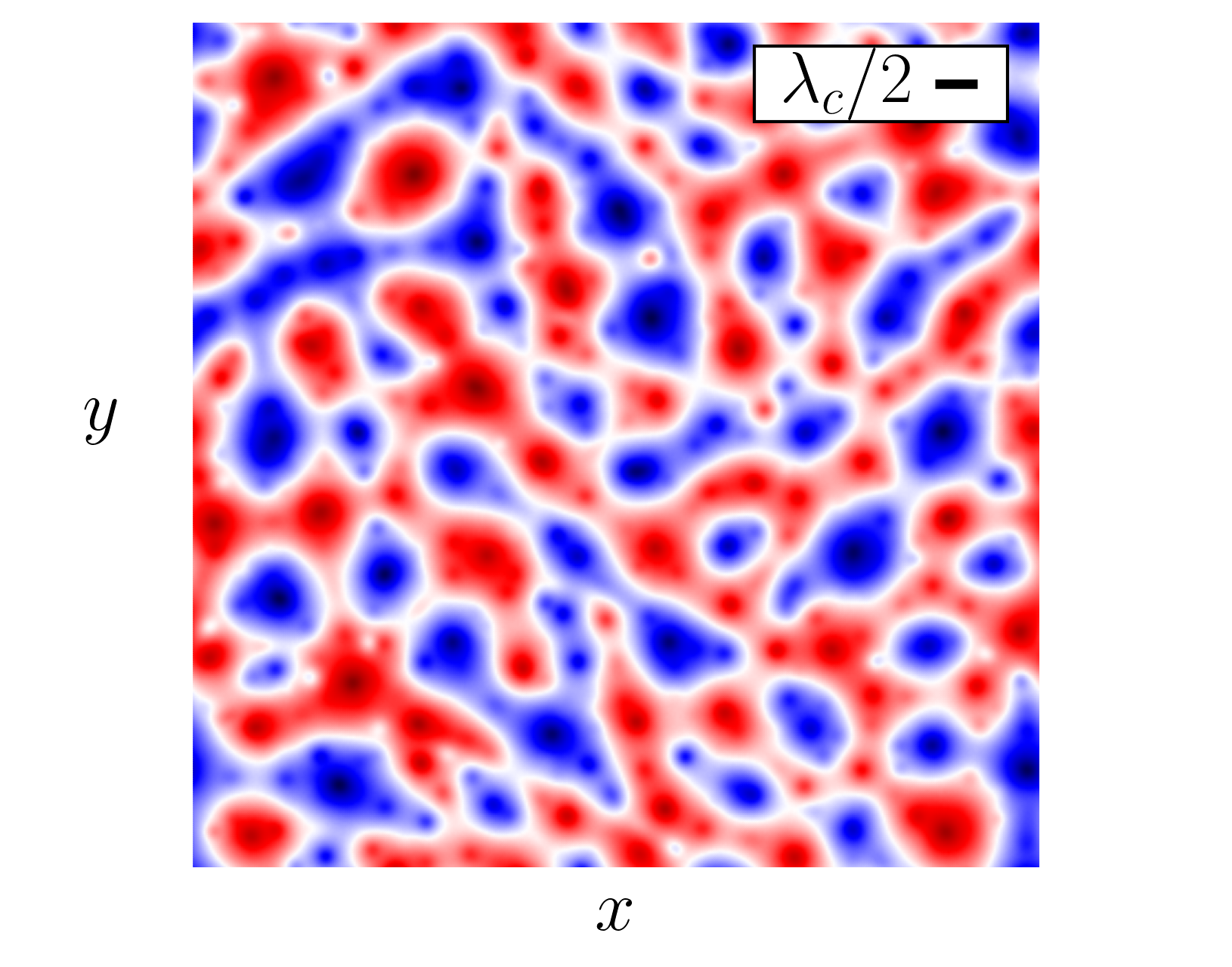}}
\subfloat[][]{\includegraphics[width=0.49\textwidth]{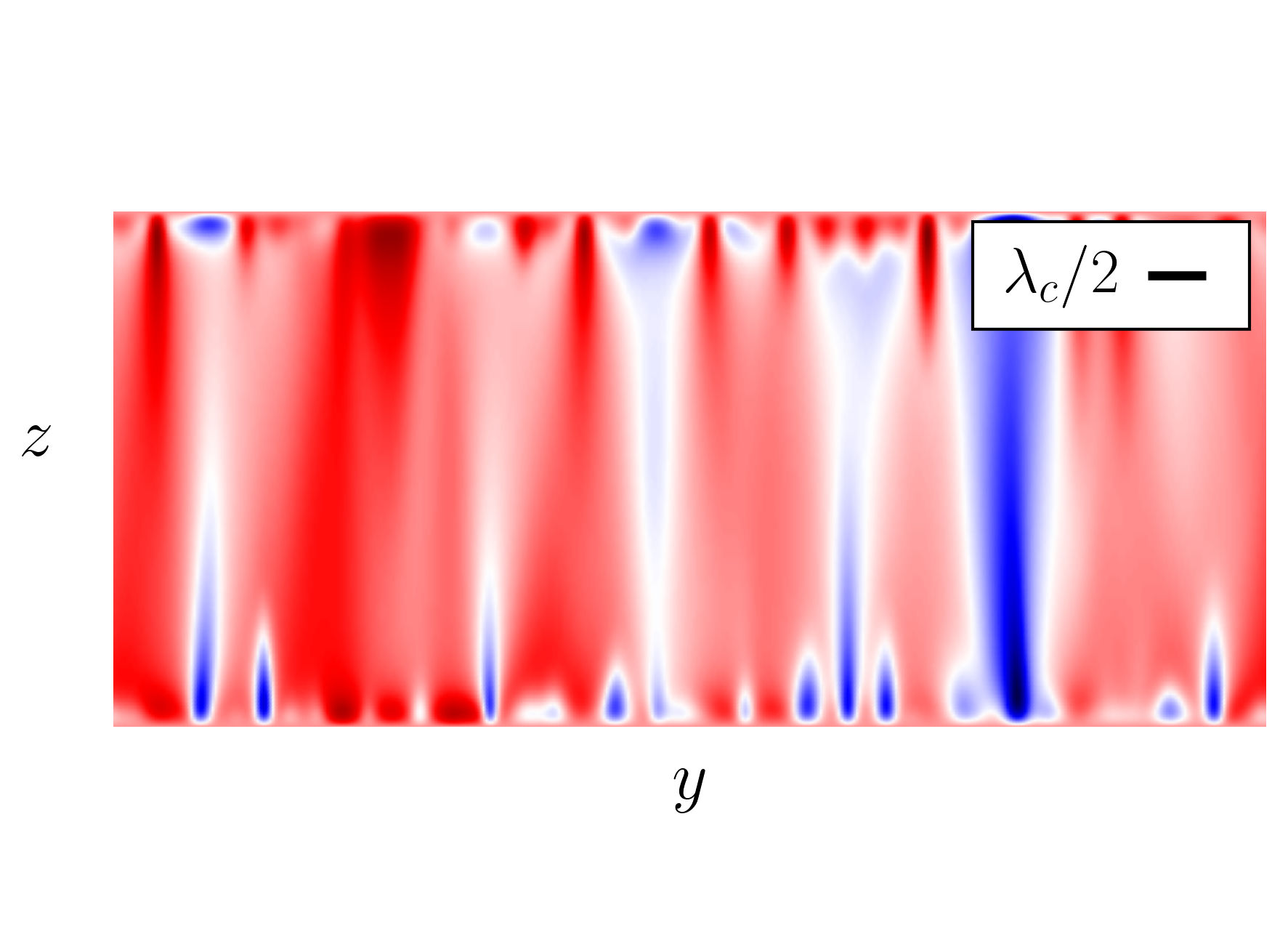}} \\
\caption{Heat transport and flow morphology in VMC. (a) Nusselt number ($Nu$) vs supercriticality ($Ra/Ra_c$); (b) compensated Nusselt number, $Nu/(Ra/Ra_c)$, vs $Ra/Ra_c$. Instantaneous visualizations of the fluctuating temperature: (c) horizontal cross section at the mid-plane ($z = 0.5$); (d) vertical cross section. The black line in (c) and (d) indicates one half of the critical horizontal wavelength, $\lambda_c/2$.} \label{F:nura}
\end{figure}

We review some of the main results on heat transport in VMC established in prior work. The boundary conditions are stress-free and electrically insulating unless stated otherwise. The Nusselt number data is shown in Fig.~\ref{F:nura}(a) for all simulations. For relatively small values of supercriticality, $Ra/Ra_c \lesssim 2$,  Fig.~\ref{F:nura}(a) shows a regime where $Nu$ is roughly independent of the imposed magnetic field strength $Q$. For larger values of supercriticality, Fig.~\ref{F:nura}(a) shows the now well-established behavior of VMC that at a fixed supercriticality $Nu$ scales more strongly as the magnetic field strength increases (larger $Q$) \citep{sC00,jmA01,mY19}. 
This same data is presented in Fig.~\ref{F:nura}(b), where it is compensated by the inviscid scaling, $Nu \sim (Ra/Ra_c)$. The inviscid scaling assumes independence on thermal diffusivity at large enough levels of supercriticality \citep{sB06}. The compensated data shows no clear convergence towards this inviscid scaling over the investigated range of parameter space. 

Instantaneous horizontal and vertical cross-sectional views of the fluctuating temperature field are shown in Fig.~\ref{F:nura}(c,d) for $Q = 10^7$ and $Ra \approx 10 Ra_c$. This case is magnetically constrained and within the `columnar' regime of VMC that consists of tall, thin plumes that extend across the entire depth of the fluid layer. The black line denotes one half of a critical wavelength, $\lambda_c/2$, and is meant to represent an $O(\ve)$ asymptotically dominant scale. Both the horizontal ($x$-$y$) and vertical ($y$-$z$) cross sections are sliced at the center of the box. In the case of the horizontal cross section of Fig.~\ref{F:nura}(c) the $O(\ve)$ scale captures the typical size of a localized cell. For the vertical cross section of Fig.~\ref{F:nura}(d), both the width of a typical thermal plume and the average height of the thermal boundary layer can also be seen to be qualitatively captured by an $O(\ve)$ scale.

\begin{figure}
\centering
\subfloat[][]{\includegraphics[width=0.45\textwidth]{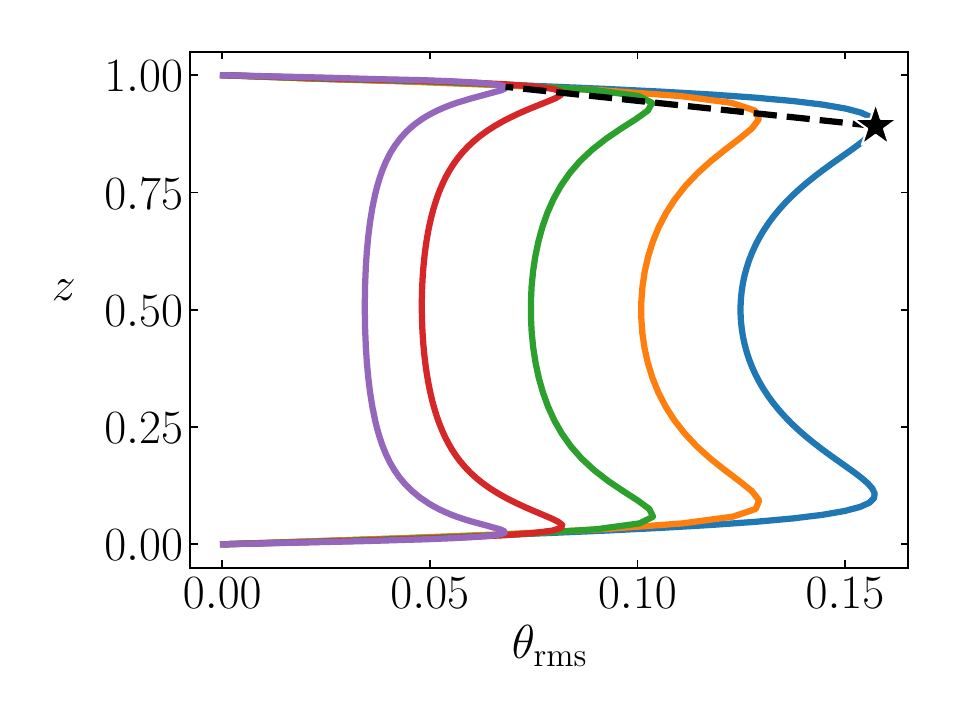}}  
\subfloat[][]{\includegraphics[width=0.45\textwidth]{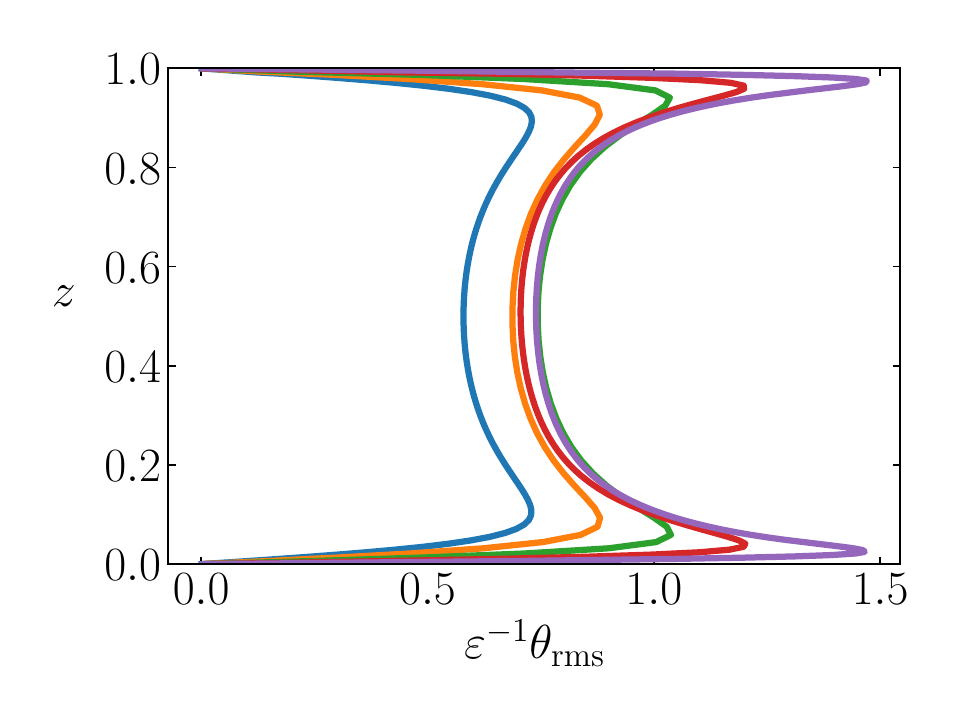}} \\
\subfloat[][]{\includegraphics[width=0.45\textwidth]{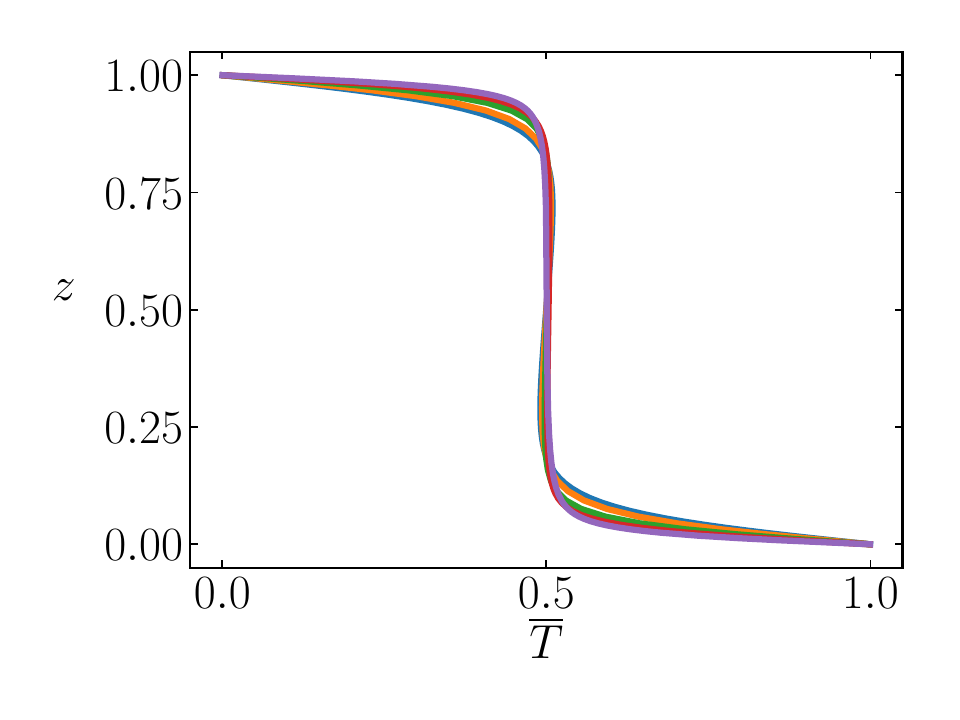}}
\caption{(a) Vertical profiles of the rms fluctuating temperature $\theta_\text{rms}$ for all values of $Q$ simulated. (b) Asymptotically rescaled profiles of $\theta_\text{rms}$. (c) Mean temperature profiles. The shrinking of the thermal boundary layer thickness $\delta_\theta$ with increasing $Q$ is emphasized with a dashed line. Supercriticality is fixed at $Ra \approx 10Ra_c$.} \label{F:tempcollapse}
\end{figure}

\begin{figure} 
\centering
\subfloat[][]{\includegraphics[width=0.49\textwidth]{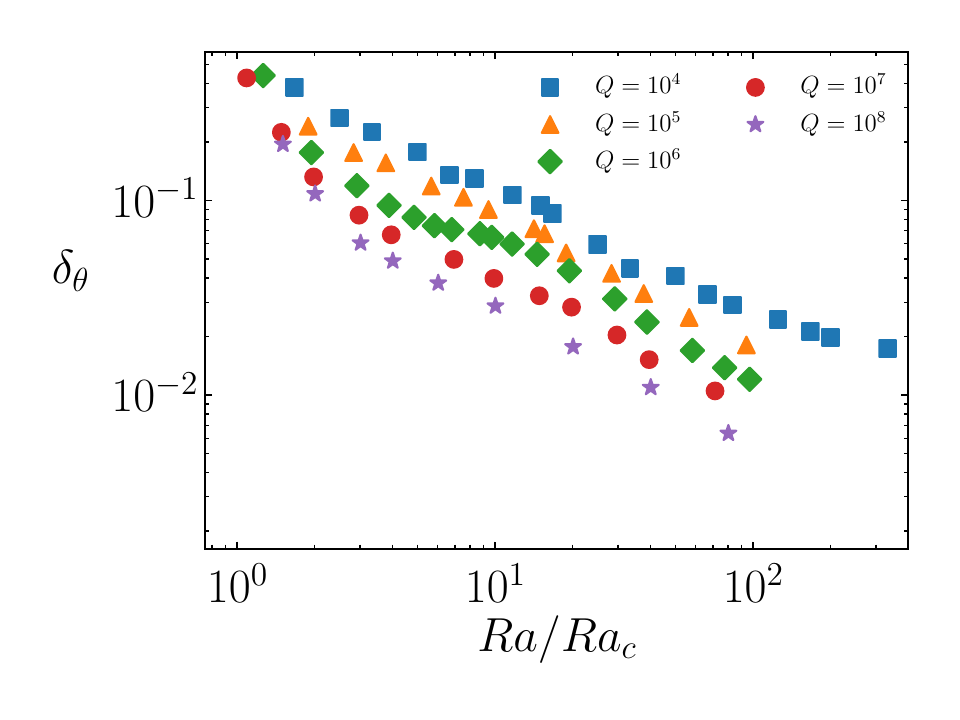}}
\subfloat[][]{\includegraphics[width=0.49\textwidth]{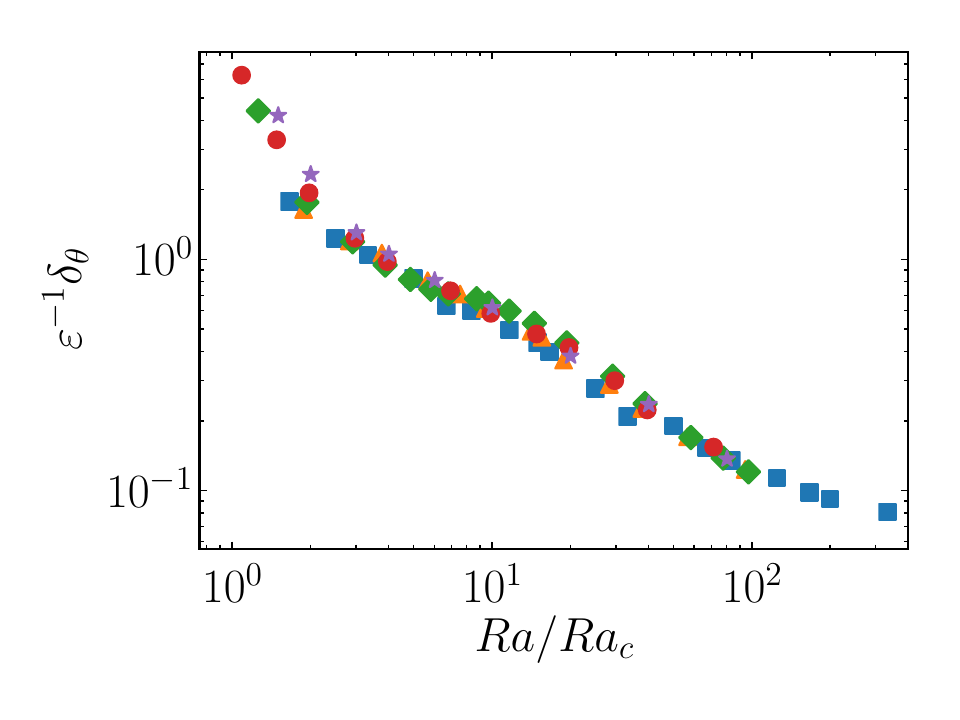}}
\caption{Thermal boundary layer thickness ($\delta_\theta$) vs $Ra/Ra_c$: (a) raw data, (b) asymptotically rescaled data.} \label{F:thermalbl}
\end{figure}

Vertical profiles of $\theta_\text{rms}$ are shown in Fig.~\ref{F:tempcollapse}(a) for a fixed supercriticality of $Ra/Ra_c \approx 10$ and all values of $Q$. Two important features of these profiles are observed: (1) the magnitudes of the rms temperature fluctuation exhibit a systematic decrease with $Q$; and (2) the thermal boundary layer thickness decreases as a function of $Q$. The first feature is well known from the linear theory of VMC -- in the limit of large $Q$ the fluctuating temperature scales as $\theta = O(\ve)$. Figure \ref{F:tempcollapse}(b) shows the profiles rescaled by this asymptotic prediction, where we find good collapse within the interior for all $Q > 10^4$. The second feature is novel, and we find that this thermal boundary layer scales as $\delta_\theta = O(\ve)$ as implied by the balance of thermal advection and diffusion. The interior scaling is unable to collapse the thermal profiles near the boundaries, indicative of possibly distinct asymptotic behavior near the boundaries. A more detailed, quantified asymptotic analysis of the boundary layer structure as a function of $Q$ is discussed in Fig.~\ref{F:thermalbl}.

Vertical profiles of the temporally averaged mean temperature are shown in Fig.~\ref{F:tempcollapse}(c) for different values of $Q$ while holding supercriticality fixed at $Ra/Ra_c \approx 10$. As with hydrodynamic convection, the mean temperature saturates at $\overline{T} \sim 0.5$ throughout most of the domain, indicating a thermally mixed convective interior. This is contrasted with rotationally constrained convection where a finite background temperature gradient persists as $Ra$ increases \citep[e.g.][]{kJ96}. Contrary to the behavior of the mean temperature throughout the interior, the boundary layer structure does appear to have a systematic dependence on the strength of the imposed magnetic field as we will discuss quantitatively in Fig.~\ref{F:thermalbl}.

We investigate the asymptotic scaling of the boundary layer height of the fluctuating temperature in Fig.~\ref{F:thermalbl}. Figure \ref{F:thermalbl}(a,b) compiles the thermal boundary layer height of the fluctuating temperature as a function of supercriticality and $Q$. We determine the height of the boundary layer $\delta_\theta$ by the location of the extremum of the fluctuating temperature rms profile. This definition of $\delta_\theta$ is widely used and has been shown by \cite{long2020thermal} to be equivalent the point where thermal advection and diffusion balance, at least for non-magnetic RBC. The location of the extremum is identified by differentiating $\theta_\text{rms}$ with respect to $z$ and interpolating the point at which $\partial_z \theta_\text{rms} = 0$. Figure \ref{F:thermalbl}(a) collects the uncompensated fluctuating temperature thermal boundary layer height $\delta_\theta$ and shows a generally monotonic decrease with $Ra/Ra_c$. For a fixed level of supercriticality, the fluctuating temperature thermal boundary layer shrinks as $Q$ increases, consistent with the qualitative plots of Fig.~\ref{F:tempcollapse}. In Fig.~ \ref{F:thermalbl}(b) we demonstrate that the $\delta_\theta = O(\ve)$ scaling collapses the data for all values of $Q$.

\begin{figure} 
\centering
\subfloat[][]{\includegraphics[width=0.49\textwidth]{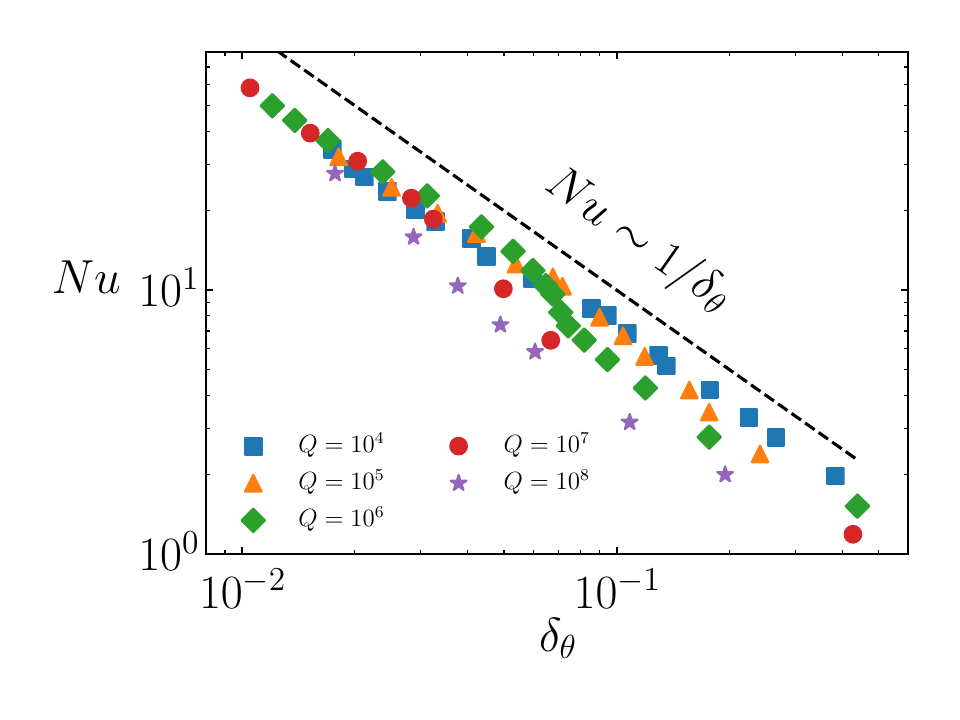} }
\subfloat[][]{\includegraphics[width=0.49\textwidth]{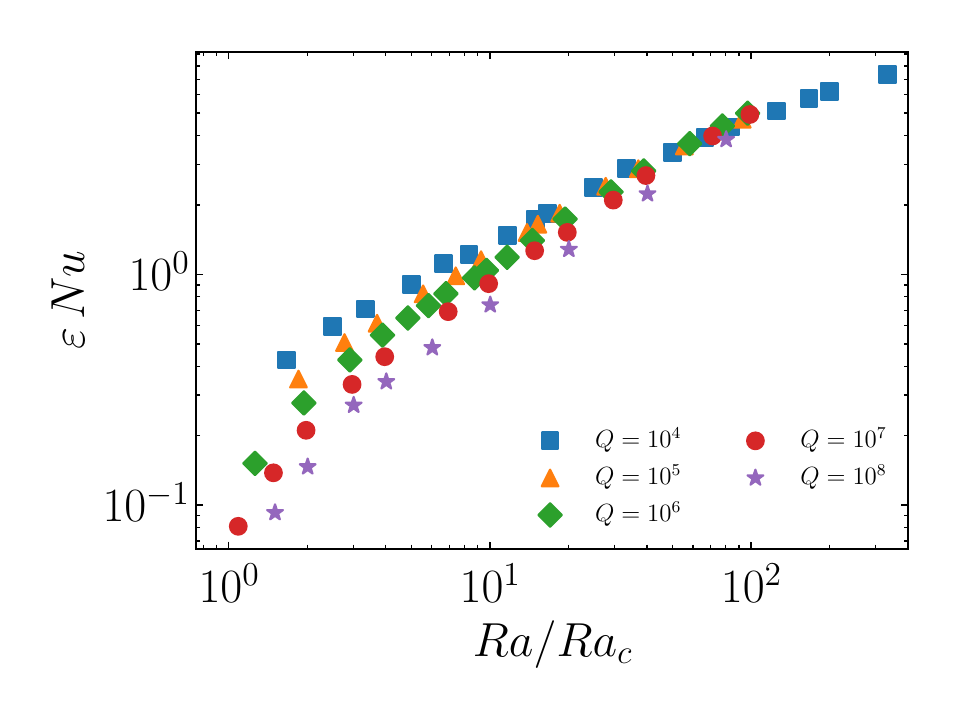}}
\caption{(a) $Nu$ as a function of the thermal boundary layer thickness $\delta_\theta$; (b) asymptotically rescaled data of $Nu$ vs $Ra/Ra_c$. A dashed line of $Nu \sim \delta_\theta^{-1}$ in (a) is shown for reference.} \label{F: nu_delta}
\end{figure}

We show $Nu$ as a function of the thermal boundary layer thickness $\delta_{{\theta}}$ in Fig.~\ref{F: nu_delta}(a). Especially for small thermal boundary layer thicknesses, when $Ra/Ra_c \gg 1$, the data appears to show that the Nusselt number scales as $\delta_{\theta}^{-1}$, indicating that the boundary is controlling most of the heat transport, as is the case in hydrodynamic convection. Given that $\delta_\theta =  O(\ve)$, it follows that, in the regime where the boundaries control most of the heat transport, $Nu = O(\ve^{-1})$. This scaling is consistent with the results of Fig.~\ref{F: nu_delta}(b). 



\begin{figure}
\centering
\subfloat[][]{\includegraphics[width=0.49\textwidth]{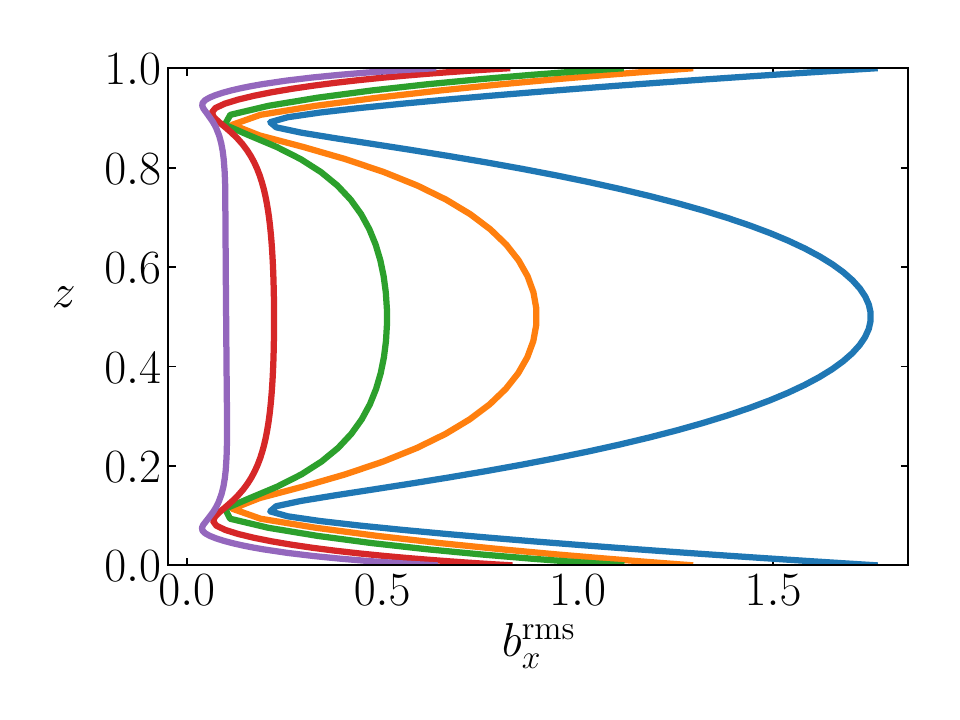}}
\subfloat[][]{\includegraphics[width=0.49\textwidth]{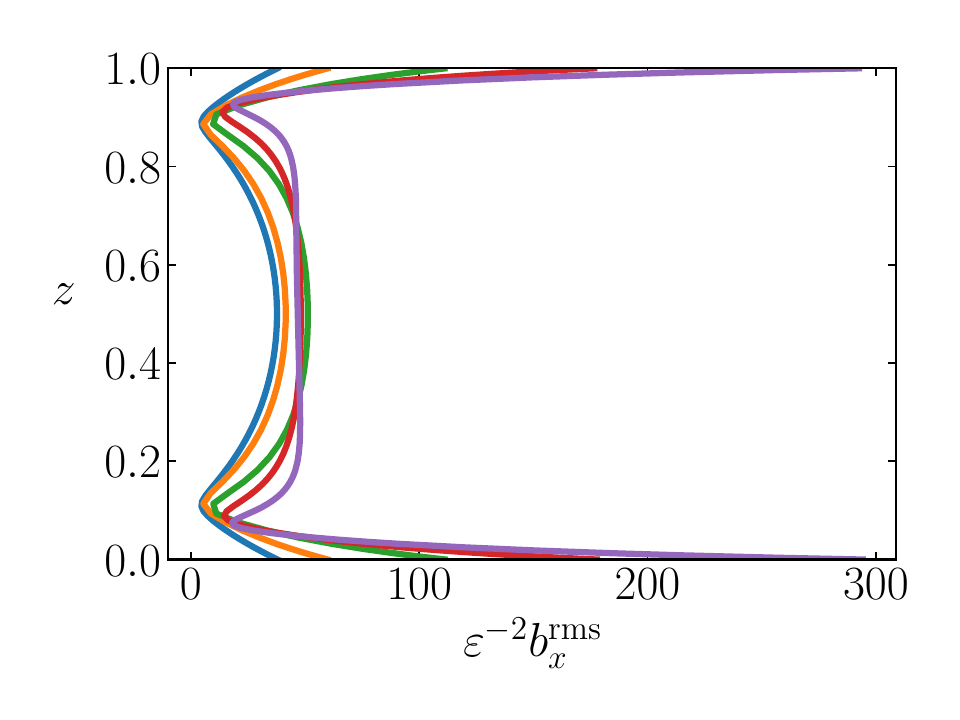}} \\
\subfloat[][]{\includegraphics[width=0.49\textwidth]{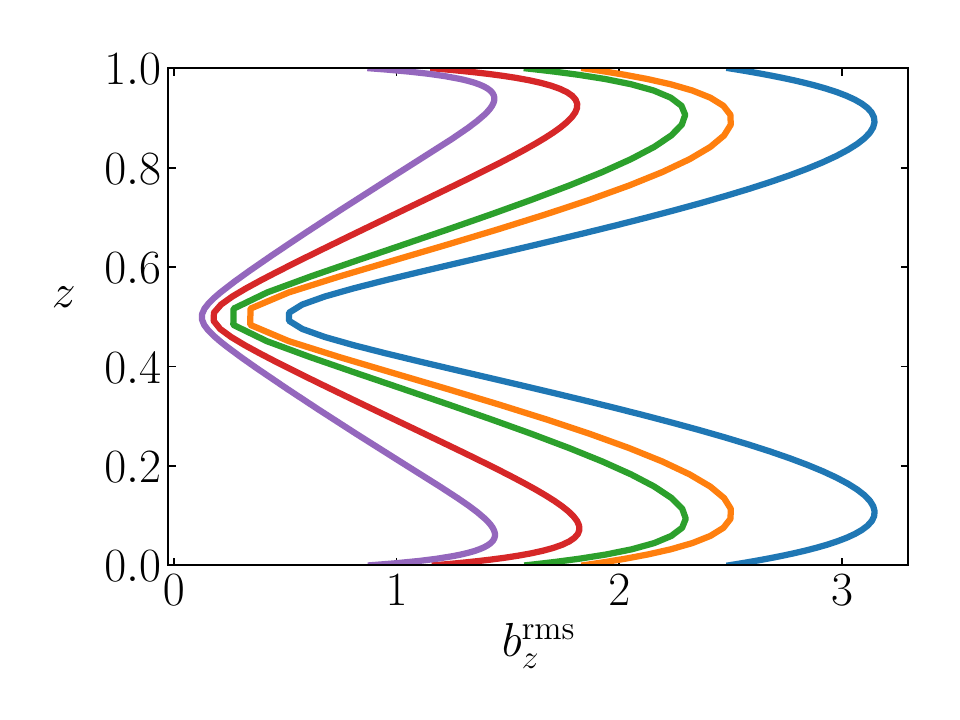}}
\subfloat[][]{\includegraphics[width=0.49\textwidth]{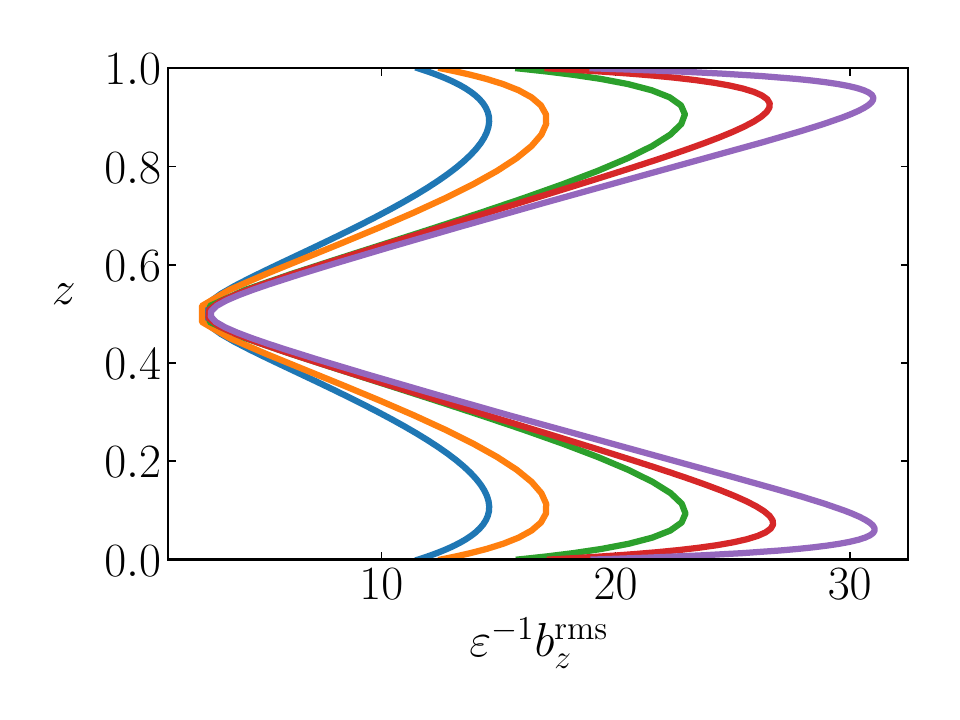}}
\caption{Vertical profiles of the rms magnetic field for all values of $Q$ simulated. Top row: horizontal field; bottom row: vertical magnetic field. (a,c) raw data; (b,d) asymptotically rescaled data. Supercriticality is held fixed at $Ra \approx 10Ra_c$. } \label{F:bcollapse}
\end{figure}

\subsection{Magnetic boundary layers}

Vertical profiles of the rms induced magnetic field $\mathbf{b}_\text{rms}$ are shown in Fig.~\ref{F:bcollapse}. Since the solutions of $b_x$ and $b_y$ are identical due to the lateral $(x,y)$ symmetry of VMC, we choose to represent the horizontal data through $b_x$ alone to avoid repetition. Raw (i.e. uncompensated) vertical profiles at $Ra/Ra_c \approx 10$ are shown for the horizontal and vertical fields in Fig.~\ref{F:bcollapse}(a,c). Magnetic boundary layers are observed in the profiles of the magnetic field as a consequence of the electrically insulating plates. In Fig.~\ref{F:bcollapse}(b,d) we show that the interior scalings $b_x = O(\ve^2)$ and $b_z = O(\ve)$ are able to collapse the interior profiles of the induced magnetic field. As with the fluctuating temperature field, the boundary layer is not as well collapsed with the interior scalings and the boundary layer height appears to shrink as a function of $Q$. 

The interior depth dependence of the induced magnetic field profiles in Fig.~\ref{F:bcollapse} can be understood by considering the dominant vertical force balance in the interior \eqref{E:zmom} alongside the solenoidal constraint. As the system becomes more strongly forced, the fluctuating temperature field appears to roughly saturate with depth in the interior. Under these conditions, equation \eqref{E:zmom} would then require that $\partial_z b_z$ be similarly uniform in $z$, implying that $b_z$ depends linearly on depth in the interior. This is supported by the interior profile of Fig.~\ref{F:bcollapse}(c). Given that the magnetic field is also divergence-free, the depth-dependence of $b_\perp$ must follow the depth-dependence of $\partial_z b_z$ and so the horizontal magnetic field should be similarly independent of $z$ throughout the interior. This is supported by the interior profile of Fig.~\ref{F:bcollapse}(a).

\begin{figure} 
\centering
\subfloat[][]{\includegraphics[width=0.49\textwidth]{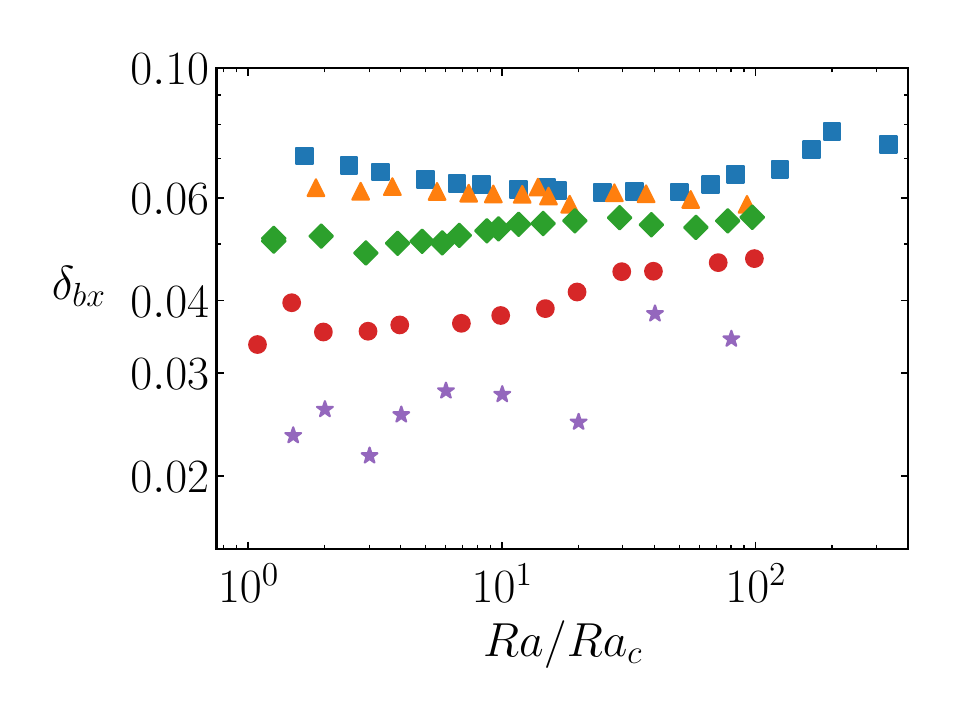}}
\subfloat[][]{\includegraphics[width=0.49\textwidth]{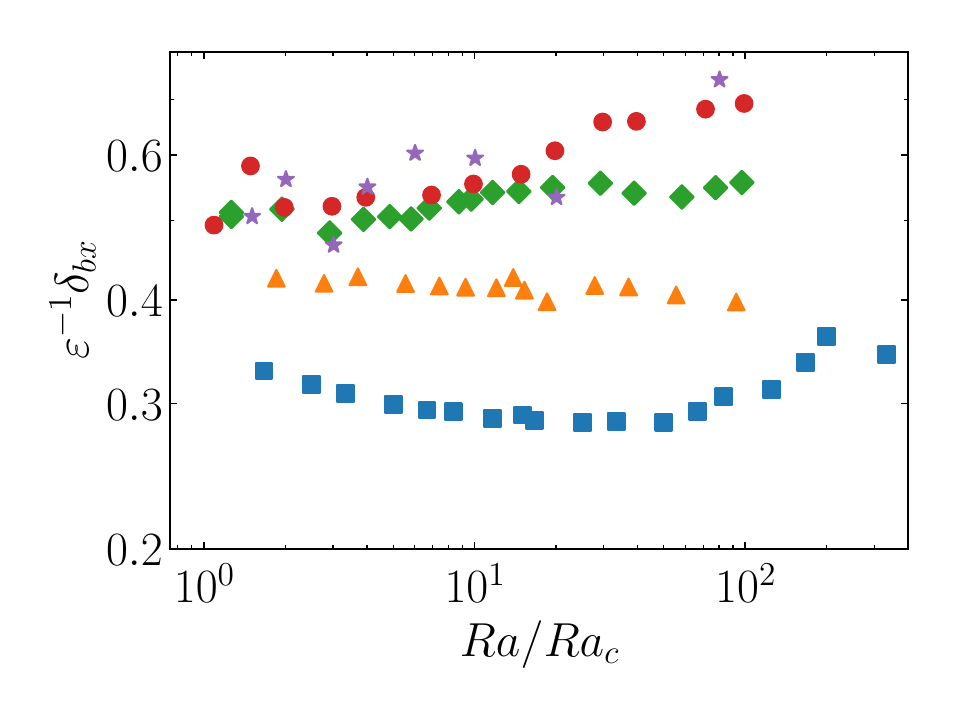}} \\
\subfloat[][]{\includegraphics[width=0.49\textwidth]{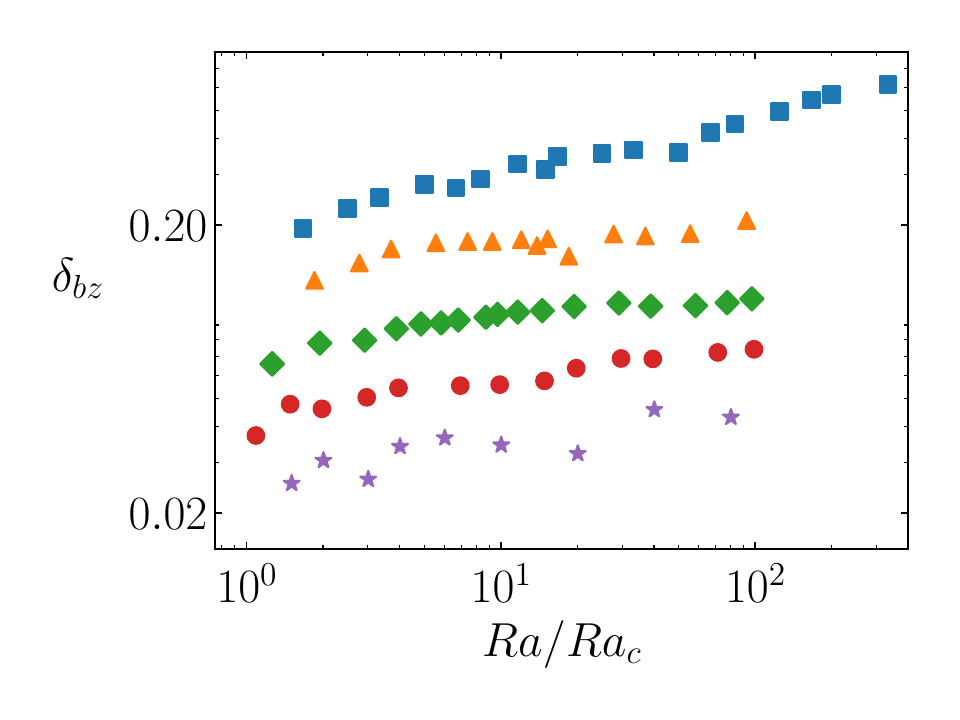}}
\subfloat[][]{\includegraphics[width=0.49\textwidth]{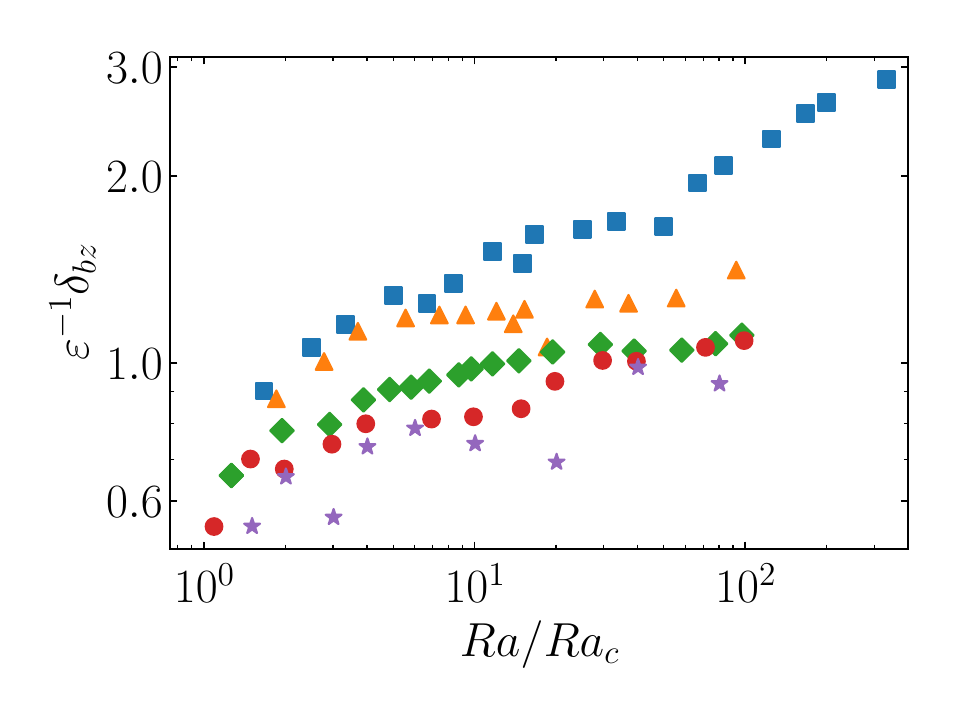}} 
\caption{Magnetic boundary layer thickness as a function of $Ra/Ra_c$ as computed from (a,b) $b_x^\text{rms}$ and (c,d) $b_z^\text{rms}$. (a,c) raw data; (b,d) rescaled data. } \label{F:bbl}
\end{figure}

Figure \ref{F:bbl} shows the boundary layer height of the horizontal and vertical induced magnetic fields. Figure \ref{F:bbl}(a,c) shows the uncompensated magnetic boundary layer height as a function of supercriticality. The magnetic boundary layer height is determined by first fitting the rms profile of the induced magnetic field to an exponential function of the form derived in the asymptotic boundary solution of (\ref{binsol}). The boundary layer height is then defined as the characteristic lengthscale from the fitted wavenumber $k_\text{fit}$, namely by equating $\delta_b = k_\text{fit}^{-1}$. We choose this definition as it leverages the linearity of the the magnetic boundary layers and, as was discussed in subsection \ref{T: b_mod}, we expect the functional form of this asymptotic boundary solution to persist in the nonlinear regime. 

 The magnetic boundary layer thicknesses $\delta_b$ appear to either plateau or slightly increase with supercriticality, though the dependence on $Ra$ is relatively weak when compared to $\delta_\theta$. In Fig.~\ref{F:bbl}(b,d) we show compensated plots assuming a $\delta_b = O(\ve)$ thickness. The compensated plots do appear to remove the systematic dependency on $Q$, though the convergence is weaker when compared to the convergence of the compensated thermal boundary layer thickness plots. Instead, the convergence in $\delta_b$ appears restricted to the largest $Q$ cases only.

\subsection{Interior and boundary layer scalings}

\begin{figure}
\centering
\subfloat[][]{\includegraphics[width=0.49\textwidth]{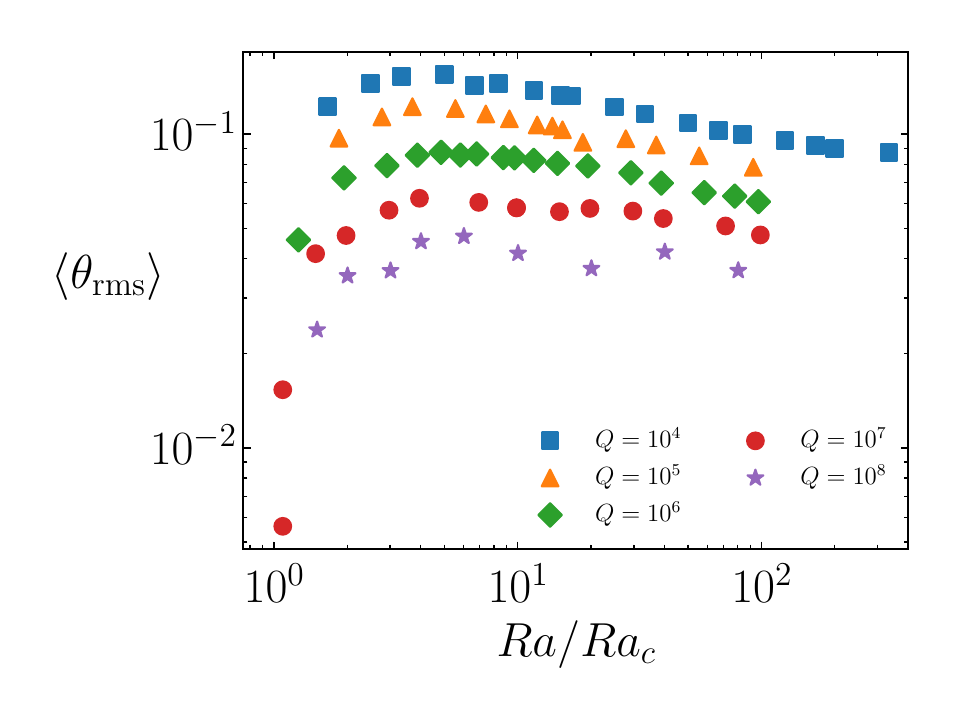}} 
\subfloat[][]{\includegraphics[width=0.49\textwidth]{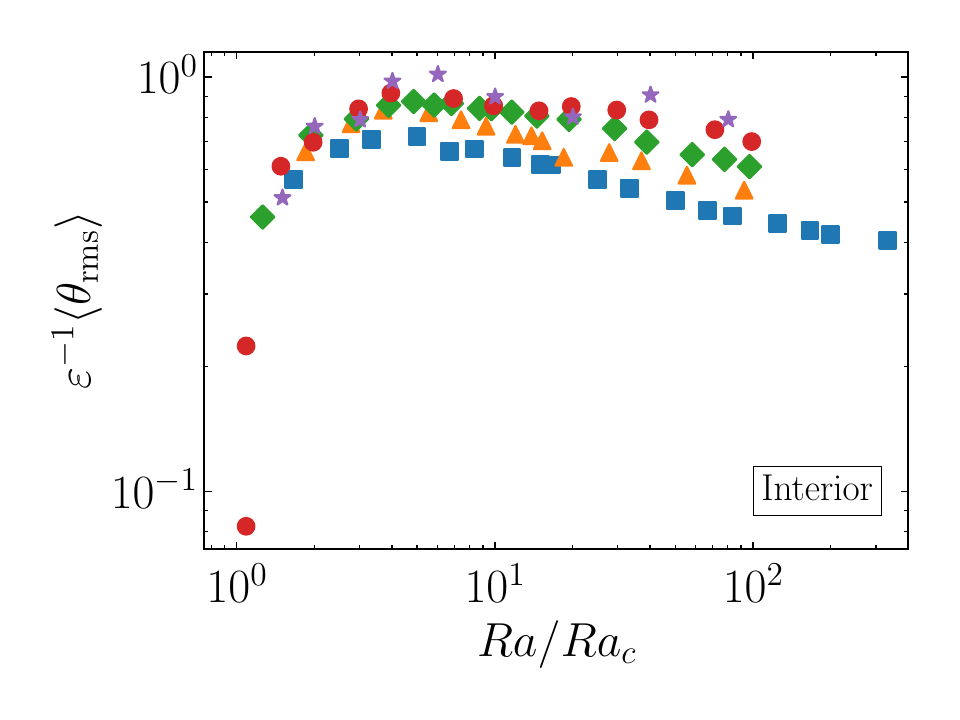}} \\
\subfloat[][]{\includegraphics[width=0.49\textwidth]{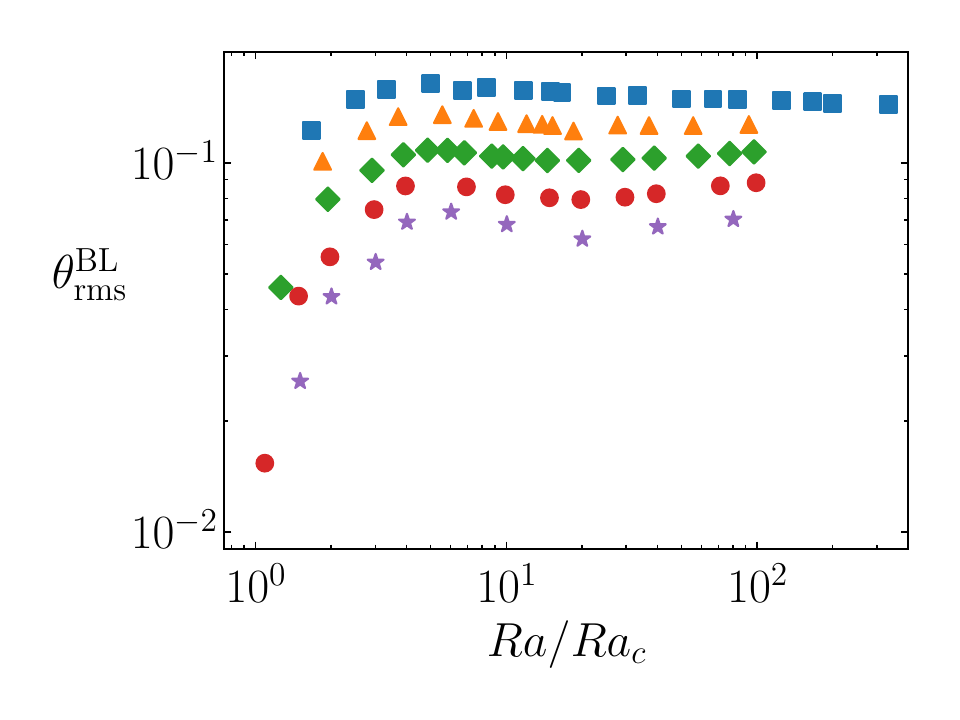}} 
\subfloat[][]{\includegraphics[width=0.49\textwidth]{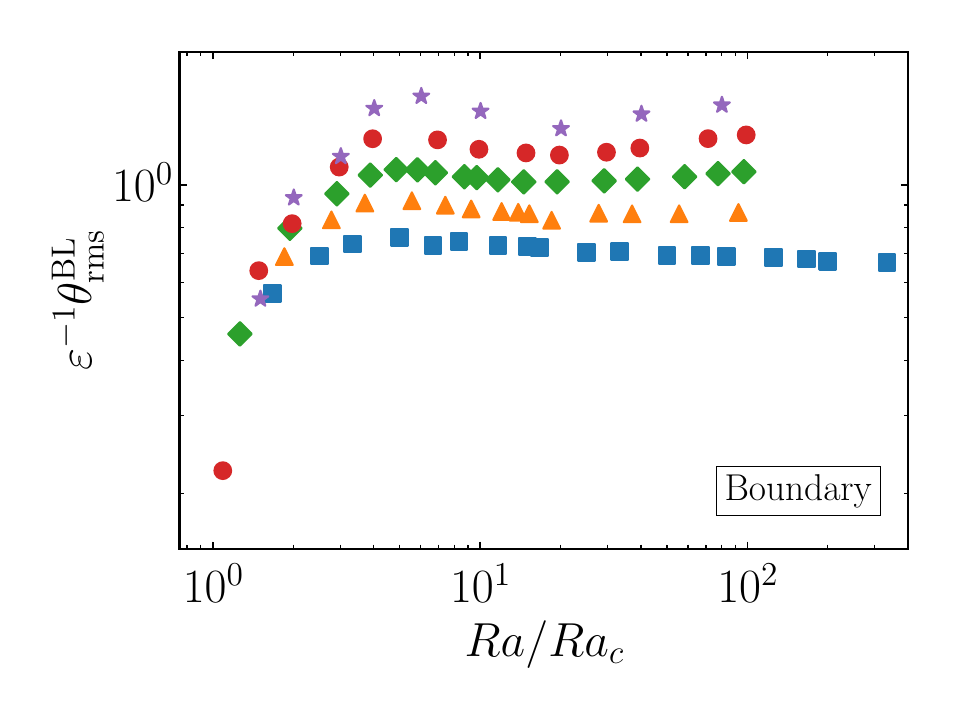}} \\
\caption{Representative amplitudes of the fluctuating temperature in (a,b) the interior and (b,d) at the boundary layer. (a,c) Raw data; (b,d) rescaled data.} \label{F:fieldint}
\end{figure}

\begin{figure}
\centering
\subfloat[][]{\includegraphics[width=0.49\textwidth]{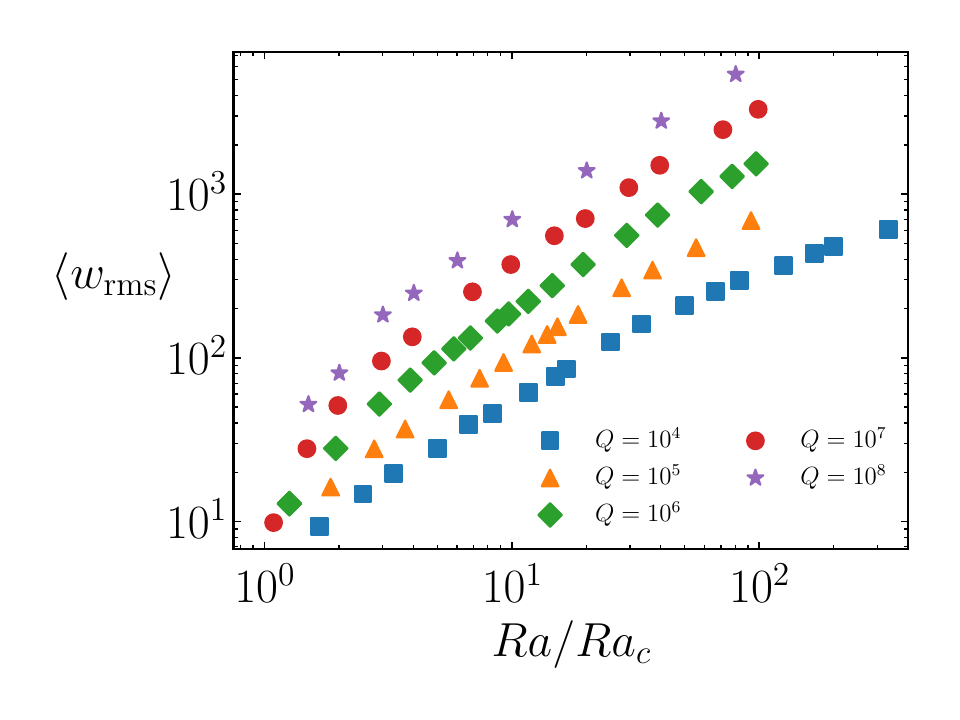}} 
\subfloat[][]{\includegraphics[width=0.49\textwidth]{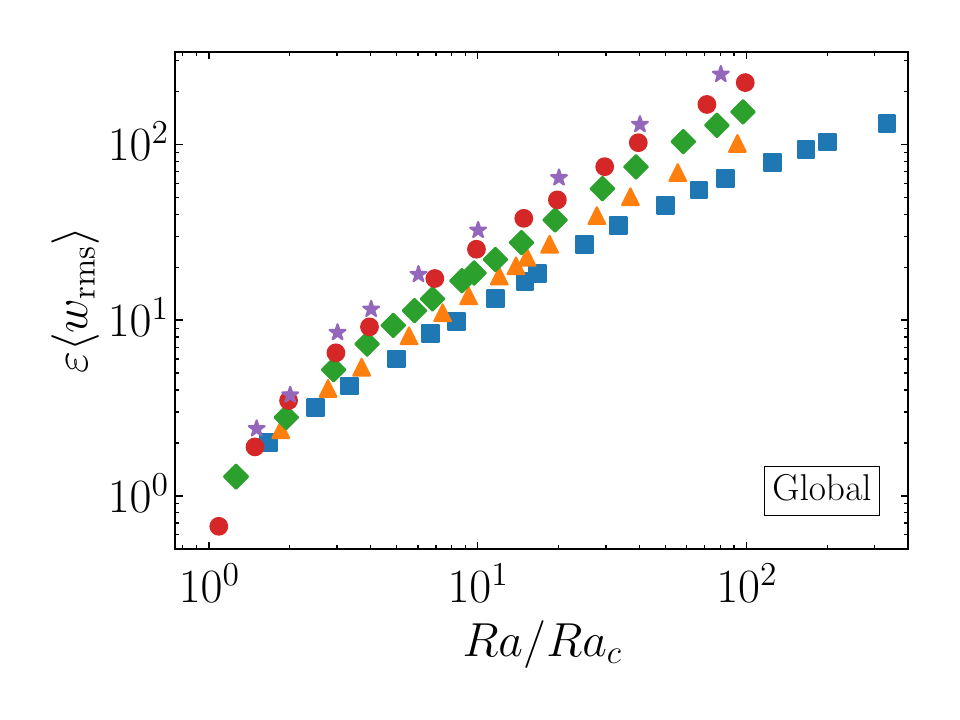}} 
\caption{Global amplitude of the vertical velocity field $\langle w_\text{rms} \rangle$ vs $Ra/Ra_c$: (a) raw data; (b) rescaled data.} \label{F: wglob}
\end{figure} 

In Fig.~\ref{F:fieldint} we plot representative amplitudes of the fluctuating temperature field. Given the development of thermal boundary layers, we restrict our vertical integration space over the bulk interior only in Fig.~\ref{F:fieldint}(a). We define the interior as the region between the two extremums of the $\theta_\text{rms}$ profile so that $z_\text{int} \in [\delta_\theta, 1-\delta_\theta]$. Besides a small transient regime for low levels of supercriticality, $Ra/Ra_c \lesssim 2$, we find that the interior values of the temperature field asymptote toward a $\theta =O(\ve)$ scaling law as seen by the compensated plots of Fig.~\ref{F:fieldint}(b). In Fig.~\ref{F:fieldint}(c), the value of $\theta_\text{rms}$ at the thermal boundary layer height is chosen as a representative amplitude of the fluctuating temperature within that region. This boundary amplitude is denoted as $\theta_\text{rms}^\text{BL}$. We find that the interior scaling of $\theta = O(\ve)$ is not able to collapse the boundary values as well as the interior, as shown by the compensated plots of Fig.~\ref{F:fieldint}(d). Instead, we suggest that $\theta = O(1)$ near the boundaries due to the development of thermal boundary layers, though for the range of $Q$ we simulate we find this scaling to be inconclusive. Given that the temperature field appears to saturate with $Ra$, this is indicative of a peculiar asymptotic property of the VMC system where the temperature field may become independent of both $Q$ and $Ra$ near the boundaries. 

Amplitudes of the vertical velocity field $\langle w_\text{rms} \rangle$ depth averaged over the global domain height is shown in Fig.~\ref{F: wglob}. Primarily due to the choice of stress-free boundary conditions, no boundary layers appear to form in the velocity fields and we therefore expect uniform asymptotic behavior with respect to height. The scaling $w =O(\ve^{-1})$ appears to accurately describe the asymptotic behavior of the vertical velocity field, especially in the large-$Q$ branches. 

\begin{figure}
\centering
\subfloat[][]{\includegraphics[width=0.49\textwidth]{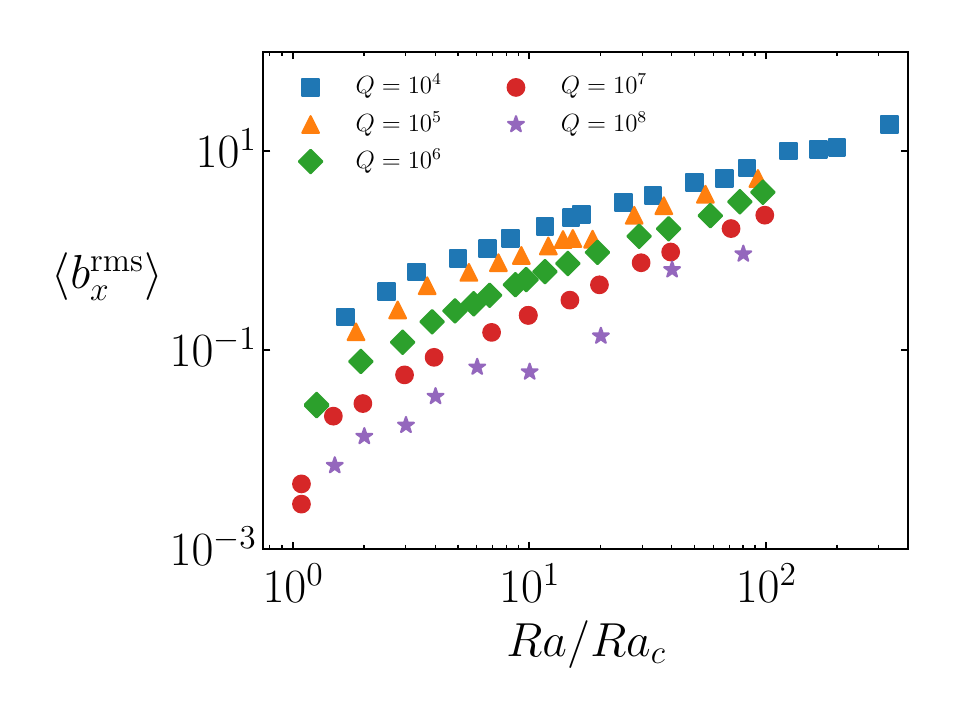}} 
\subfloat[][]{\includegraphics[width=0.49\textwidth]{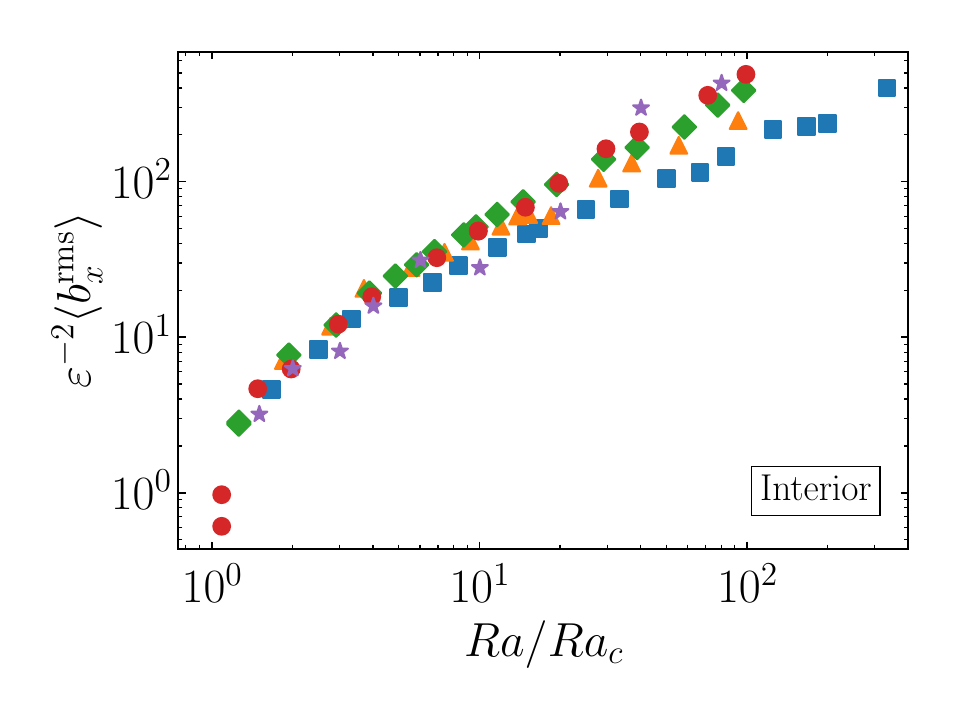}} \\
\subfloat[][]{\includegraphics[width=0.49\textwidth]{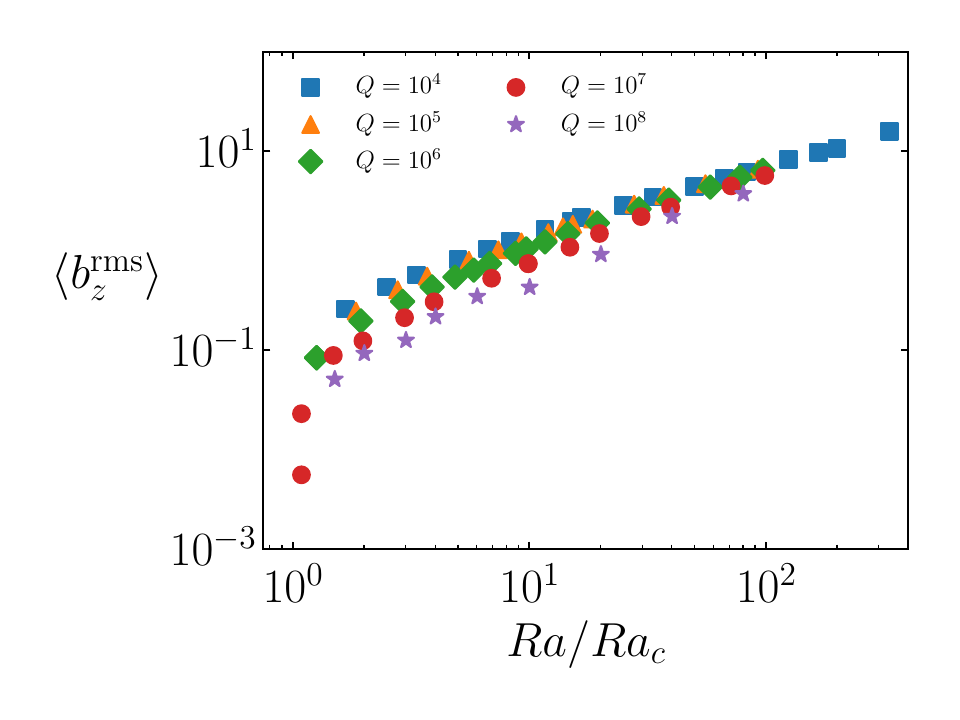}} 
\subfloat[][]{\includegraphics[width=0.49\textwidth]{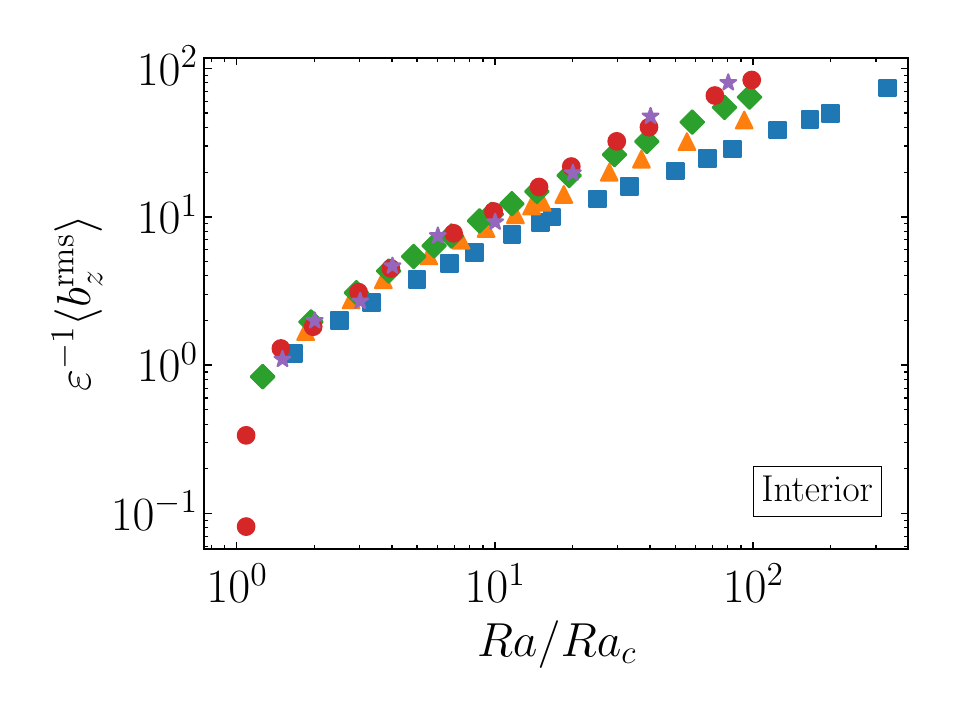}} 
\caption{Interior amplitude of (a,b) the horizontal magnetic field and (c,d) the vertical magnetic field. (a,c) raw data; (b,d) rescaled data.} \label{F:b_int}
\end{figure}

Figure \ref{F:b_int} shows representative amplitudes of the magnetic field from $\langle \mathbf{b}_\text{rms} \rangle$. We define the interior as the region between the two extremums of the rms magnetic field profiles. Uncompensated values of the horizontal and vertical magnetic field amplitudes are shown in Fig.~\ref{F:b_int}(a,c), respectively. Using the interior scalings of $b_x = O(\ve^2)$ and $b_z = O(\ve)$, we are able to collapse the data onto a single unifying curve as shown in Fig.~\ref{F:b_int}(b,d).

\begin{figure}
\centering
\subfloat[][]{\includegraphics[width=0.49\textwidth]{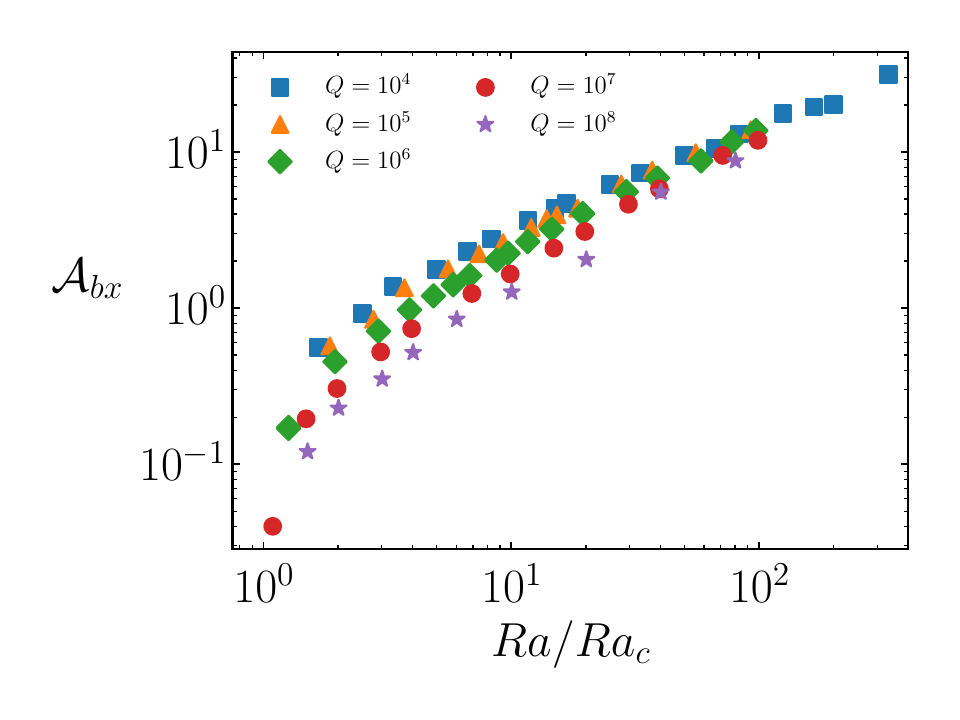}} 
\subfloat[][]{\includegraphics[width=0.49\textwidth]{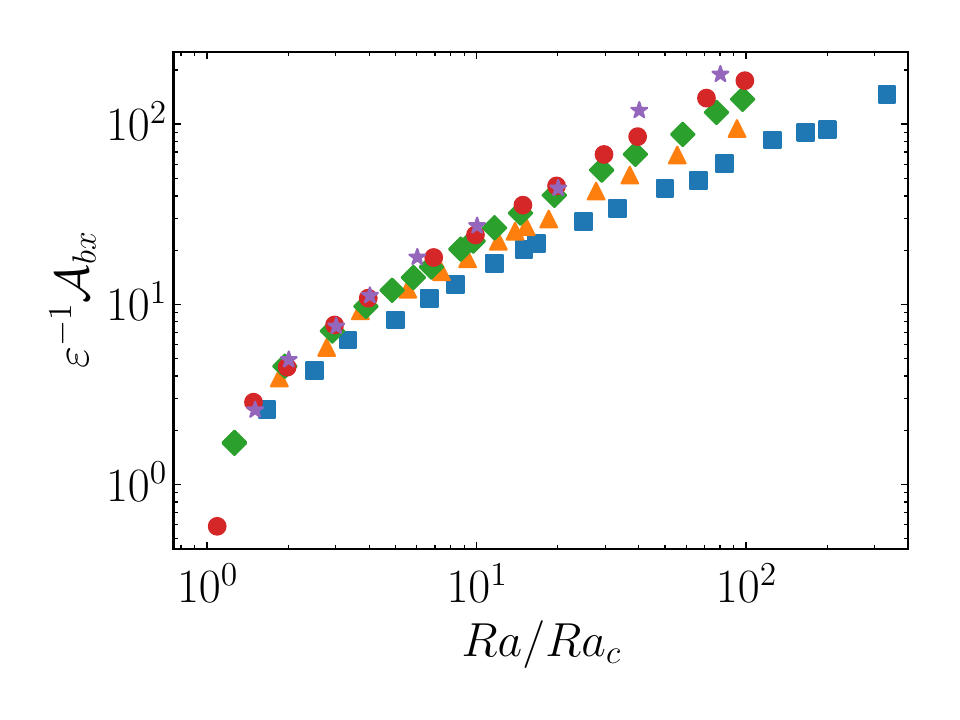}} \\
\subfloat[][]{\includegraphics[width=0.49\textwidth]{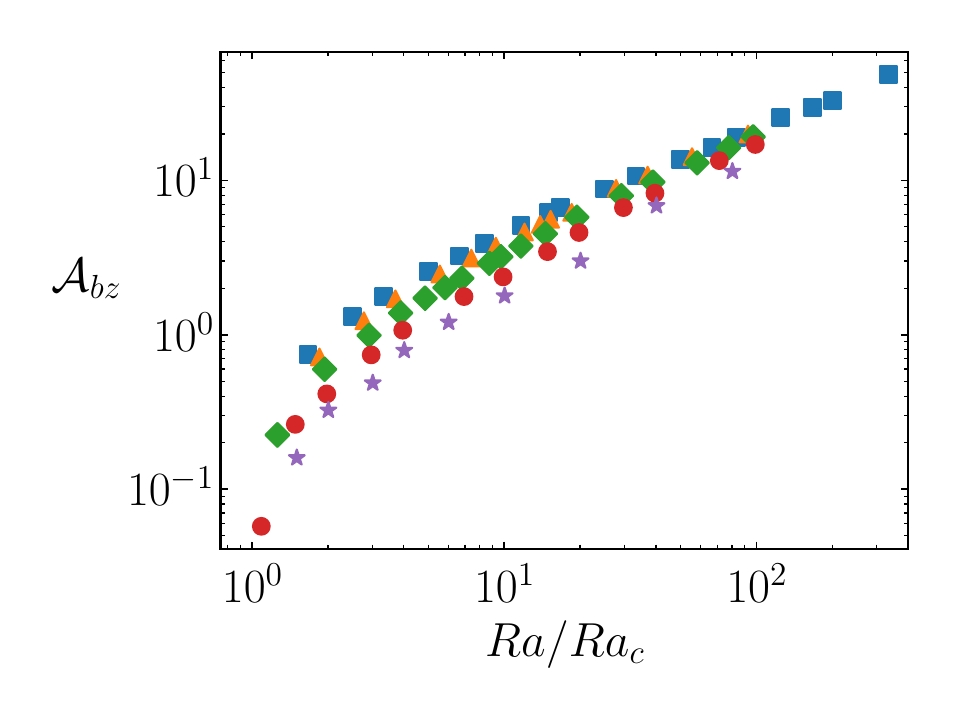}} 
\subfloat[][]{\includegraphics[width=0.49\textwidth]{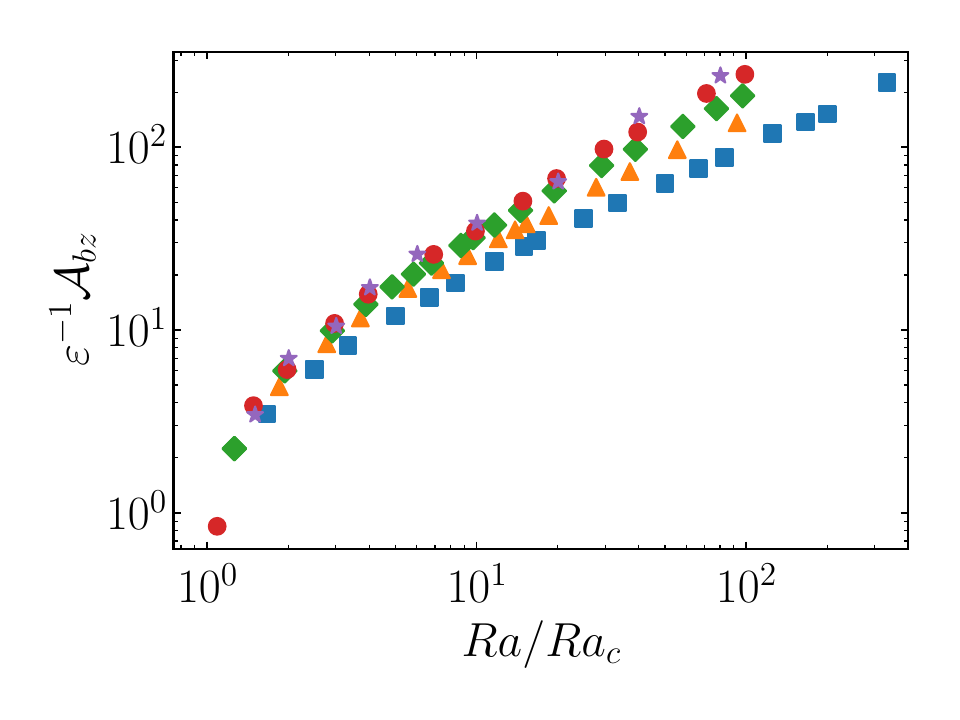}} \\
\caption{Boundary layer amplitude of (a,b) the horizontal magnetic field and (c,d) the vertical magnetic field. Values are obtained by fitting an exponential function near the bottom boundary of the horizontal rms vertical profile of the respective field.} \label{F:b_bl}
\end{figure}

Figure \ref{F:b_bl} shows the boundary layer amplitude of the horizontal and vertical magnetic fields as a function of supercriticality. As with defining the magnetic boundary layer height, the amplitude was obtained by fitting an exponential function to the $\mathbf{b}_\text{rms}$ profiles. The form of the exponential is informed by the asymptotic solution derived in equation (\ref{binsol}). Systematic dependencies on the strength of the imposed magnetic field can be seen in the uncompensated plots of the horizontal and vertical component of the induced magnetic field in Fig.~\ref{F:b_bl}(a) and (c), respectively. We observe a noticeably smaller spread in the horizontal magnetic field boundary amplitude as a function of $Q$ when compared to the spread found in the interior amplitudes of Fig.~\ref{F:b_int}(a). As argued earlier, a consequence of the $O(\ve)$ magnetic boundary layer coupled with the solenoidal constraint is that $b_\perp$ and $b_z$ must scale isotropically near the boundaries. Moreover, the matching principle requires $b_z = O(\ve)$ near the boundaries, thereby implying that $b_x = O(\ve)$. The compensated plots of Fig.~\ref{F:b_bl}(b,d) appear to support the isotropic amplitude scaling.

\subsection{Influence of mechanical boundary conditions} \label{S: bcM}

\begin{figure}
\centering
\subfloat[][]{\includegraphics[width=0.45\textwidth]{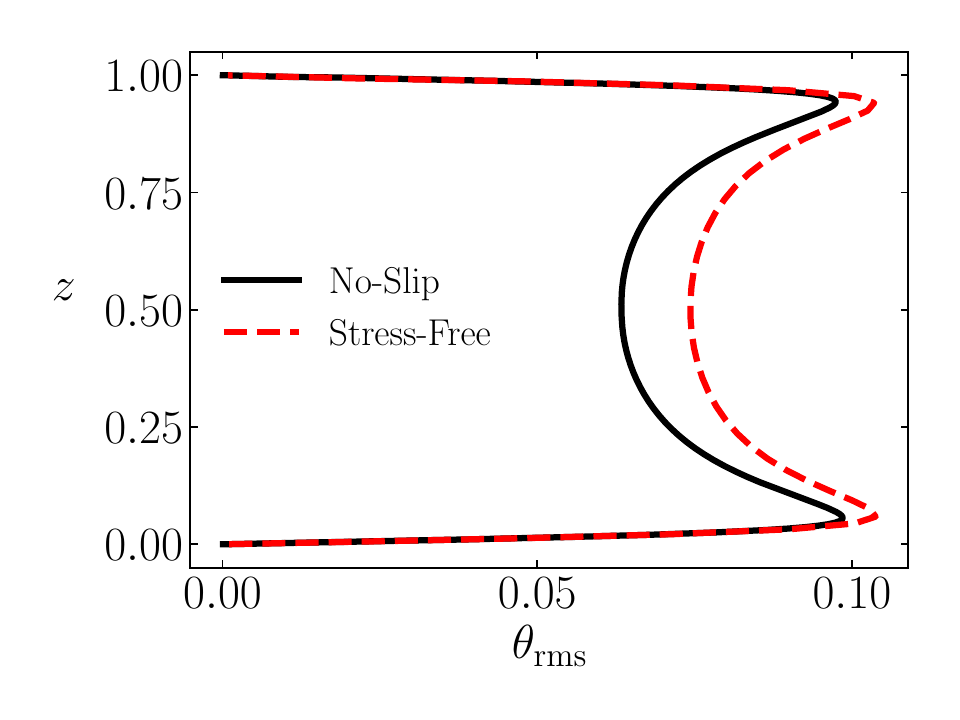}}
\subfloat[][]{\includegraphics[width=0.45\textwidth]{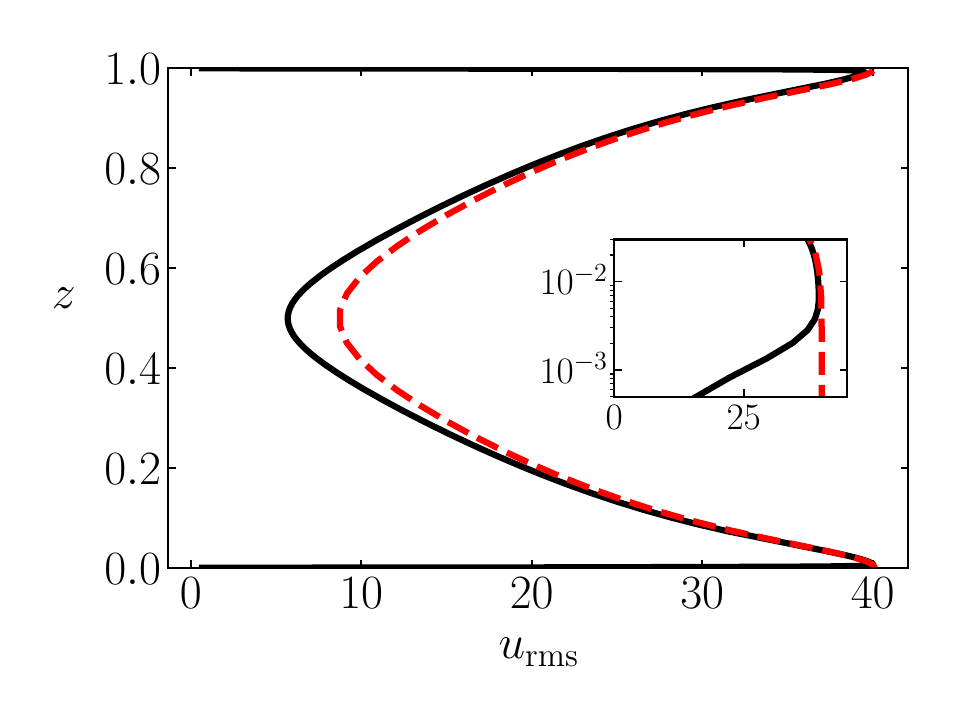}} \\
\subfloat[][]{\includegraphics[width=0.45\textwidth]{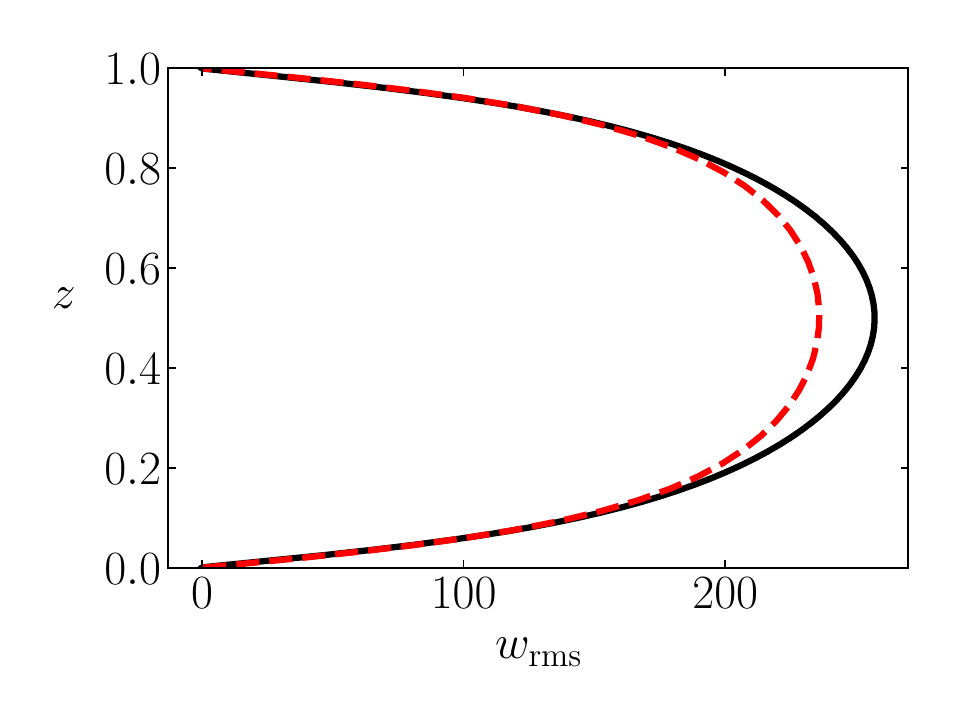}}
\caption{Comparison of stress-free (dashed red) and no-slip (solid black) boundary conditions. Profiles of rms quantities: (a) fluctuating temperature; (b) $x$-component of the velocity field; and (c) vertical velocity field. The parameters are fixed to $Q = 10^6$ and $Ra = 10^8$.} \label{F:sfns_profiles}
\end{figure}

To test the sensitivity of our principal conclusions to the mechanical boundary conditions, a subset of simulations at fixed $Q=10^6$ with no-slip boundary conditions were carried out. For this value of $Q$, the critical parameters for stress-free and no-slip boundary conditions are $(Ra^\text{SF}_c, k^\text{SF}_c) = (1.0281\times10^7, 18.9823)$ and $(Ra^\text{NS}_c, k^\text{NS}_c) = (1.0315\times10^7, 18.9676)$, respectively. The difference in critical parameters is therefore less than 1\%. Figure \ref{F:sfns_profiles} shows vertical rms profiles of the flow field using no-slip mechanical (solid-line) and stress-free (dashed-line) boundary conditions with $Ra = 10^8$, i.e.~$Ra/Ra_c \approx 9.7$. The profiles are broadly similar in shape and amplitude for both sets of boundary conditions. Small differences in the amplitude of each variable are most notable in the fluid interior. Note that despite these differences in amplitude, the thermal boundary layer thickness is nearly identical for both cases, as shown in Fig.~\ref{F:sfns_profiles}(a).

As shown in the inset of Fig.~\ref{F:sfns_profiles}(b), the horizontal velocity field develops Hartmann boundary layers which are significantly thinner than the thermal boundary layers \cite{uB01}. A balance between the viscous force and the Lorentz force would suggest a characteristic scaling
\be
\nabla^2 u_\perp \sim Q \partial_z b_\perp \quad \Rightarrow \quad \delta_u \sim \sqrt{ \frac{\delta_b u_\perp}{Q b_\perp}} = O(\ve^{-3}),
\ee
which corresponds to the classical Hartmann boundary layer thickness $O(Q^{-1/2})$ \citep[e.g.][]{pD01}. In the final equality we use the boundary-modified amplitudes of $b_\perp = \delta_b = O(\ve)$, along with $u_\perp = O(1)$. Although we have not explored varying $Q$ for no-slip boundary conditions, the consistency with the asymptotic scalings and the results shown in Fig.~\ref{F:sfns_profiles} suggests that Hartmann boundary layers play no role in the heat transport in the $Q \rightarrow \infty$ limit.

\begin{figure}
\centering
\subfloat[][]{\includegraphics[width=0.47\textwidth]{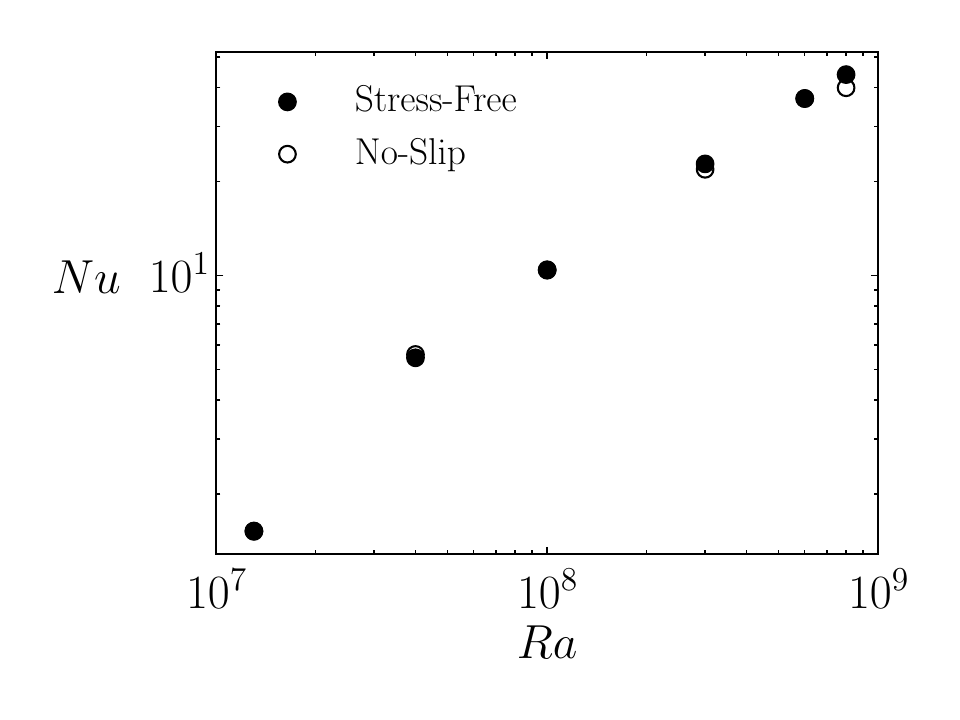}} \quad
\subfloat[][]{\includegraphics[width=0.47\textwidth]{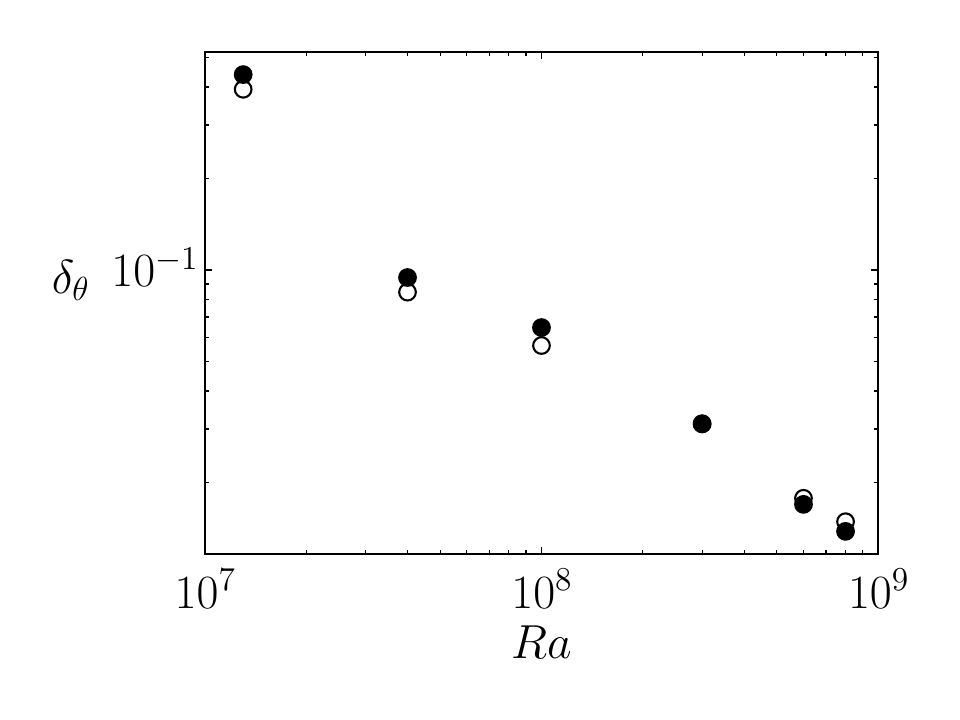}} 
\caption{Comparison of stress-free (solid symbols) and no-slip (open symbols) boundary conditions for $Q = 10^6$: (a) $Nu$ vs $Ra$; (b) $\delta_\theta$ vs $Ra$. } \label{F:sfns}
\end{figure}

The Nusselt number and thermal boundary layer thickness are shown in Fig.~\ref{F:sfns} for varying $Ra$. Both $Nu$ and $\delta_\theta$ are nearly identical, suggesting an insensitivity to the mechanical boundary conditions. Figure \ref{F:sfns_profiles} demonstrates that both $\theta$ and $w$ are primarily unaffected by the choice of mechanical boundary conditions, therefore the insensitivity of $Nu$ and $\delta_\theta$ may not be surprising given the definition of $Nu$ in equation (\ref{nudef}) coupled with the expected relation $Nu \sim \delta_\theta^{-1}$. Small discrepancies may be attributable to the slight differences in the critical Rayleigh numbers. Figure \ref{F:sfns} further supports the claim that the dynamics are insensitive to the mechanical boundary conditions chosen for this problem.

\subsection{Influence of electromagnetic boundary conditions} \label{S: bcE}

 \begin{figure}
\centering
\subfloat[][]{\includegraphics[width=0.48\textwidth]{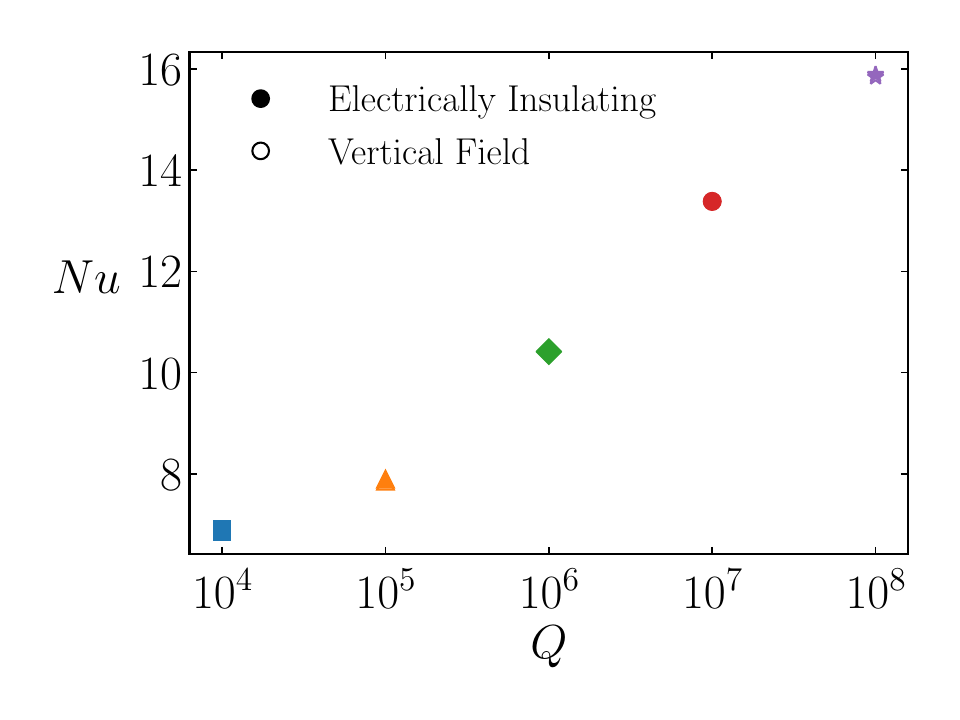}} \quad
\subfloat[][]{\includegraphics[width=0.48\textwidth]{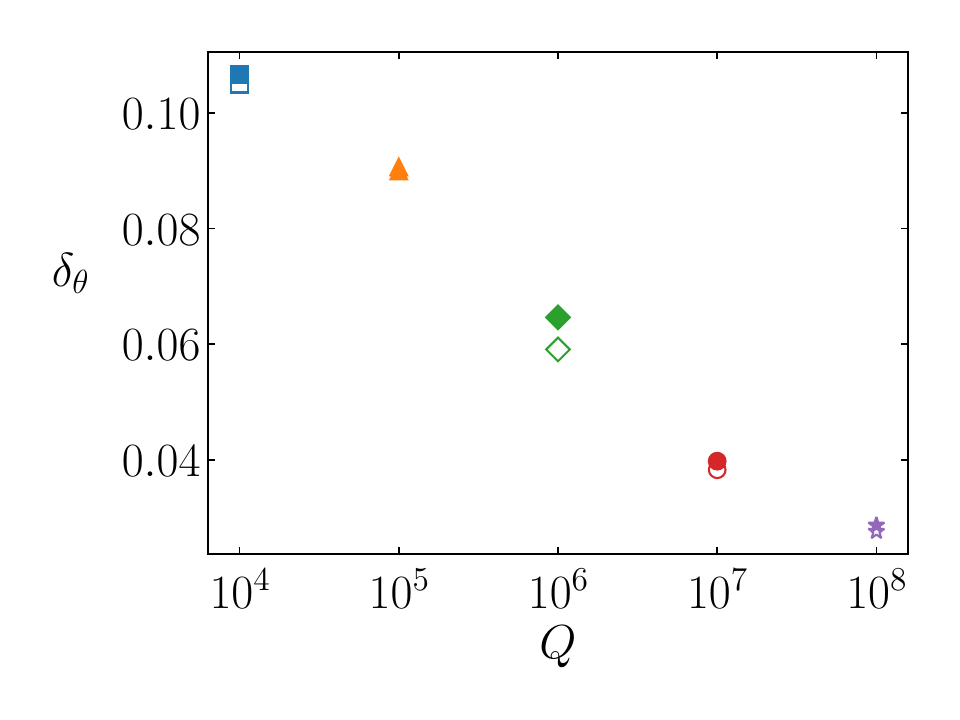}} 
\caption{Comparison of electrically insulating (solid symbols) and vertical field (open symbols) electromagnetic boundary conditions. (a) $Nu$ vs $Q$; (b) $\delta_\theta$ vs $Q$. The supercriticality is fixed at $Ra \approx 10Ra_c$. } \label{F:vf_ei}
\end{figure}

A subset of simulations were conducted using vertical field magnetic boundary conditions where the magnetic field is made to satisfy (\ref{vf}). Heat transport data for fixed supercriticality $Ra \approx 10 Ra_c$ is shown in Fig.~\ref{F:vf_ei}(a) as a function of $Q$. The data shows an insensitivity to the change in electromagnetic boundary conditions. A consequence of this finding is that the thermal boundary layer thickness is almost identical between electrically insulating and vertical field magnetic boundary conditions, as confirmed in Fig.~\ref{F:vf_ei}(b).  As opposed to the linear magnetic boundary layers formed with electrically insulating boundary conditions, the development of thermal boundary layers is a nonlinear phenomenon occurring from the balance of thermal advection and diffusion. This causes the thermal boundary layer to be magnetically controlled by the asymptotic scaling of the vertical velocity field. This balance is independent of the electromagnetic boundary conditions chosen and thus the equality of $Nu$ and $\delta_\theta$ between vertical field and electrically insulating boundary conditions may be expected.

 \begin{figure}
\centering
\subfloat[][]{\includegraphics[width=0.45\textwidth]{\detokenize{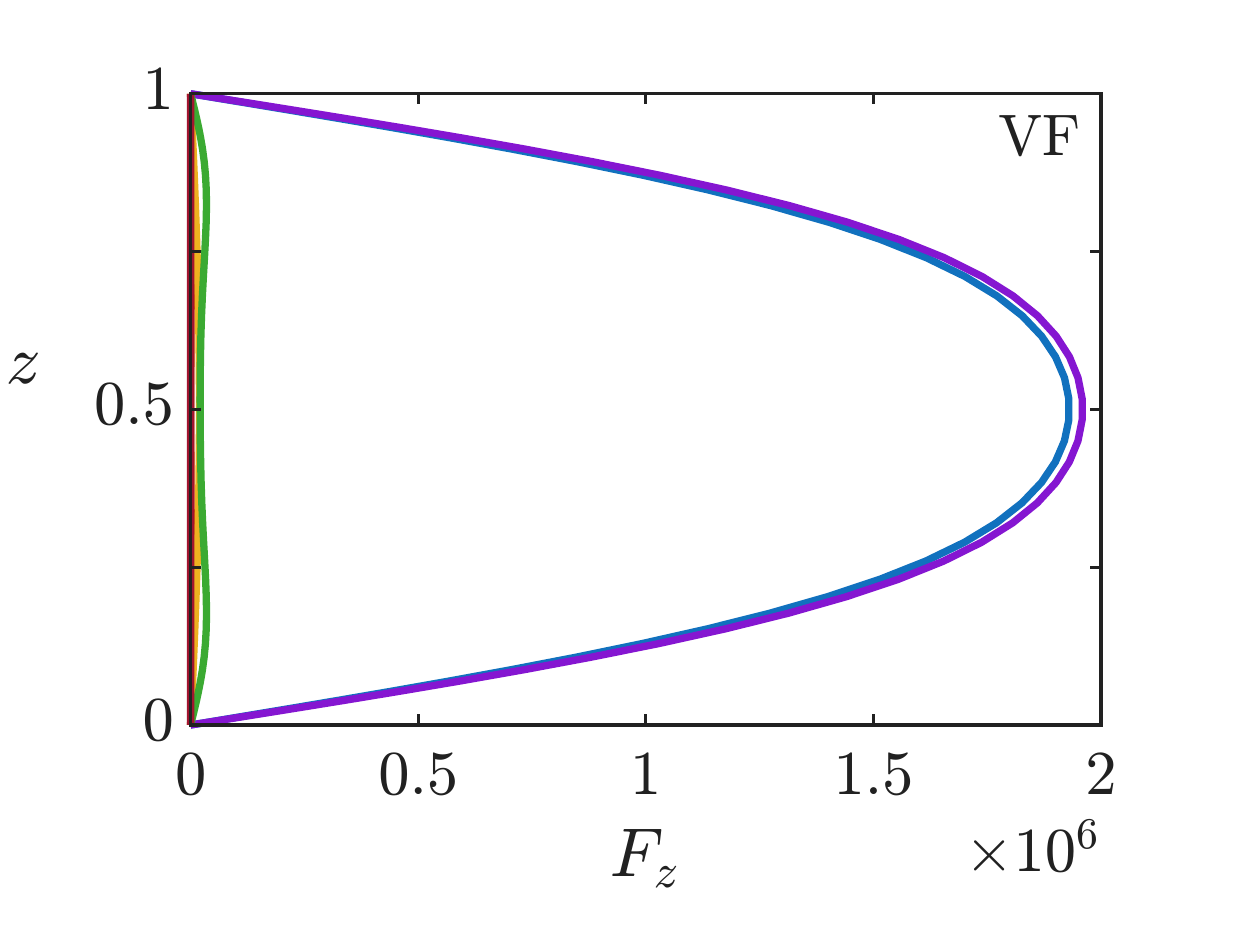}}}
\subfloat[][]{\includegraphics[width=0.45\textwidth]{\detokenize{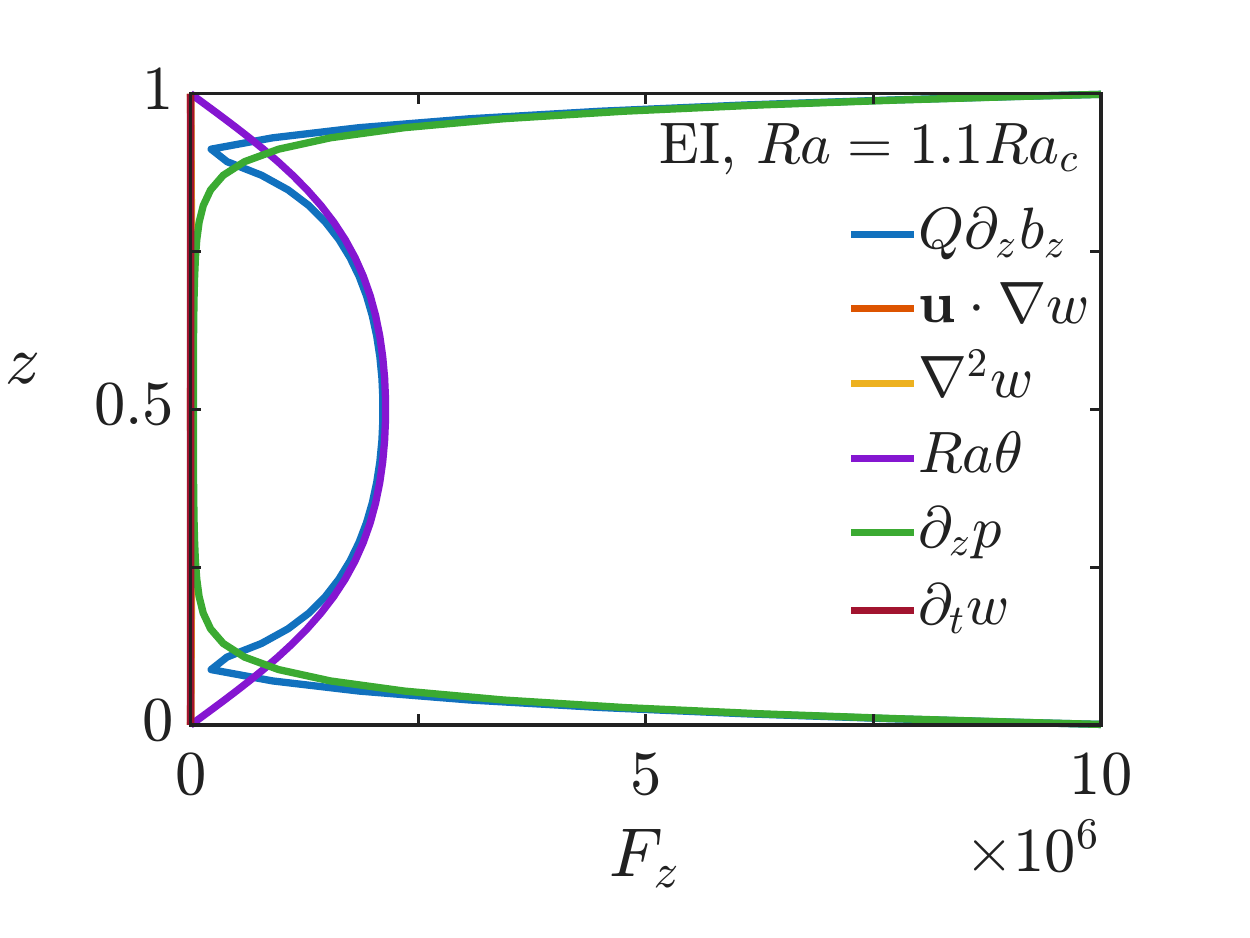}}} \\
\subfloat[][]{\includegraphics[width=0.45\textwidth]{\detokenize{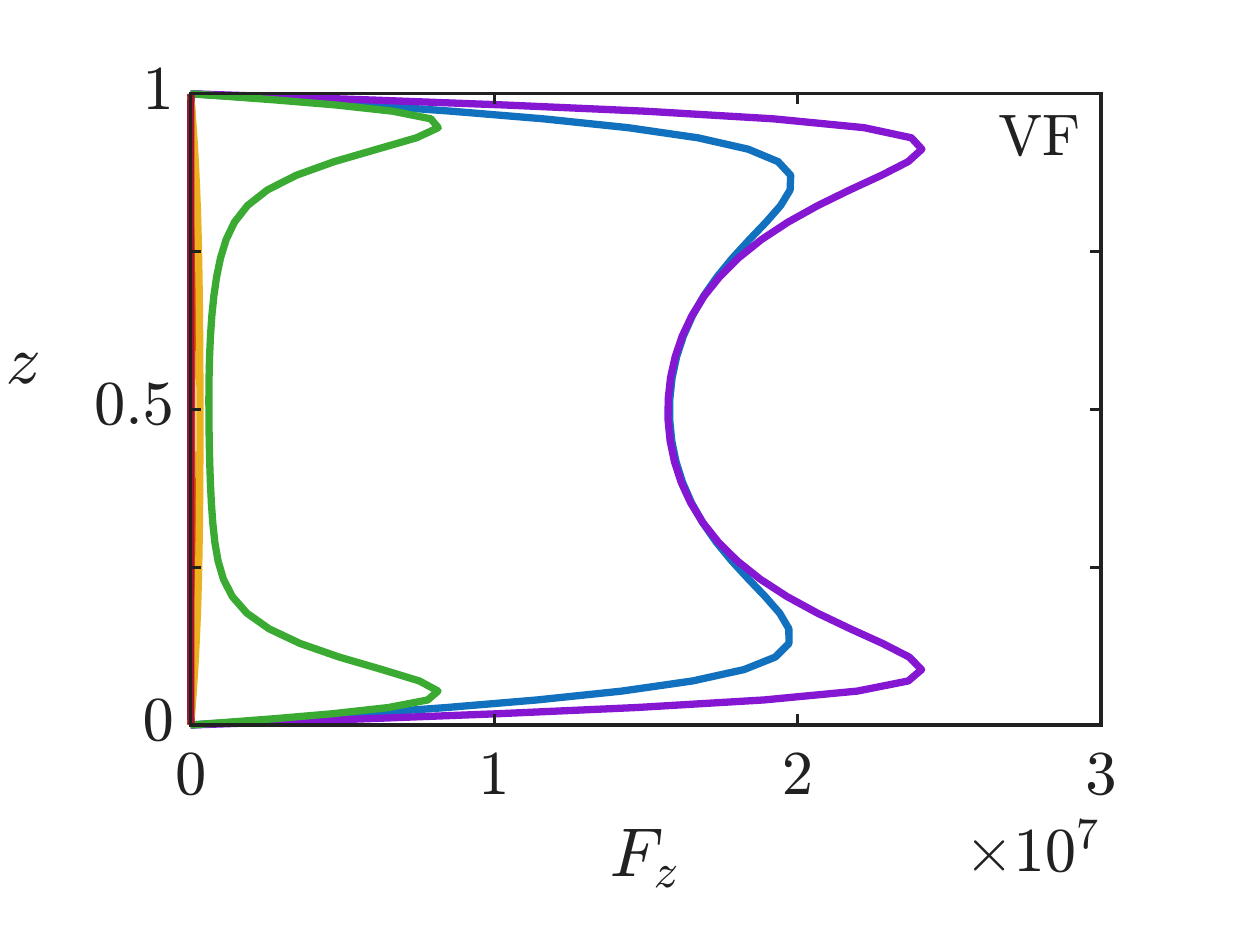}}}
\subfloat[][]{\includegraphics[width=0.45\textwidth]{\detokenize{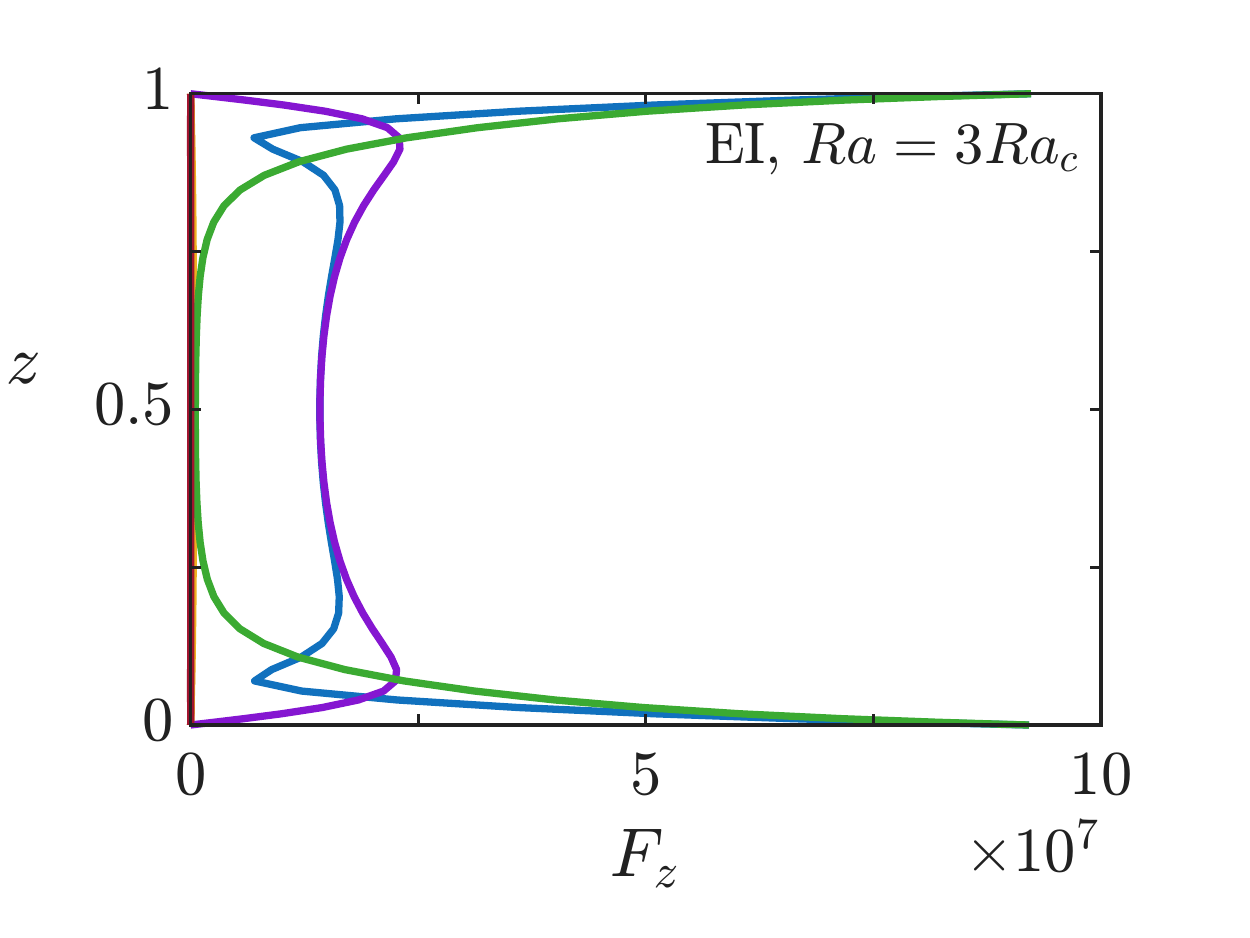}}} 
\caption{Horizontal rms of the instantaneous vertical force balances for (a,c) vertical field (VF) and (b,d) electrically insulating (EI) boundary conditions for $Q = 10^7$. Subfigures (a,b) show $Ra = 1.1Ra_c$, subfigures (c,d) show $Ra = 3Ra_c$.} \label{F:forces_onset}
\end{figure}

Figure \ref{F:forces_onset} shows instantaneous vertical profiles of the horizontal rms force balances near the onset of convection for $Q = 10^7$ with $Ra = 1.1Ra_c$ in Fig.~\ref{F:forces_onset}(a,b) and $Ra = 3Ra_c$ in Fig.~\ref{F:forces_onset}(c,d). We note that the hydrostatic balance has been removed from the force balances. Figure \ref{F:forces_onset}(a) shows the dominant balance achieved using vertical field (VF) boundary conditions. As predicted by Ref.~\cite{pM99}, the Lorentz force balances buoyancy uniformly throughout the entire domain. Fig.~\ref{F:forces_onset}(b) shows the vertical force balances using electrically insulating (EI) boundary conditions. The EI interior dynamics are virtually indistinguishable from the state achieved using VF boundary conditions where buoyancy balances the Lorentz force. Near the boundaries, pressure balances the Lorentz force in localized $O(\ve)$ boundary layers. As discussed earlier, this balance is achieved due to the increased asymptotic amplitude of the horizontal magnetic field near the boundaries, as shown in Fig.~\ref{F:b_bl}. The increased amplitude of the horizontal magnetic field near the boundaries arises as a consequence of the flow isotropy and the solenoidal condition.

For $Ra = 3 Ra_c$, Fig.~\ref{F:forces_onset}(c,d) demonstrate the development of thermal boundary layers under both VF and EI boundary conditions. As argued earlier, a balance of thermal advection and diffusion would cause $\delta_\theta = O(\ve)$ irrespective of the electromagnetic boundary conditions. The invariance of $\delta_\theta$ to the choice of magnetic boundary condition is further supported by Fig.~\ref{F:vf_ei}(b). In both Fig.~\ref{F:forces_onset}(c,d) we observe an increase in pressure near the boundaries. In the case of EI this is to be expected. With VF electromagnetic boundary conditions, the increase of pressure near the boundaries can be explained by the development of asymptotic thermal boundary layers as the coupling between buoyancy and the Lorentz force throughout the entirety of the domain would imply the existence of magnetic boundary layers of the same asymptotic thickness. As such, the arguments developed for the increase in pressure near the boundaries under EI boundary conditions may extend to the asymptotic description of nonlinear VF dynamics. Moreover, a distinct force balance near the boundaries could necessitate a different asymptotic amplitude to the horizontal magnetic field. A more exhaustive analysis using VF conditions would need to be done to understand the impact of the thermal boundary layers on the flow dynamics, but our results suggest that the fully nonlinear asymptotic state achieved with VF is similar to the case of EI where the pressure gradient is a dominant force near the boundaries.

\section{Conclusions} \label{S:conclude}

Previous studies have shown that the Nusselt number in magnetoconvection with a vertical field (VMC) increases more strongly with Rayleigh number as the strength of the imposed magnetic field ($Q$) is increased, with no clear evidence that the system approaches an asymptote and becomes independent of $Q$ \citep{mY19}. Like non-magnetic RBC, the fluid interior in VMC becomes isothermal for sufficiently large supercriticality such that the conductive heat transport across the thermal boundary layers control the heat transport. The present work has shown that the thermal boundary layer thickness in VMC exhibits an $O(Q^{-1/6})$ scaling in the limit $Q \rightarrow \infty$. This scaling is interpreted as arising naturally from the anisotropic nature of VMC -- thin layers are necessary for satisfying the imposed boundary conditions. Thus, for a fixed value of the supercriticality, $Ra/Ra_c$, heat transport will necessarily increase as $Q$ is increased. These findings suggest that heat transport in VMC does not asymptote to a constant value. 

Data from our simulations was compared with asymptotic theory. The linear asymptotic theory of \cite{pM99} was extended to encapsulate the effects of electrically-insulating boundary conditions. The interior solutions are characterized by asymptotic scalings that are consistent with \cite{pM99}. The presence of boundary layers require modification of the scaling of the magnetic field components as well as the fluctuating temperature field. In particular, the solenoidal constraint implies that the three components of the induced magnetic field must follow identical isotropic asymptotic scalings in the boundary layer. These modified scalings have implications on heat transport, namely that $Nu = O(Q^{1/6})$ in the regime where the flow is magnetically constrained and when heat transport is limited by conduction across the thermal boundary layers. Furthermore, the boundary layer scalings modify the dominant vertical force balance to be between the pressure gradient and the Lorentz force in the linear regime, prior to the development of thermal boundary layers. In the nonlinear regime, the dominant force balance is between the pressure gradient, the Lorentz force, and the buoyancy force.

A subset of simulations were conducted to investigate the influence of boundary conditions. Both no-slip boundary conditions and vertical magnetic field boundary conditions were tested. Though limited in the parameter space, these simulations show that heat transport is essentially independent of the choice of mechanical and electromagnetic boundary conditions. Hartmann boundary layers, with thickness $O(Q^{-1/2})$, are evident in the no-slip simulations but play no dynamical role. The use of vertical field conditions leads to differences in the vertical profiles of the induced magnetic field, but only within the vicinity of the boundary, and $O(Q^{-1/6})$ boundary layers may still be present as a consequence of the development of asymptotic thermal boundary layers. These observations indicate that the thermal boundary layers are uninfluenced by the choice of boundary conditions on other fields, indicating that the principal conclusions of this study are generalizable. 

Like VMC, convection in the presence of uniform background rotation (rotating Rayleigh-B\'enard convection, RRBC) with a non-zero component parallel to gravity represents an example of constrained fluid dynamics. Both systems are stabilized by the constraining force and small perturbations away from the leading order force balance become necessary for convection to occur. Motions along the direction of the constraining force are preferred which leads to anisotropy in the flow field. Linear theory shows that in both cases the viscous force is necessary to select the most unstable wavelength \citep{sC61}. Asymptotic analysis of the governing equations, relying on expansions in small parameters, has been helpful for improving understanding for VMC and RRBC, particularly in the latter case since a fully nonlinear multimodal model is possible \citep{kJ99,mS06}. This reduced model shows excellent agreement with direct numerical simulations and has paved the way for a deeper understanding of RRBC \citep{sS14,aK25}. However, rotationally dependent thermal boundary layers do not appear in RRBC when stress free mechanical boundary conditions are used. In addition, the heat transport in RRBC is controlled by the fluid interior given that interior temperature gradient saturates at a finite value for sufficiently constrained motions \citep{kJ96,kJ12}. As a result the Nusselt number approaches an asymptote as the influence of the Coriolis force increases \citep{aK25}. We have shown that there is no indication of a similar trend in VMC though the reason for this difference is not fully understood. Ultimately, simulations at yet larger values of $Q$ may be necessary to fully unravel the details of this problem.


The present study has focused on the case of an imposed magnetic field, though it is possible that similar boundary layers form when the magnetic field is self-sustaining, i.e.~dynamo generated. Whether such boundary layers form will depend on the type of dynamo. Dynamos can be classified based on the degree of scale separation between the magnetic and velocity fields. Small-scale dynamos generate magnetic field with length scales that are comparable to the length scales present within the velocity field \citep[e.g.][]{mY21}, whereas large-scale dynamos generate magnetic fields with scales that are considerably larger than the corresponding velocity field that drives the magnetic field \citep{mY22}. Some physical ingredients, such as system rotation, are known to be conducive to the generation of large scale magnetic field \citep{sC72,aS74,mY22}. Large scale field with a strong vertical component would be necessary to generate boundary layers similar to those studied here. Unfortunately such field does not form in a plane layer geometry, thus requiring the use of a global spherical geometry. An investigation of the details of boundary layers in rotating spherical dynamos with strong magnetic field would be of interest. In particular, an analysis of the field strength in the vicinity of the boundary layer, and the associated balances that are obtainable. Much remains to be understood when both rotation and magnetic field constrain convection \citep[e.g.][]{sH22,sH25}.

\newpage
\section*{Funding}
This research was supported with funding from the National Science Foundation through grant EAR-1945270. T.A. was supported by Kuwait University. 
The Anvil supercomputer at Purdue University was made available through allocation PHY180013 from the Advanced Cyberinfrastructure Coordination Ecosystem: Services \& Support (ACCESS) program, which is supported by National Science Foundation grants \#2138259, \#2138286, \#2138307, \#2137603, and \#2138296. This work also utilized the Alpine high performance computing resource at the University of Colorado Boulder. Alpine is jointly funded by the University of Colorado Boulder, the University of Colorado Anschutz, and Colorado State University.

\section*{Declaration of interests}
The authors report no conflict of interest.

\section*{References}
\bibliography{journal_abbreviations,References}

\providecommand{\jfm}{J. Fluid Mech.~}\providecommand{\apj}{Astrophys.
  J.~}\providecommand{\jpp}{J. Plasma Phys.~}\providecommand{\mnras}{Mon. Not.
  Roy. Astron. Soc. }\providecommand{\jgr}{J. Geophys.
  Res.~}\providecommand{\araa}{Annu. Rev. Astron.
  Astrophys.~}\providecommand{\areps}{Annu. Rev. Earth Planet.
  Sci.~}\providecommand{\icarus}{Icarus }\providecommand{\aap}{Astron.
  Astrophys.~}\providecommand{\physscr}{Phys. Scripta
  }\providecommand{\ssr}{Space Sci. Rev.~}\providecommand{\pnas}{Proc. Nat.
  Acad. Sci.~}\providecommand{\prsa}{Proc. R. Soc.
  A~}\providecommand{\ncom}{Nat. Comm.~}\providecommand{\njp}{New J.
  Phys.~}\providecommand{\prf}{Phys. Rev. Fluids }\providecommand{\prr}{Phys.
  Rev. Res. }\providecommand{\pre}{Phys. Rev. E }\providecommand{\pepi}{Phys.
  Earth Planet. Int.~}\providecommand{\gji}{Geophys. J.
  Int.~}\providecommand{\grl}{Geophys. Res. Lett.~}
\begin{thebibliography}{48}%
\makeatletter
\providecommand \@ifxundefined [1]{%
 \@ifx{#1\undefined}
}%
\providecommand \@ifnum [1]{%
 \ifnum #1\expandafter \@firstoftwo
 \else \expandafter \@secondoftwo
 \fi
}%
\providecommand \@ifx [1]{%
 \ifx #1\expandafter \@firstoftwo
 \else \expandafter \@secondoftwo
 \fi
}%
\providecommand \natexlab [1]{#1}%
\providecommand \enquote  [1]{``#1''}%
\providecommand \bibnamefont  [1]{#1}%
\providecommand \bibfnamefont [1]{#1}%
\providecommand \citenamefont [1]{#1}%
\providecommand \href@noop [0]{\@secondoftwo}%
\providecommand \href [0]{\begingroup \@sanitize@url \@href}%
\providecommand \@href[1]{\@@startlink{#1}\@@href}%
\providecommand \@@href[1]{\endgroup#1\@@endlink}%
\providecommand \@sanitize@url [0]{\catcode `\\12\catcode `\$12\catcode
  `\&12\catcode `\#12\catcode `\^12\catcode `\_12\catcode `\%12\relax}%
\providecommand \@@startlink[1]{}%
\providecommand \@@endlink[0]{}%
\providecommand \url  [0]{\begingroup\@sanitize@url \@url }%
\providecommand \@url [1]{\endgroup\@href {#1}{\urlprefix }}%
\providecommand \urlprefix  [0]{URL }%
\providecommand \Eprint [0]{\href }%
\providecommand \doibase [0]{https://doi.org/}%
\providecommand \selectlanguage [0]{\@gobble}%
\providecommand \bibinfo  [0]{\@secondoftwo}%
\providecommand \bibfield  [0]{\@secondoftwo}%
\providecommand \translation [1]{[#1]}%
\providecommand \BibitemOpen [0]{}%
\providecommand \bibitemStop [0]{}%
\providecommand \bibitemNoStop [0]{.\EOS\space}%
\providecommand \EOS [0]{\spacefactor3000\relax}%
\providecommand \BibitemShut  [1]{\csname bibitem#1\endcsname}%
\let\auto@bib@innerbib\@empty
\bibitem [{\citenamefont {Jones}(2011)}]{cJ11b}%
  \BibitemOpen
  \bibfield  {author} {\bibinfo {author} {\bibfnamefont {C.~A.}\ \bibnamefont
  {Jones}},\ }\bibfield  {title} {\bibinfo {title} {Planetary magnetic fields
  and fluid dynamos},\ }\href@noop {} {\bibfield  {journal} {\bibinfo
  {journal} {Annu. Rev. Fluid Mech.}\ }\textbf {\bibinfo {volume} {43}},\
  \bibinfo {pages} {583} (\bibinfo {year} {2011})}\BibitemShut {NoStop}%
\bibitem [{\citenamefont {Soderlund}\ \emph {et~al.}(2025)\citenamefont
  {Soderlund}, \citenamefont {Stanley}, \citenamefont {Cao}, \citenamefont
  {Calkins},\ and\ \citenamefont {Browning}}]{kS25}%
  \BibitemOpen
  \bibfield  {author} {\bibinfo {author} {\bibfnamefont {K.~M.}\ \bibnamefont
  {Soderlund}}, \bibinfo {author} {\bibfnamefont {S.}~\bibnamefont {Stanley}},
  \bibinfo {author} {\bibfnamefont {H.}~\bibnamefont {Cao}}, \bibinfo {author}
  {\bibfnamefont {M.~A.}\ \bibnamefont {Calkins}},\ and\ \bibinfo {author}
  {\bibfnamefont {M.~K.}\ \bibnamefont {Browning}},\ }\bibfield  {title}
  {\bibinfo {title} {Puzzles in planetary dynamos: Implications for planetary
  interiors},\ }\href@noop {} {\bibfield  {journal} {\bibinfo  {journal}
  {\areps}\ }\textbf {\bibinfo {volume} {53}} (\bibinfo {year}
  {2025})}\BibitemShut {NoStop}%
\bibitem [{\citenamefont {Cioni}\ \emph {et~al.}(2000)\citenamefont {Cioni},
  \citenamefont {Chaumat},\ and\ \citenamefont {Sommeria}}]{sC00}%
  \BibitemOpen
  \bibfield  {author} {\bibinfo {author} {\bibfnamefont {S.}~\bibnamefont
  {Cioni}}, \bibinfo {author} {\bibfnamefont {S.}~\bibnamefont {Chaumat}},\
  and\ \bibinfo {author} {\bibfnamefont {J.}~\bibnamefont {Sommeria}},\
  }\bibfield  {title} {\bibinfo {title} {{Effect of a vertical magnetic field
  on turbulent Rayleigh-B\'enard convection}},\ }\href@noop {} {\bibfield
  {journal} {\bibinfo  {journal} {Phys. Rev. E}\ }\textbf {\bibinfo {volume}
  {62}} (\bibinfo {year} {2000})}\BibitemShut {NoStop}%
\bibitem [{\citenamefont {Aurnou}\ and\ \citenamefont {Olson}(2001)}]{jmA01}%
  \BibitemOpen
  \bibfield  {author} {\bibinfo {author} {\bibfnamefont {J.~M.}\ \bibnamefont
  {Aurnou}}\ and\ \bibinfo {author} {\bibfnamefont {P.}~\bibnamefont {Olson}},\
  }\bibfield  {title} {\bibinfo {title} {{Experiments on Rayleigh-B\'enard
  convection, magnetoconvection, and rotating magnetoconvection in liquid
  gallium}},\ }\href@noop {} {\bibfield  {journal} {\bibinfo  {journal} {J.
  Fluid Mech.}\ }\textbf {\bibinfo {volume} {430}},\ \bibinfo {pages} {283}
  (\bibinfo {year} {2001})}\BibitemShut {NoStop}%
\bibitem [{\citenamefont {Burr}\ and\ \citenamefont {M\"uller}(2001)}]{uB01}%
  \BibitemOpen
  \bibfield  {author} {\bibinfo {author} {\bibfnamefont {U.}~\bibnamefont
  {Burr}}\ and\ \bibinfo {author} {\bibfnamefont {U.}~\bibnamefont
  {M\"uller}},\ }\bibfield  {title} {\bibinfo {title} {{Rayleigh-B\'enard in
  liquid metal layers under the influence of a vertical magnetic field}},\
  }\href@noop {} {\bibfield  {journal} {\bibinfo  {journal} {Phys. Fluids}\
  }\textbf {\bibinfo {volume} {13}} (\bibinfo {year} {2001})}\BibitemShut
  {NoStop}%
\bibitem [{\citenamefont {Z{\"u}rner}\ \emph {et~al.}(2016)\citenamefont
  {Z{\"u}rner}, \citenamefont {Liu}, \citenamefont {Krasnov},\ and\
  \citenamefont {Schumacher}}]{tZ16}%
  \BibitemOpen
  \bibfield  {author} {\bibinfo {author} {\bibfnamefont {T.}~\bibnamefont
  {Z{\"u}rner}}, \bibinfo {author} {\bibfnamefont {W.}~\bibnamefont {Liu}},
  \bibinfo {author} {\bibfnamefont {D.}~\bibnamefont {Krasnov}},\ and\ \bibinfo
  {author} {\bibfnamefont {J.}~\bibnamefont {Schumacher}},\ }\bibfield  {title}
  {\bibinfo {title} {Heat and momentum transfer for magnetoconvection in a
  vertical external magnetic field},\ }\href@noop {} {\bibfield  {journal}
  {\bibinfo  {journal} {Phys. Rev. E}\ }\textbf {\bibinfo {volume} {94}},\
  \bibinfo {pages} {043108} (\bibinfo {year} {2016})}\BibitemShut {NoStop}%
\bibitem [{\citenamefont {Liu}\ \emph {et~al.}(2018)\citenamefont {Liu},
  \citenamefont {Krasnov},\ and\ \citenamefont {Schumacher}}]{wL18}%
  \BibitemOpen
  \bibfield  {author} {\bibinfo {author} {\bibfnamefont {W.}~\bibnamefont
  {Liu}}, \bibinfo {author} {\bibfnamefont {D.}~\bibnamefont {Krasnov}},\ and\
  \bibinfo {author} {\bibfnamefont {J.}~\bibnamefont {Schumacher}},\ }\bibfield
   {title} {\bibinfo {title} {Wall modes in magnetoconvection at high hartmann
  numbers},\ }\href@noop {} {\bibfield  {journal} {\bibinfo  {journal} {J.
  Fluid Mech.}\ }\textbf {\bibinfo {volume} {849}} (\bibinfo {year}
  {2018})}\BibitemShut {NoStop}%
\bibitem [{\citenamefont {Yan}\ \emph {et~al.}(2019)\citenamefont {Yan},
  \citenamefont {Calkins}, \citenamefont {Maffei}, \citenamefont {Julien},
  \citenamefont {Tobias},\ and\ \citenamefont {Marti}}]{mY19}%
  \BibitemOpen
  \bibfield  {author} {\bibinfo {author} {\bibfnamefont {M.}~\bibnamefont
  {Yan}}, \bibinfo {author} {\bibfnamefont {M.~A.}\ \bibnamefont {Calkins}},
  \bibinfo {author} {\bibfnamefont {S.}~\bibnamefont {Maffei}}, \bibinfo
  {author} {\bibfnamefont {K.}~\bibnamefont {Julien}}, \bibinfo {author}
  {\bibfnamefont {S.~M.}\ \bibnamefont {Tobias}},\ and\ \bibinfo {author}
  {\bibfnamefont {P.}~\bibnamefont {Marti}},\ }\bibfield  {title} {\bibinfo
  {title} {Heat transfer and flow regimes in quasi-static magnetoconvection
  with a vertical magnetic field},\ }\href@noop {} {\bibfield  {journal}
  {\bibinfo  {journal} {\jfm}\ }\textbf {\bibinfo {volume} {877}},\ \bibinfo
  {pages} {1186} (\bibinfo {year} {2019})}\BibitemShut {NoStop}%
\bibitem [{\citenamefont {Akhmedagaev}\ \emph {et~al.}(2020)\citenamefont
  {Akhmedagaev}, \citenamefont {Zikanov}, \citenamefont {Krasnov},\ and\
  \citenamefont {Schumacher}}]{rA20}%
  \BibitemOpen
  \bibfield  {author} {\bibinfo {author} {\bibfnamefont {R.}~\bibnamefont
  {Akhmedagaev}}, \bibinfo {author} {\bibfnamefont {O.}~\bibnamefont
  {Zikanov}}, \bibinfo {author} {\bibfnamefont {D.}~\bibnamefont {Krasnov}},\
  and\ \bibinfo {author} {\bibfnamefont {J.}~\bibnamefont {Schumacher}},\
  }\bibfield  {title} {\bibinfo {title} {{Turbulent Rayleigh--B{\'e}nard
  convection in a strong vertical magnetic field}},\ }\href@noop {} {\bibfield
  {journal} {\bibinfo  {journal} {\jfm}\ }\textbf {\bibinfo {volume} {895}},\
  \bibinfo {pages} {R4} (\bibinfo {year} {2020})}\BibitemShut {NoStop}%
\bibitem [{\citenamefont {Vogt}\ \emph {et~al.}(2021)\citenamefont {Vogt},
  \citenamefont {Horn},\ and\ \citenamefont {Aurnou}}]{tV21}%
  \BibitemOpen
  \bibfield  {author} {\bibinfo {author} {\bibfnamefont {T.}~\bibnamefont
  {Vogt}}, \bibinfo {author} {\bibfnamefont {S.}~\bibnamefont {Horn}},\ and\
  \bibinfo {author} {\bibfnamefont {J.~M.}\ \bibnamefont {Aurnou}},\ }\bibfield
   {title} {\bibinfo {title} {Oscillatory thermal--inertial flows in liquid
  metal rotating convection},\ }\href@noop {} {\bibfield  {journal} {\bibinfo
  {journal} {\jfm}\ }\textbf {\bibinfo {volume} {911}} (\bibinfo {year}
  {2021})}\BibitemShut {NoStop}%
\bibitem [{\citenamefont {Xu}\ \emph {et~al.}(2023)\citenamefont {Xu},
  \citenamefont {Horn},\ and\ \citenamefont {Aurnou}}]{yX23}%
  \BibitemOpen
  \bibfield  {author} {\bibinfo {author} {\bibfnamefont {Y.}~\bibnamefont
  {Xu}}, \bibinfo {author} {\bibfnamefont {S.}~\bibnamefont {Horn}},\ and\
  \bibinfo {author} {\bibfnamefont {J.~M.}\ \bibnamefont {Aurnou}},\ }\bibfield
   {title} {\bibinfo {title} {Transition from wall modes to multimodality in
  liquid gallium magnetoconvection},\ }\href@noop {} {\bibfield  {journal}
  {\bibinfo  {journal} {\prf}\ }\textbf {\bibinfo {volume} {8}},\ \bibinfo
  {pages} {103503} (\bibinfo {year} {2023})}\BibitemShut {NoStop}%
\bibitem [{\citenamefont {Bhattacharya}\ \emph {et~al.}(2023)\citenamefont
  {Bhattacharya}, \citenamefont {Boeck}, \citenamefont {Krasnov},\ and\
  \citenamefont {Schumacher}}]{sB23}%
  \BibitemOpen
  \bibfield  {author} {\bibinfo {author} {\bibfnamefont {S.}~\bibnamefont
  {Bhattacharya}}, \bibinfo {author} {\bibfnamefont {T.}~\bibnamefont {Boeck}},
  \bibinfo {author} {\bibfnamefont {D.}~\bibnamefont {Krasnov}},\ and\ \bibinfo
  {author} {\bibfnamefont {J.}~\bibnamefont {Schumacher}},\ }\bibfield  {title}
  {\bibinfo {title} {Effects of strong fringing magnetic fields on turbulent
  thermal convection},\ }\href@noop {} {\bibfield  {journal} {\bibinfo
  {journal} {\jfm}\ }\textbf {\bibinfo {volume} {964}},\ \bibinfo {pages} {A31}
  (\bibinfo {year} {2023})}\BibitemShut {NoStop}%
\bibitem [{\citenamefont {Cresswell}\ \emph {et~al.}(2023)\citenamefont
  {Cresswell}, \citenamefont {Anders}, \citenamefont {Brown}, \citenamefont
  {Oishi},\ and\ \citenamefont {Vasil}}]{iC23}%
  \BibitemOpen
  \bibfield  {author} {\bibinfo {author} {\bibfnamefont {I.~G.}\ \bibnamefont
  {Cresswell}}, \bibinfo {author} {\bibfnamefont {E.~H.}\ \bibnamefont
  {Anders}}, \bibinfo {author} {\bibfnamefont {B.~P.}\ \bibnamefont {Brown}},
  \bibinfo {author} {\bibfnamefont {J.~S.}\ \bibnamefont {Oishi}},\ and\
  \bibinfo {author} {\bibfnamefont {G.~M.}\ \bibnamefont {Vasil}},\ }\bibfield
  {title} {\bibinfo {title} {Force balances in strong-field magnetoconvection
  simulations},\ }\href@noop {} {\bibfield  {journal} {\bibinfo  {journal}
  {\prf}\ }\textbf {\bibinfo {volume} {8}},\ \bibinfo {pages} {093503}
  (\bibinfo {year} {2023})}\BibitemShut {NoStop}%
\bibitem [{\citenamefont {Bader}\ and\ \citenamefont {Zhu}(2023)}]{shB23}%
  \BibitemOpen
  \bibfield  {author} {\bibinfo {author} {\bibfnamefont {S.~H.}\ \bibnamefont
  {Bader}}\ and\ \bibinfo {author} {\bibfnamefont {X.}~\bibnamefont {Zhu}},\
  }\bibfield  {title} {\bibinfo {title} {Scaling relations in quasi-static
  magnetoconvection with a strong vertical magnetic field},\ }\href@noop {}
  {\bibfield  {journal} {\bibinfo  {journal} {\jfm}\ }\textbf {\bibinfo
  {volume} {976}},\ \bibinfo {pages} {A4} (\bibinfo {year} {2023})}\BibitemShut
  {NoStop}%
\bibitem [{\citenamefont {Teimurazov}\ \emph {et~al.}(2024)\citenamefont
  {Teimurazov}, \citenamefont {McCormack}, \citenamefont {Linkmann},\ and\
  \citenamefont {Shishkina}}]{aT24}%
  \BibitemOpen
  \bibfield  {author} {\bibinfo {author} {\bibfnamefont {A.}~\bibnamefont
  {Teimurazov}}, \bibinfo {author} {\bibfnamefont {M.}~\bibnamefont
  {McCormack}}, \bibinfo {author} {\bibfnamefont {M.}~\bibnamefont
  {Linkmann}},\ and\ \bibinfo {author} {\bibfnamefont {O.}~\bibnamefont
  {Shishkina}},\ }\bibfield  {title} {\bibinfo {title} {{Unifying heat
  transport model for the transition between buoyancy-dominated and
  Lorentz-force-dominated regimes in quasistatic magnetoconvection}},\
  }\href@noop {} {\bibfield  {journal} {\bibinfo  {journal} {\jfm}\ }\textbf
  {\bibinfo {volume} {980}},\ \bibinfo {pages} {R3} (\bibinfo {year}
  {2024})}\BibitemShut {NoStop}%
\bibitem [{\citenamefont {Nicoski}\ \emph {et~al.}(2022)\citenamefont
  {Nicoski}, \citenamefont {Yan},\ and\ \citenamefont {Calkins}}]{jN22}%
  \BibitemOpen
  \bibfield  {author} {\bibinfo {author} {\bibfnamefont {J.~A.}\ \bibnamefont
  {Nicoski}}, \bibinfo {author} {\bibfnamefont {M.}~\bibnamefont {Yan}},\ and\
  \bibinfo {author} {\bibfnamefont {M.~A.}\ \bibnamefont {Calkins}},\
  }\bibfield  {title} {\bibinfo {title} {Quasistatic magnetoconvection with a
  tilted magnetic field},\ }\href@noop {} {\bibfield  {journal} {\bibinfo
  {journal} {\prf}\ }\textbf {\bibinfo {volume} {7}},\ \bibinfo {pages}
  {043504} (\bibinfo {year} {2022})}\BibitemShut {NoStop}%
\bibitem [{\citenamefont {Calkins}\ \emph {et~al.}(2023)\citenamefont
  {Calkins}, \citenamefont {AlRefae}, \citenamefont {Hernandez}, \citenamefont
  {Yan},\ and\ \citenamefont {Maffei}}]{chmc23}%
  \BibitemOpen
  \bibfield  {author} {\bibinfo {author} {\bibfnamefont {M.~A.}\ \bibnamefont
  {Calkins}}, \bibinfo {author} {\bibfnamefont {T.}~\bibnamefont {AlRefae}},
  \bibinfo {author} {\bibfnamefont {A.}~\bibnamefont {Hernandez}}, \bibinfo
  {author} {\bibfnamefont {M.}~\bibnamefont {Yan}},\ and\ \bibinfo {author}
  {\bibfnamefont {S.}~\bibnamefont {Maffei}},\ }\bibfield  {title} {\bibinfo
  {title} {Numerical investigation of quasistatic magnetoconvection with an
  imposed horizontal magnetic field},\ }\href@noop {} {\bibfield  {journal}
  {\bibinfo  {journal} {\prf}\ }\textbf {\bibinfo {volume} {8}},\ \bibinfo
  {pages} {123501} (\bibinfo {year} {2023})}\BibitemShut {NoStop}%
\bibitem [{\citenamefont {Moffatt}\ and\ \citenamefont {Dormy}(2019)}]{kM19}%
  \BibitemOpen
  \bibfield  {author} {\bibinfo {author} {\bibfnamefont {K.}~\bibnamefont
  {Moffatt}}\ and\ \bibinfo {author} {\bibfnamefont {E.}~\bibnamefont
  {Dormy}},\ }\href@noop {} {\emph {\bibinfo {title} {Self-exciting fluid
  dynamos}}},\ Vol.~\bibinfo {volume} {59}\ (\bibinfo  {publisher} {Cambridge
  University Press},\ \bibinfo {year} {2019})\BibitemShut {NoStop}%
\bibitem [{\citenamefont {Ossendrijver}(2003)}]{mO03}%
  \BibitemOpen
  \bibfield  {author} {\bibinfo {author} {\bibfnamefont {M.}~\bibnamefont
  {Ossendrijver}},\ }\bibfield  {title} {\bibinfo {title} {The solar dynamo},\
  }\href@noop {} {\bibfield  {journal} {\bibinfo  {journal} {Astron. Astrophys.
  Rev.}\ }\textbf {\bibinfo {volume} {11}},\ \bibinfo {pages} {287} (\bibinfo
  {year} {2003})}\BibitemShut {NoStop}%
\bibitem [{\citenamefont {Chandrasekhar}(1961)}]{sC61}%
  \BibitemOpen
  \bibfield  {author} {\bibinfo {author} {\bibfnamefont {S.}~\bibnamefont
  {Chandrasekhar}},\ }\href@noop {} {\emph {\bibinfo {title} {Hydrodynamic and
  Hydromagnetic Stability}}}\ (\bibinfo  {publisher} {Oxford University
  Press},\ \bibinfo {address} {U.K.},\ \bibinfo {year} {1961})\BibitemShut
  {NoStop}%
\bibitem [{\citenamefont {Matthews}(1999)}]{pM99}%
  \BibitemOpen
  \bibfield  {author} {\bibinfo {author} {\bibfnamefont {P.~C.}\ \bibnamefont
  {Matthews}},\ }\bibfield  {title} {\bibinfo {title} {Asymptotic solutions for
  nonlinear magnetoconvection},\ }\href@noop {} {\bibfield  {journal} {\bibinfo
   {journal} {J. Fluid Mech.}\ }\textbf {\bibinfo {volume} {387}},\ \bibinfo
  {pages} {397} (\bibinfo {year} {1999})}\BibitemShut {NoStop}%
\bibitem [{\citenamefont {Julien}\ \emph
  {et~al.}(1999{\natexlab{a}})\citenamefont {Julien}, \citenamefont
  {Knobloch},\ and\ \citenamefont {Tobias}}]{kJ99c}%
  \BibitemOpen
  \bibfield  {author} {\bibinfo {author} {\bibfnamefont {K.}~\bibnamefont
  {Julien}}, \bibinfo {author} {\bibfnamefont {E.}~\bibnamefont {Knobloch}},\
  and\ \bibinfo {author} {\bibfnamefont {S.~M.}\ \bibnamefont {Tobias}},\
  }\bibfield  {title} {\bibinfo {title} {Strongly nonlinear magnetoconvection
  in three dimensions},\ }\href@noop {} {\bibfield  {journal} {\bibinfo
  {journal} {Physica D}\ }\textbf {\bibinfo {volume} {128}},\ \bibinfo {pages}
  {105} (\bibinfo {year} {1999}{\natexlab{a}})}\BibitemShut {NoStop}%
\bibitem [{\citenamefont {Roberts}(2007)}]{roberts2007magnetohydrodynamics}%
  \BibitemOpen
  \bibfield  {author} {\bibinfo {author} {\bibfnamefont {P.~H.}\ \bibnamefont
  {Roberts}},\ }\bibfield  {title} {\bibinfo {title} {Magnetohydrodynamics},\
  }in\ \href@noop {} {\emph {\bibinfo {booktitle} {Encyclopedia of Geomagnetism
  and Paleomagnetism}}}\ (\bibinfo  {publisher} {Springer},\ \bibinfo {year}
  {2007})\ pp.\ \bibinfo {pages} {639--654}\BibitemShut {NoStop}%
\bibitem [{\citenamefont {Grossmann}\ and\ \citenamefont {Lohse}(2000)}]{sG00}%
  \BibitemOpen
  \bibfield  {author} {\bibinfo {author} {\bibfnamefont {S.}~\bibnamefont
  {Grossmann}}\ and\ \bibinfo {author} {\bibfnamefont {D.}~\bibnamefont
  {Lohse}},\ }\bibfield  {title} {\bibinfo {title} {Scaling in thermal
  convection: a unifying theory},\ }\href@noop {} {\bibfield  {journal}
  {\bibinfo  {journal} {J. Fluid Mech.}\ }\textbf {\bibinfo {volume} {407}},\
  \bibinfo {pages} {27} (\bibinfo {year} {2000})}\BibitemShut {NoStop}%
\bibitem [{\citenamefont {Zuerner}\ \emph {et~al.}(2016)\citenamefont
  {Zuerner}, \citenamefont {Liu}, \citenamefont {Krasnov},\ and\ \citenamefont
  {Schumacher}}]{zu16}%
  \BibitemOpen
  \bibfield  {author} {\bibinfo {author} {\bibfnamefont {T.}~\bibnamefont
  {Zuerner}}, \bibinfo {author} {\bibfnamefont {W.}~\bibnamefont {Liu}},
  \bibinfo {author} {\bibfnamefont {D.}~\bibnamefont {Krasnov}},\ and\ \bibinfo
  {author} {\bibfnamefont {J.}~\bibnamefont {Schumacher}},\ }\bibfield  {title}
  {\bibinfo {title} {Heat and momentum transfer for magnetoconvection in a
  vertical external magnetic field},\ }\href@noop {} {\bibfield  {journal}
  {\bibinfo  {journal} {\pre}\ }\textbf {\bibinfo {volume} {94}},\ \bibinfo
  {pages} {043108} (\bibinfo {year} {2016})}\BibitemShut {NoStop}%
\bibitem [{\citenamefont {Z{\"u}rner}(2020)}]{zu20}%
  \BibitemOpen
  \bibfield  {author} {\bibinfo {author} {\bibfnamefont {T.}~\bibnamefont
  {Z{\"u}rner}},\ }\bibfield  {title} {\bibinfo {title} {Refined mean field
  model of heat and momentum transfer in magnetoconvection},\ }\href@noop {}
  {\bibfield  {journal} {\bibinfo  {journal} {Phys. Fluids}\ }\textbf {\bibinfo
  {volume} {32}} (\bibinfo {year} {2020})}\BibitemShut {NoStop}%
\bibitem [{\citenamefont {Bhattacharyya}(2006)}]{sB06}%
  \BibitemOpen
  \bibfield  {author} {\bibinfo {author} {\bibfnamefont {S.~N.}\ \bibnamefont
  {Bhattacharyya}},\ }\bibfield  {title} {\bibinfo {title} {{Scaling in
  magnetohydrodynamic convection at high Rayleigh number}},\ }\href@noop {}
  {\bibfield  {journal} {\bibinfo  {journal} {Phys. Rev. E}\ }\textbf {\bibinfo
  {volume} {74}},\ \bibinfo {pages} {035301} (\bibinfo {year}
  {2006})}\BibitemShut {NoStop}%
\bibitem [{\citenamefont {Jones}\ and\ \citenamefont {Roberts}(2000)}]{cJ00b}%
  \BibitemOpen
  \bibfield  {author} {\bibinfo {author} {\bibfnamefont {C.~A.}\ \bibnamefont
  {Jones}}\ and\ \bibinfo {author} {\bibfnamefont {P.~H.}\ \bibnamefont
  {Roberts}},\ }\bibfield  {title} {\bibinfo {title} {Convection-driven dynamos
  in a rotating plane layer},\ }\href@noop {} {\bibfield  {journal} {\bibinfo
  {journal} {J. Fluid Mech.}\ }\textbf {\bibinfo {volume} {404}},\ \bibinfo
  {pages} {311} (\bibinfo {year} {2000})}\BibitemShut {NoStop}%
\bibitem [{\citenamefont {Marti}\ \emph {et~al.}(2016)\citenamefont {Marti},
  \citenamefont {Calkins},\ and\ \citenamefont {Julien}}]{pM16}%
  \BibitemOpen
  \bibfield  {author} {\bibinfo {author} {\bibfnamefont {P.}~\bibnamefont
  {Marti}}, \bibinfo {author} {\bibfnamefont {M.~A.}\ \bibnamefont {Calkins}},\
  and\ \bibinfo {author} {\bibfnamefont {K.}~\bibnamefont {Julien}},\
  }\bibfield  {title} {\bibinfo {title} {A computationally efficent spectral
  method for modeling core dynamics},\ }\href@noop {} {\bibfield  {journal}
  {\bibinfo  {journal} {Geochem. Geophys. Geosys.}\ }\textbf {\bibinfo {volume}
  {17}},\ \bibinfo {pages} {3031} (\bibinfo {year} {2016})}\BibitemShut
  {NoStop}%
\bibitem [{\citenamefont {Bender}\ and\ \citenamefont {Orszag}(2010)}]{cB10}%
  \BibitemOpen
  \bibfield  {author} {\bibinfo {author} {\bibfnamefont {C.~M.}\ \bibnamefont
  {Bender}}\ and\ \bibinfo {author} {\bibfnamefont {S.~A.}\ \bibnamefont
  {Orszag}},\ }\href@noop {} {\emph {\bibinfo {title} {Advanced mathematical
  methods for scientists and engineers I: Asymptotic methods and perturbation
  theory}}}\ (\bibinfo  {publisher} {Springer},\ \bibinfo {address} {New
  York},\ \bibinfo {year} {2010})\BibitemShut {NoStop}%
\bibitem [{\citenamefont {Heard}\ and\ \citenamefont {Veronis}(1971)}]{wH71}%
  \BibitemOpen
  \bibfield  {author} {\bibinfo {author} {\bibfnamefont {W.~B.}\ \bibnamefont
  {Heard}}\ and\ \bibinfo {author} {\bibfnamefont {G.}~\bibnamefont
  {Veronis}},\ }\bibfield  {title} {\bibinfo {title} {Asymptotic treatment of
  the stability of a rotating layer of fluid with rigid boundaries},\
  }\href@noop {} {\bibfield  {journal} {\bibinfo  {journal} {Geophys. Fluid
  Dyn.}\ }\textbf {\bibinfo {volume} {2}},\ \bibinfo {pages} {299} (\bibinfo
  {year} {1971})}\BibitemShut {NoStop}%
\bibitem [{\citenamefont {Calkins}\ \emph {et~al.}(2015)\citenamefont
  {Calkins}, \citenamefont {Hale}, \citenamefont {Julien}, \citenamefont
  {Nieves}, \citenamefont {Driggs},\ and\ \citenamefont {Marti}}]{mC15c}%
  \BibitemOpen
  \bibfield  {author} {\bibinfo {author} {\bibfnamefont {M.~A.}\ \bibnamefont
  {Calkins}}, \bibinfo {author} {\bibfnamefont {K.}~\bibnamefont {Hale}},
  \bibinfo {author} {\bibfnamefont {K.}~\bibnamefont {Julien}}, \bibinfo
  {author} {\bibfnamefont {D.}~\bibnamefont {Nieves}}, \bibinfo {author}
  {\bibfnamefont {D.}~\bibnamefont {Driggs}},\ and\ \bibinfo {author}
  {\bibfnamefont {P.}~\bibnamefont {Marti}},\ }\bibfield  {title} {\bibinfo
  {title} {The asymptotic equivalence of fixed heat flux and fixed temperature
  thermal boundary conditions for rapidly rotating convection},\ }\href@noop {}
  {\bibfield  {journal} {\bibinfo  {journal} {J. Fluid Mech.}\ }\textbf
  {\bibinfo {volume} {784}} (\bibinfo {year} {2015})}\BibitemShut {NoStop}%
\bibitem [{\citenamefont {Julien}\ \emph {et~al.}(2016)\citenamefont {Julien},
  \citenamefont {Aurnou}, \citenamefont {Calkins}, \citenamefont {Knobloch},
  \citenamefont {Marti}, \citenamefont {Stellmach},\ and\ \citenamefont
  {Vasil}}]{kJ16}%
  \BibitemOpen
  \bibfield  {author} {\bibinfo {author} {\bibfnamefont {K.}~\bibnamefont
  {Julien}}, \bibinfo {author} {\bibfnamefont {J.~M.}\ \bibnamefont {Aurnou}},
  \bibinfo {author} {\bibfnamefont {M.~A.}\ \bibnamefont {Calkins}}, \bibinfo
  {author} {\bibfnamefont {E.}~\bibnamefont {Knobloch}}, \bibinfo {author}
  {\bibfnamefont {P.}~\bibnamefont {Marti}}, \bibinfo {author} {\bibfnamefont
  {S.}~\bibnamefont {Stellmach}},\ and\ \bibinfo {author} {\bibfnamefont
  {G.~M.}\ \bibnamefont {Vasil}},\ }\bibfield  {title} {\bibinfo {title} {{A
  nonlinear model for rotationally constrained convection with Ekman
  pumping}},\ }\href@noop {} {\bibfield  {journal} {\bibinfo  {journal} {J.
  Fluid Mech.}\ }\textbf {\bibinfo {volume} {798}},\ \bibinfo {pages} {50}
  (\bibinfo {year} {2016})}\BibitemShut {NoStop}%
\bibitem [{\citenamefont {Chini}\ and\ \citenamefont
  {Cox}(2009)}]{chini2009large}%
  \BibitemOpen
  \bibfield  {author} {\bibinfo {author} {\bibfnamefont {G.~P.}\ \bibnamefont
  {Chini}}\ and\ \bibinfo {author} {\bibfnamefont {S.~M.}\ \bibnamefont
  {Cox}},\ }\bibfield  {title} {\bibinfo {title} {{Large Rayleigh number
  thermal convection: heat flux predictions and strongly nonlinear
  solutions}},\ }\href@noop {} {\bibfield  {journal} {\bibinfo  {journal}
  {Phys. Fluids}\ }\textbf {\bibinfo {volume} {21}} (\bibinfo {year}
  {2009})}\BibitemShut {NoStop}%
\bibitem [{\citenamefont {Long}\ \emph {et~al.}(2020)\citenamefont {Long},
  \citenamefont {Mound}, \citenamefont {Davies},\ and\ \citenamefont
  {Tobias}}]{long2020thermal}%
  \BibitemOpen
  \bibfield  {author} {\bibinfo {author} {\bibfnamefont {R.~S.}\ \bibnamefont
  {Long}}, \bibinfo {author} {\bibfnamefont {J.~E.}\ \bibnamefont {Mound}},
  \bibinfo {author} {\bibfnamefont {C.~J.}\ \bibnamefont {Davies}},\ and\
  \bibinfo {author} {\bibfnamefont {S.~M.}\ \bibnamefont {Tobias}},\ }\bibfield
   {title} {\bibinfo {title} {Thermal boundary layer structure in convection
  with and without rotation},\ }\href@noop {} {\bibfield  {journal} {\bibinfo
  {journal} {\prf}\ }\textbf {\bibinfo {volume} {5}},\ \bibinfo {pages}
  {113502} (\bibinfo {year} {2020})}\BibitemShut {NoStop}%
\bibitem [{\citenamefont {Julien}\ \emph {et~al.}(1996)\citenamefont {Julien},
  \citenamefont {Legg}, \citenamefont {McWilliams},\ and\ \citenamefont
  {Werne}}]{kJ96}%
  \BibitemOpen
  \bibfield  {author} {\bibinfo {author} {\bibfnamefont {K.}~\bibnamefont
  {Julien}}, \bibinfo {author} {\bibfnamefont {S.}~\bibnamefont {Legg}},
  \bibinfo {author} {\bibfnamefont {J.}~\bibnamefont {McWilliams}},\ and\
  \bibinfo {author} {\bibfnamefont {J.}~\bibnamefont {Werne}},\ }\bibfield
  {title} {\bibinfo {title} {{Rapidly rotating turbulent Rayleigh-B\'enard
  convection}},\ }\href@noop {} {\bibfield  {journal} {\bibinfo  {journal} {J.
  Fluid Mech.}\ }\textbf {\bibinfo {volume} {322}},\ \bibinfo {pages} {243}
  (\bibinfo {year} {1996})}\BibitemShut {NoStop}%
\bibitem [{\citenamefont {Davidson}(2001)}]{pD01}%
  \BibitemOpen
  \bibfield  {author} {\bibinfo {author} {\bibfnamefont {P.~A.}\ \bibnamefont
  {Davidson}},\ }\href@noop {} {\emph {\bibinfo {title} {An Introduction to
  Magnetohydrodynamics}}}\ (\bibinfo  {publisher} {Cambridge University
  Press},\ \bibinfo {address} {Cambridge},\ \bibinfo {year} {2001})\BibitemShut
  {NoStop}%
\bibitem [{\citenamefont {Julien}\ \emph
  {et~al.}(1999{\natexlab{b}})\citenamefont {Julien}, \citenamefont
  {Knobloch},\ and\ \citenamefont {Werne}}]{kJ99}%
  \BibitemOpen
  \bibfield  {author} {\bibinfo {author} {\bibfnamefont {K.}~\bibnamefont
  {Julien}}, \bibinfo {author} {\bibfnamefont {E.}~\bibnamefont {Knobloch}},\
  and\ \bibinfo {author} {\bibfnamefont {J.}~\bibnamefont {Werne}},\ }\bibfield
   {title} {\bibinfo {title} {Reduced equations for rotationally constrained
  convection},\ }in\ \href@noop {} {\emph {\bibinfo {booktitle} {Turbulence and
  Shear Flows - I}}},\ \bibinfo {editor} {edited by\ \bibinfo {editor}
  {\bibfnamefont {S.}~\bibnamefont {Banerjee}}\ and\ \bibinfo {editor}
  {\bibfnamefont {J.~K.}\ \bibnamefont {Eaton}}}\ (\bibinfo  {publisher}
  {Begell House},\ \bibinfo {address} {New York},\ \bibinfo {year} {1999})\
  pp.\ \bibinfo {pages} {101--106}\BibitemShut {NoStop}%
\bibitem [{\citenamefont {Sprague}\ \emph {et~al.}(2006)\citenamefont
  {Sprague}, \citenamefont {Julien}, \citenamefont {Knobloch},\ and\
  \citenamefont {Werne}}]{mS06}%
  \BibitemOpen
  \bibfield  {author} {\bibinfo {author} {\bibfnamefont {M.}~\bibnamefont
  {Sprague}}, \bibinfo {author} {\bibfnamefont {K.}~\bibnamefont {Julien}},
  \bibinfo {author} {\bibfnamefont {E.}~\bibnamefont {Knobloch}},\ and\
  \bibinfo {author} {\bibfnamefont {J.}~\bibnamefont {Werne}},\ }\bibfield
  {title} {\bibinfo {title} {Numerical simulation of an asymptotically reduced
  system for rotationally constrained convection},\ }\href@noop {} {\bibfield
  {journal} {\bibinfo  {journal} {J. Fluid Mech.}\ }\textbf {\bibinfo {volume}
  {551}},\ \bibinfo {pages} {141} (\bibinfo {year} {2006})}\BibitemShut
  {NoStop}%
\bibitem [{\citenamefont {Stellmach}\ \emph {et~al.}(2014)\citenamefont
  {Stellmach}, \citenamefont {Lischper}, \citenamefont {Julien}, \citenamefont
  {Vasil}, \citenamefont {Cheng}, \citenamefont {Ribeiro}, \citenamefont
  {King},\ and\ \citenamefont {Aurnou}}]{sS14}%
  \BibitemOpen
  \bibfield  {author} {\bibinfo {author} {\bibfnamefont {S.}~\bibnamefont
  {Stellmach}}, \bibinfo {author} {\bibfnamefont {M.}~\bibnamefont {Lischper}},
  \bibinfo {author} {\bibfnamefont {K.}~\bibnamefont {Julien}}, \bibinfo
  {author} {\bibfnamefont {G.}~\bibnamefont {Vasil}}, \bibinfo {author}
  {\bibfnamefont {J.~S.}\ \bibnamefont {Cheng}}, \bibinfo {author}
  {\bibfnamefont {A.}~\bibnamefont {Ribeiro}}, \bibinfo {author} {\bibfnamefont
  {E.~M.}\ \bibnamefont {King}},\ and\ \bibinfo {author} {\bibfnamefont
  {J.~M.}\ \bibnamefont {Aurnou}},\ }\bibfield  {title} {\bibinfo {title}
  {Approaching the asymptotic regime of rapidly rotating convection: boundary
  layers versus interior dynamics},\ }\href@noop {} {\bibfield  {journal}
  {\bibinfo  {journal} {Phys. Rev. Lett.}\ }\textbf {\bibinfo {volume} {113}},\
  \bibinfo {pages} {254501} (\bibinfo {year} {2014})}\BibitemShut {NoStop}%
\bibitem [{\citenamefont {van Kan}\ \emph {et~al.}(2025)\citenamefont {van
  Kan}, \citenamefont {Julien}, \citenamefont {Miquel},\ and\ \citenamefont
  {Knobloch}}]{aK25}%
  \BibitemOpen
  \bibfield  {author} {\bibinfo {author} {\bibfnamefont {A.}~\bibnamefont {van
  Kan}}, \bibinfo {author} {\bibfnamefont {K.}~\bibnamefont {Julien}}, \bibinfo
  {author} {\bibfnamefont {B.}~\bibnamefont {Miquel}},\ and\ \bibinfo {author}
  {\bibfnamefont {E.}~\bibnamefont {Knobloch}},\ }\bibfield  {title} {\bibinfo
  {title} {{Bridging the Rossby number gap in rapidly rotating thermal
  convection}},\ }\href@noop {} {\bibfield  {journal} {\bibinfo  {journal}
  {\jfm}\ }\textbf {\bibinfo {volume} {1010}},\ \bibinfo {pages} {A42}
  (\bibinfo {year} {2025})}\BibitemShut {NoStop}%
\bibitem [{\citenamefont {Julien}\ \emph {et~al.}(2012)\citenamefont {Julien},
  \citenamefont {Rubio}, \citenamefont {Grooms},\ and\ \citenamefont
  {Knobloch}}]{kJ12}%
  \BibitemOpen
  \bibfield  {author} {\bibinfo {author} {\bibfnamefont {K.}~\bibnamefont
  {Julien}}, \bibinfo {author} {\bibfnamefont {A.~M.}\ \bibnamefont {Rubio}},
  \bibinfo {author} {\bibfnamefont {I.}~\bibnamefont {Grooms}},\ and\ \bibinfo
  {author} {\bibfnamefont {E.}~\bibnamefont {Knobloch}},\ }\bibfield  {title}
  {\bibinfo {title} {{Statistical and physical balances in low Rossby number
  Rayleigh-B\'enard convection}},\ }\href@noop {} {\bibfield  {journal}
  {\bibinfo  {journal} {Geophys. Astrophys. Fluid Dyn.}\ }\textbf {\bibinfo
  {volume} {106}},\ \bibinfo {pages} {392} (\bibinfo {year}
  {2012})}\BibitemShut {NoStop}%
\bibitem [{\citenamefont {Yan}\ \emph {et~al.}(2021)\citenamefont {Yan},
  \citenamefont {Tobias},\ and\ \citenamefont {Calkins}}]{mY21}%
  \BibitemOpen
  \bibfield  {author} {\bibinfo {author} {\bibfnamefont {M.}~\bibnamefont
  {Yan}}, \bibinfo {author} {\bibfnamefont {S.~M.}\ \bibnamefont {Tobias}},\
  and\ \bibinfo {author} {\bibfnamefont {M.~A.}\ \bibnamefont {Calkins}},\
  }\bibfield  {title} {\bibinfo {title} {{Scaling behaviour of small-scale
  dynamos driven by Rayleigh--B{\'e}nard convection}},\ }\href@noop {}
  {\bibfield  {journal} {\bibinfo  {journal} {J. Fluid Mech.}\ }\textbf
  {\bibinfo {volume} {915}},\ \bibinfo {pages} {A15} (\bibinfo {year}
  {2021})}\BibitemShut {NoStop}%
\bibitem [{\citenamefont {Yan}\ and\ \citenamefont {Calkins}(2022)}]{mY22}%
  \BibitemOpen
  \bibfield  {author} {\bibinfo {author} {\bibfnamefont {M.}~\bibnamefont
  {Yan}}\ and\ \bibinfo {author} {\bibfnamefont {M.~A.}\ \bibnamefont
  {Calkins}},\ }\bibfield  {title} {\bibinfo {title} {Strong large scale
  magnetic fields in rotating convection-driven dynamos: The important role of
  magnetic diffusion},\ }\href@noop {} {\bibfield  {journal} {\bibinfo
  {journal} {\prr}\ }\textbf {\bibinfo {volume} {4}},\ \bibinfo {pages}
  {L012026} (\bibinfo {year} {2022})}\BibitemShut {NoStop}%
\bibitem [{\citenamefont {Childress}\ and\ \citenamefont
  {Soward}(1972)}]{sC72}%
  \BibitemOpen
  \bibfield  {author} {\bibinfo {author} {\bibfnamefont {S.}~\bibnamefont
  {Childress}}\ and\ \bibinfo {author} {\bibfnamefont {A.~M.}\ \bibnamefont
  {Soward}},\ }\bibfield  {title} {\bibinfo {title} {Convection-driven
  hydromagnetic dynamo},\ }\href@noop {} {\bibfield  {journal} {\bibinfo
  {journal} {Phys. Rev. Lett.}\ }\textbf {\bibinfo {volume} {29}},\ \bibinfo
  {pages} {837} (\bibinfo {year} {1972})}\BibitemShut {NoStop}%
\bibitem [{\citenamefont {Soward}(1974)}]{aS74}%
  \BibitemOpen
  \bibfield  {author} {\bibinfo {author} {\bibfnamefont {A.~M.}\ \bibnamefont
  {Soward}},\ }\bibfield  {title} {\bibinfo {title} {A convection-driven
  dynamo: I. the weak field case},\ }\href@noop {} {\bibfield  {journal}
  {\bibinfo  {journal} {Phil. Trans. R. Soc. Lond. A}\ }\textbf {\bibinfo
  {volume} {275}},\ \bibinfo {pages} {611} (\bibinfo {year}
  {1974})}\BibitemShut {NoStop}%
\bibitem [{\citenamefont {Horn}\ and\ \citenamefont {Aurnou}(2022)}]{sH22}%
  \BibitemOpen
  \bibfield  {author} {\bibinfo {author} {\bibfnamefont {S.}~\bibnamefont
  {Horn}}\ and\ \bibinfo {author} {\bibfnamefont {J.~M.}\ \bibnamefont
  {Aurnou}},\ }\bibfield  {title} {\bibinfo {title} {{The Elbert range of
  magnetostrophic convection. I. Linear theory}},\ }\href@noop {} {\bibfield
  {journal} {\bibinfo  {journal} {\prsa}\ }\textbf {\bibinfo {volume} {478}},\
  \bibinfo {pages} {20220313} (\bibinfo {year} {2022})}\BibitemShut {NoStop}%
\bibitem [{\citenamefont {Horn}\ and\ \citenamefont {Aurnou}(2025)}]{sH25}%
  \BibitemOpen
  \bibfield  {author} {\bibinfo {author} {\bibfnamefont {S.}~\bibnamefont
  {Horn}}\ and\ \bibinfo {author} {\bibfnamefont {J.~M.}\ \bibnamefont
  {Aurnou}},\ }\bibfield  {title} {\bibinfo {title} {{The Elbert range of
  magnetostrophic convection. II. Comparing linear theory to nonlinear low-Rm
  simulations}},\ }\href@noop {} {\bibfield  {journal} {\bibinfo  {journal}
  {\prsa}\ }\textbf {\bibinfo {volume} {481}},\ \bibinfo {pages} {20240016}
  (\bibinfo {year} {2025})}\BibitemShut {NoStop}%
\end{thebibliography}%

\end{document}